\documentclass[11pt,a4paper]{article}
\pdfoutput=1
\usepackage[margin=1.5cm,includefoot,footskip=30pt]{geometry}
\usepackage[ascii]{inputenc}
\usepackage[bookmarksnumbered,linktocpage,hypertexnames=false,colorlinks=true,linkcolor=blue,urlcolor=blue,citecolor=blue,anchorcolor=green,breaklinks=true,pdfusetitle]{hyperref}
\usepackage{bbm,braket,microtype,mathrsfs,amsmath,amssymb,xcolor,amsthm,amsfonts,latexsym,mathtools,graphicx,enumitem,booktabs,bm,xspace,float,mathdots,caption,subcaption,ellipsis,enumitem,mleftright,stmaryrd,overpic}
\usepackage{authblk}
\usepackage[longend,ruled,linesnumbered,nokwfunc]{algorithm2e}
\SetKwInput{Input}{Input}
\SetKwInput{Output}{Output}
\SetKwFor{Loop}{loop}{}{end loop}
\usepackage[T1]{fontenc}
\usepackage{textcomp}
\usepackage[english]{babel}
\usepackage[capitalize,nameinlink]{cleveref}
\newtheorem{thm}{Theorem}[section]\crefname{thm}{Theorem}{Theorems}
\newtheorem{lem}[thm]{Lemma}
\newtheorem{prp}[thm]{Proposition}
\newtheorem{cor}[thm]{Corollary}

\newtheorem*{thm*}{Theorem}
\newtheorem*{prp*}{Proposition}
\newtheorem{res}{Result}\crefname{res}{Result}{Results}
\theoremstyle{definition}
\newtheorem{dfn}[thm]{Definition}
\newtheorem{rem}[thm]{Remark}
\newtheorem{exa}[thm]{Example}\crefname{exa}{Example}{Examples}
\DeclareMathOperator{\tr}{tr}
\DeclareMathOperator{\poly}{poly}

\DeclareMathOperator{\diag}{diag}

\DeclareMathOperator{\supp}{supp}
\DeclareMathOperator{\rank}{rank}
\DeclareMathOperator{\Lie}{Lie}
\DeclareMathOperator{\End}{End}
\DeclareMathOperator{\U}{U}
\DeclareMathOperator{\GL}{GL}
\DeclareMathOperator{\SL}{SL}

\DeclareMathOperator{\PD}{PD}

\DeclareMathOperator{\argmin}{arg\,min}
\DeclareMathOperator{\Mat}{Mat}
\DeclareMathOperator{\Herm}{Herm}
\DeclareMathOperator{\phys}{phys}
\renewcommand{\vec}[1]{\mathbold{#1}}

\newcommand{\proj}[1]{\ket{#1}\!\bra{#1}}
\renewcommand{\Re}{\operatorname{Re}}
\newcommand{\R}{\mathbbm R}
\newcommand{\C}{\mathbbm C}
\newcommand{\N}{\mathbbm N}
\newcommand{\Z}{\mathbbm Z}
\newcommand{\Q}{\mathbbm Q}

\newcommand{\HH}{\mathcal H}
\newcommand{\PP}{\mathbb{P}}

\newcommand{\ot}{\otimes}
\newcommand{\op}{\oplus}
\newcommand{\eps}{\varepsilon}

\newcommand{\grad}{\nabla}

\renewcommand{\vec}[1]{\boldsymbol{#1}}
\newcommand{\ketbra}[2]{\mathinner{|{#1}\rangle\,\langle{#2}|}}

\newcommand{\mcf}[1]{#1_{\mathrm{min}}}
\DeclareMathOperator{\conv}{conv}
\DeclareMathOperator{\adj}{adj}
\newcommand{\inpsize}[1]{\langle#1\rangle}
\newcommand{\Cstar}{\C^*}
\DeclarePairedDelimiter\abs{\lvert}{\rvert}
\DeclarePairedDelimiter\norm{\lVert}{\rVert_2}
\DeclarePairedDelimiter\opnorm{\lVert}{\rVert_{\infty}}

\allowdisplaybreaks[4]
\numberwithin{equation}{section}
\graphicspath{{./figures/}}
\addto\captionsenglish{}

\begin{document}
\title{The minimal canonical form of a tensor network}
\author[1]{Arturo Acuaviva}
\author[2]{Visu Makam}
\author[3]{Harold Nieuwboer}
\author[1]{David P\'erez-Garc\'ia}
\author[]{Friedrich Sittner}
\author[4]{Michael Walter}
\author[5]{Freek Witteveen}
\affil[1]{Departamento de An\'alisis Matem\'atico y Matem\'atica Aplicada, Universidad Complutense de Madrid, Spain}
\affil[2]{Radix Trading Europe B.V., Amsterdam, The Netherlands}
\affil[3]{Korteweg-de Vries Institute for Mathematics and QuSoft, University of Amsterdam, The Netherlands}
\affil[4]{Faculty of Computer Science, Ruhr University Bochum, Germany}
\affil[5]{Department of Mathematical Sciences and QMATH, University of Copenhagen, Denmark}
\date{}
\maketitle
\vspace{-13mm}
\begin{abstract}
Tensor networks have a gauge degree of freedom on the virtual degrees of freedom that are contracted.
A \emph{canonical form} is a choice of fixing this degree of freedom.
For matrix product states, choosing a canonical form is a powerful tool, both for theoretical and numerical purposes.
On the other hand, for tensor networks in dimension two or greater there is only limited understanding of the gauge symmetry.
Here we introduce a new canonical form, the \emph{minimal canonical form}, which applies to projected entangled pair states (PEPS) in any dimension, and prove a corresponding fundamental theorem.
Already for matrix product states this gives a new canonical form, while in higher dimensions it is the first rigorous definition of a canonical form valid for any choice of tensor.
We show that two tensors have the same minimal canonical forms if and only if they are gauge equivalent up to taking limits; moreover, this is the case if and only if they give the same quantum state for \emph{any} geometry.
In particular, this implies that the latter problem is decidable -- in contrast to the well-known undecidability for PEPS on grids.
We also provide rigorous algorithms for computing minimal canonical forms.
To achieve this we draw on geometric invariant theory and recent progress in theoretical computer science in non-commutative group optimization.
\end{abstract}
\vspace{-4mm}
\tableofcontents

\section{Introduction}
Tensor networks are a fruitful area of interconnection between quantum information theory and quantum many-body physics.
On the one hand, tensor network states are rich enough to approximate with high accuracy most states which are relevant in condensed matter physics, such as Gibbs states and ground states.
On the other hand, tensor networks are sufficiently simple that they enable one to manipulate complex quantum states, both numerically and theoretically.
For the purpose of numerics, one can design variational optimization algorithms to simulate strongly interacting quantum systems.
On the other side of the spectrum, tensor networks have been a powerful theoretical method to obtain simple characterizations of complex global phenomena like topological order.

\begin{figure}
\centering
\begin{subfigure}[t]{0.45\textwidth}
\begin{overpic}[width=0.9\textwidth,grid=false]{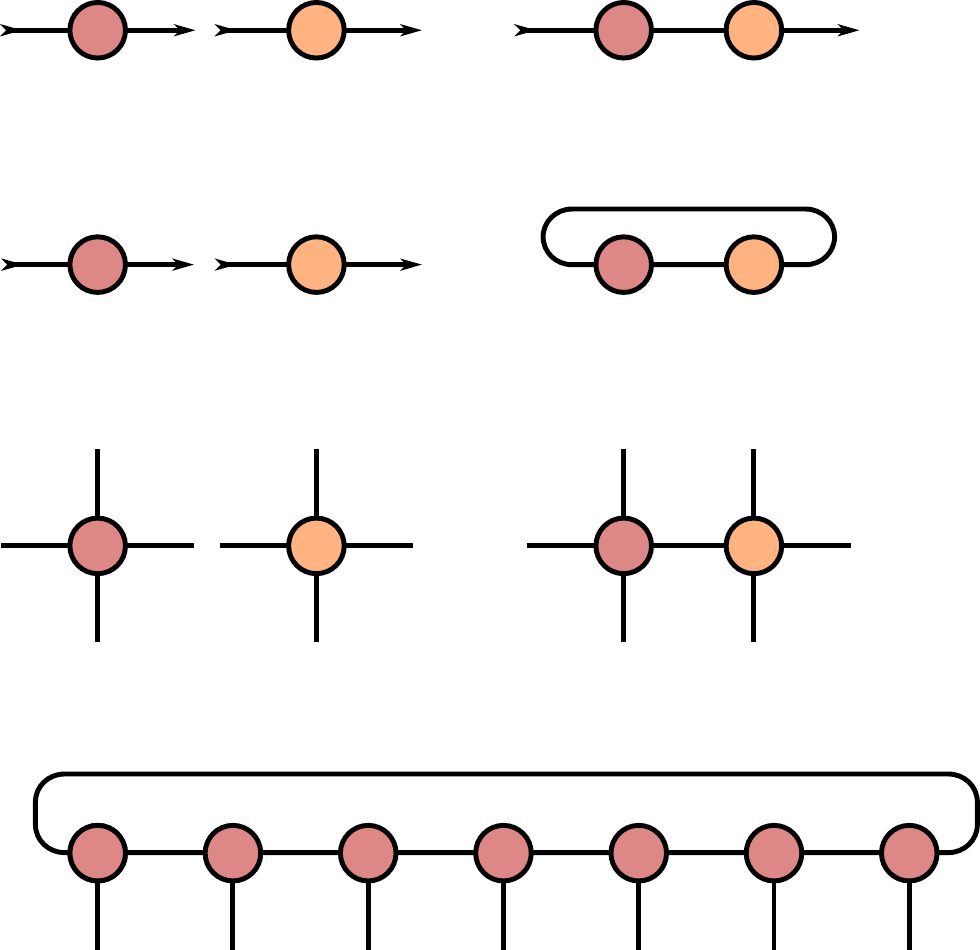}
\put(8,84){$A_{ij}$} \put(31,84){$B_{lk}$}
\put(3,96){\footnotesize{$i$}} \put(15,96){\footnotesize{$j$}}
\put(25,96){\footnotesize{$l$}} \put(37,96){\footnotesize{$k$}}
\put(62,84){$\sum_j A_{ij}B_{jk}$}
\put(57,96){\footnotesize{$i$}} \put(82,96){\footnotesize{$k$}}
\put(47,93){$\Rightarrow$}

\put(8,60){$A_{ij}$} \put(31,60){$B_{ji}$}
\put(3,72){\footnotesize{$i$}} \put(15,72){\footnotesize{$j$}}
\put(25,72){\footnotesize{$k$}} \put(37,72){\footnotesize{$l$}}
\put(55,60){$\tr[AB] = \sum_{i,j} A_{ij}B_{ji}$}
\put(47,69){$\Rightarrow$}

\put(3,43){\footnotesize{$i_1$}} \put(15,43){\footnotesize{$j$}}
\put(25,43){\footnotesize{$j$}} \put(37,43){\footnotesize{$k_2$}}
\put(57,43){\footnotesize{$i_1$}}
\put(11,50){\footnotesize{$i_2$}} \put(11,32){\footnotesize{$i_3$}}
\put(33,50){\footnotesize{$k_1$}} \put(33,32){\footnotesize{$k_3$}}
\put(65,50){\footnotesize{$i_2$}} \put(65,32){\footnotesize{$i_3$}}
\put(78,50){\footnotesize{$k_1$}} \put(78,32){\footnotesize{$k_3$}}
\put(82,43){\footnotesize{$k_2$}}
\put(8,25){$A_{i_1 i_2 i_3 j}$} \put(31,25){$B_{jk_1 k_2 k_3}$}
\put(56,25){$\sum_j A_{i_1 i_2 i_3 j}B_{jk_1 k_2 k_3}$}
\put(47,40,5){$\Rightarrow$}
\end{overpic}
\vspace*{0.5cm}
\caption{\emph{Graphical notation:} A reminder of the graphical notation for tensor network contractions.
If the tensors are interpreted as matrices, arrows indicate the direction of multiplication.
The examples include matrix multiplication, the trace of a product of matrices, and in the bottom row, a matrix product state.}
\label{fig:tensor-network-reminder}
\end{subfigure}
\hspace*{0.5cm}
\begin{subfigure}[t]{0.45\textwidth}
\centering
\begin{overpic}[width=0.6\textwidth,grid=false]{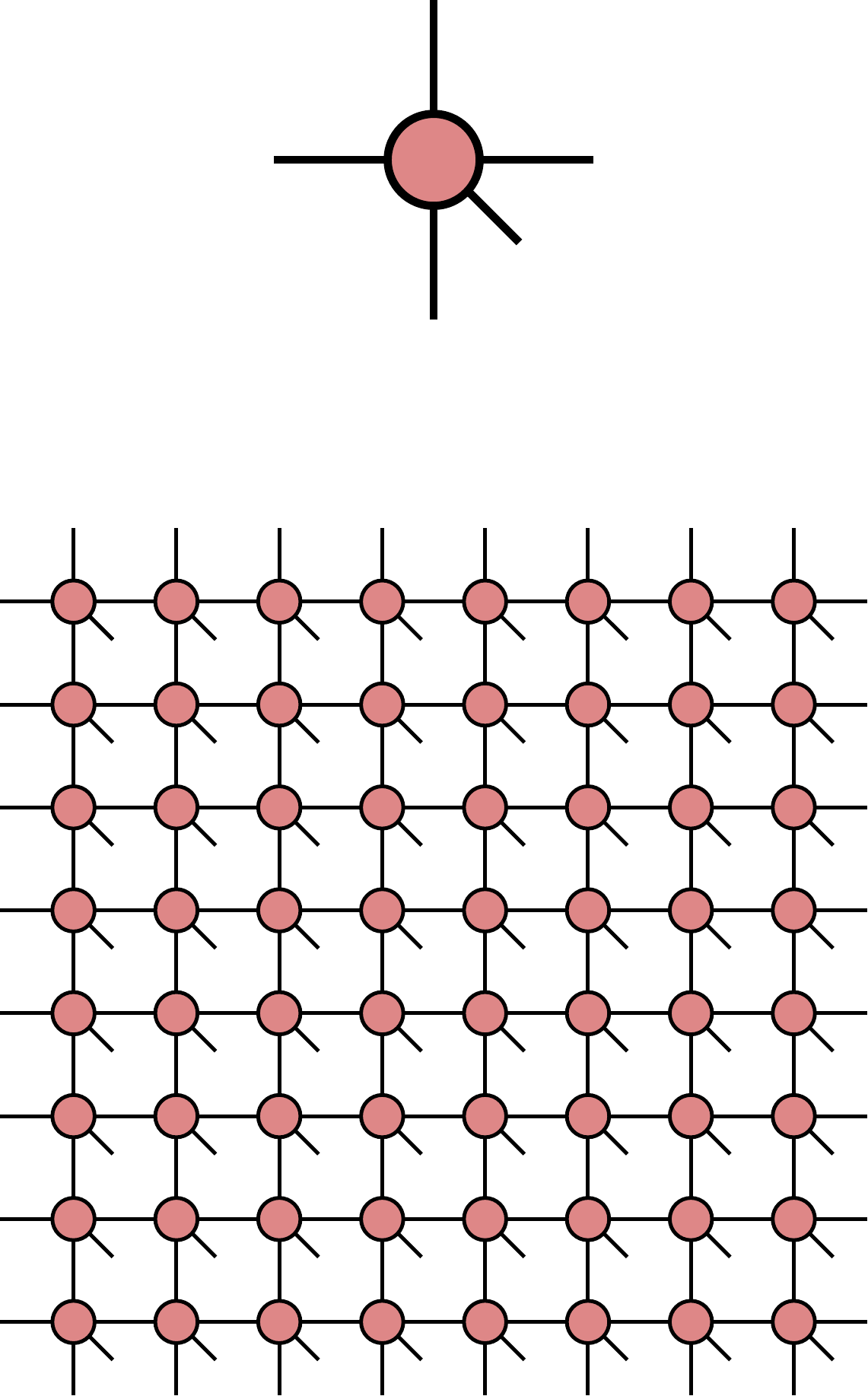}
\put(10,87){\footnotesize{$\C^{D_1}$}} \put(45,87){\footnotesize{$\C^{D_1}$}}
\put(29.5,102){\footnotesize{$\C^{D_2}$}} \put(29.5,71){\footnotesize{$\C^{D_2}$}}
\put(38.5,77.5){\footnotesize{$\C^{d}$}}
\put(17,77){\large{$T$}}
\put(66,30){$\ket{T_{n,m}}$}
\end{overpic}
\vspace*{0.5cm}
\caption{\emph{Uniform PEPS in 2D:}
A tensor $T$ gives rise to states~$\ket{T_{n,m}}$ on a periodic $n\times m$ lattice by placing $T$ at the sites and contracting with periodic boundary conditions.}
\label{fig:two-d-peps}
\end{subfigure}
\caption{}
\label{fig:tensor-network}
\end{figure}

Roughly speaking a tensor network is defined by a set of tensors with two types of indices: virtual ones, whose dimension is called the \emph{bond dimension}, and physical ones, associated to the different subsystems of a quantum many-body system.
These tensors generate a state (called a tensor network state) in the physical Hilbert spaces of the system by contracting the virtual indices on a given graph, typically a lattice associated to the interaction pattern of a Hamiltonian.
The graphical notation for tensor network contractions is briefly reviewed in \cref{fig:tensor-network-reminder}.

The success of tensor network states as a numerical variational family dates back to the pioneering paper \cite{white:DMRG}, where the Density Matrix Renormalization Group (DMRG) algorithm was proposed as a way to approximate ground states of one-dimensional systems.
Nowadays, this algorithm is seen as a way to minimize energy over the manifold of \emph{Matrix Product States} (MPS), the first and most well-known family of tensor networks.
From the perspective of quantum information theory, one may also see MPS as pairs of maximally entangled states to which locally a projection operator is applied.
This allowed the generalization of the construction to more complex scenarios, including higher dimensions \cite{verstraete:dmrg-mps, verstraete:2D-dmrg}.
There, the associated objects are called \emph{Projected Entangled Pair States} (PEPS), precisely due to the perspective of applying projectors to a configuration of maximally entangles states.
By now, there can be no doubt that this is one of the most important and powerful paradigms in numerical simulation of quantum systems \cite{jimenez2021quantum, robaina2021simulating, shi2022discovery, zheng2017stripe}, a recent highlight being the classical simulation \cite{pan2022simulation} of the Google \emph{quantum supremacy experiment} \cite{arute2019quantum}.

On the theoretical side tensor networks allow one to give local characterizations, in terms of their defining tensors, of global properties of interest, such as symmetries or topological order.
The pioneering work \cite{fannes:FCS}, independently from the DMRG proposal \cite{white:DMRG}, started this line of research.
One of the first milestones was the cohomology-based classification of one-dimensional symmetry-protected topological (SPT) phases \cite{chen:1d-phases-rg, pollmann2012symmetry, schuch:mps-phases}.
Today, this is an active area of investigation, see for instance the recent review \cite{cirac2021matrix} for details on the current state of the art.
For instance, tensor networks are used for the characterization of topological order and topological phase transitions in higher spatial dimensions.
Other important theoretical results concern rigorous approximation bounds, showing rigorously that classes of physically relevant states such as ground states and Gibbs states can be approximated accurately by PEPS.

Recently, due to their nice numerical and analytical properties, tensor networks have started to permeate other areas.
Prominent examples are quantum gravity \cite{hayden2016holographic, pastawski2015holographic} and machine learning \cite{stoudenmire2016supervised, cichocki2017tensor}, as well as (hybrid) classical simulation of quantum circuits \cite{peng2020simulating,napp2022efficient}.

An important feature both in theory and practice is the \textbf{gauge symmetry} of a tensor network.
By inserting matrices on the virtual bonds of a tensor in such way that they cancel when the network is contracted, one modifies the local tensors while leaving the many-body state unchanged, see \cref{fig:gauge invariance}.
In this context one desires:
(1)~a \textbf{fundamental theorem} that guarantees the gauge symmetry is the \emph{only} freedom in tensors to give rise to the same states, and
(2)~a \textbf{canonical form}, which \emph{fixes} this gauge degree of freedom in a natural way.
Sometimes, both come together: some fundamental theorems only apply to tensors in a canonical form.

To make this more concrete, we consider PEPS in one spatial dimension, i.e., MPS.
One key reason which make MPS easier to work with than, e.g., 2D PEPS, is that there are canonical forms with good theoretical properties and an associated fundamental theorem~\cite{cirac2021matrix}.
We focus on the \emph{uniform} (or \emph{translation-invariant}) case, where one places the same 3-tensor~$T$ on each site and contracts with periodic boundary conditions, resulting in a many-body quantum state $\ket{T_n}$ for any system size~$n$.
One may view~$T$ as a tripartite quantum state on one physical and two virtual Hilbert spaces, the latter of bond dimension~$D$.
It is always possible (after blocking sites together and setting irrelevant off-diagonals to zero) to choose a gauge such that the reduced state on one of the two virtual Hilbert space is maximally mixed.\footnote{As we will see in \cref{dfn:lrcan}, strictly speaking this is only true independently in each of the diagonal blocks which remain in the canonical form. There is a proportionality constant that can be different in each one of those blocks.}
The result is called a \emph{left or right canonical form} and it is unique up to \emph{unitary} gauge symmetries.
It has the following virtues:
\begin{enumerate}[label=(\Alph*)]
\item\label{it:compare}
It satisfies a \textbf{fundamental theorem}: two tensors $T$ and $T'$ give rise to the same states on any number of sites, meaning~$\ket{T_n} = \ket{T'_n}$ for all~$n$, if and only if they have a common canonical form.
\item\label{it:symmetry}
It allows \textbf{lifting symmetries}:
if $T$ is in canonical form, any global symmetry $U^{\ot n} \ket{T_n} = \ket{T_n}$ for all~$n$ can be implemented by a \emph{unitary} gauge symmetry on~$T$.
This is key to classifying phases of matter and when studying entanglement spectra/Hamiltonians, to upgrade virtual to physical degrees of freedom.
\item\label{it:truncate}
It provides a way to \textbf{truncate}, which is key for efficient accurate numerics:
given a tensor $T$ with bond dimension $D$, it allows finding a tensor $T'$ of bond dimension $D' < D$ such that $\ket{T'_n} \approx \ket{T_n}$ for all $n$.
\end{enumerate}
Clearly, it would be of great use to extend the theory of canonical forms to tensor networks in two or more spatial dimensions!
However, it is known that there are significant obstructions.
For example~\cite{scarpa2020projected,schuch2020periodic}:
\begin{enumerate}[label={$(\lightning)$},ref={$(\lightning)$}]
\item\label{it:undecidable}
The following problem is \textbf{undecidable}:
Given a PEPS tensor $T$, decide if the associated states~$\ket{T_{n,m}}$ vanish on periodic lattices of any size $n \times m$.
\end{enumerate}
This suggests there should not exist any useful (computable) canonical form generalizing~\ref{it:compare}, since by comparing the canonical forms of~$T$ and the zero tensor one could otherwise decide whether $\ket{T_{n,m}} = 0$ for all $n$ and $m$.
Indeed, before our work, no canonical form was known for PEPS tensor networks in two or more dimensions that applied to general tensors and rigorously satisfied properties such as the above.

On the other hand, a fundamental theorem is known if one restricts, e.g., to the class of \emph{normal}~tensors~\cite{molnar2018normal}.
Moreover, heuristic approaches for canonical forms \cite{evenbly2018gauge, kalis2012fate,lubasch2014unifying,phien2015fast,phien2015infinite} and the truncation problem~\ref{it:truncate} are successfully used in practice to trade off efficient computation and approximation accuracy~\cite{ran2020tensor}.

\subsection{Summary of results: a canonical form in any dimension and a fundamental theorem}
In this work we introduce a new canonical form for general PEPS in arbitrary spatial dimension.
It rigorously satisfies a number of desirable properties --  particularly a \emph{fundamental theorem}.
The obstruction~\ref{it:undecidable} is overcome by the following twist:
roughly speaking, the canonical form captures when two tensors give rise to the same quantum states not just on the torus, but on any surface!
This is achieved by pioneering the application of \emph{geometric invariant theory}, an area of mathematics that studies symmetries, to tensor network theory and drawing on recent theoretical computer science research in \emph{non-commutative group optimization}.%
\footnote{Geometric invariant theory has already been used in quantum information in other contexts, such as in the study of multipartite entanglement \cite{klyachko2002coherent,verstraete2003normal,gour2010all,bryan2018existence}, or in the quantum marginal problem \cite{klyachko2004quantum,daftuar2005quantum,christandl2006spectra,klyachko2006quantum,walter2013entanglement,walter2014multipartite}, but not in the area of tensor networks to the best of our knowledge.}
\begin{figure}
\centering
\begin{subfigure}{0.5\textwidth}
\centering
\begin{overpic}[width=0.95\textwidth,grid=false]{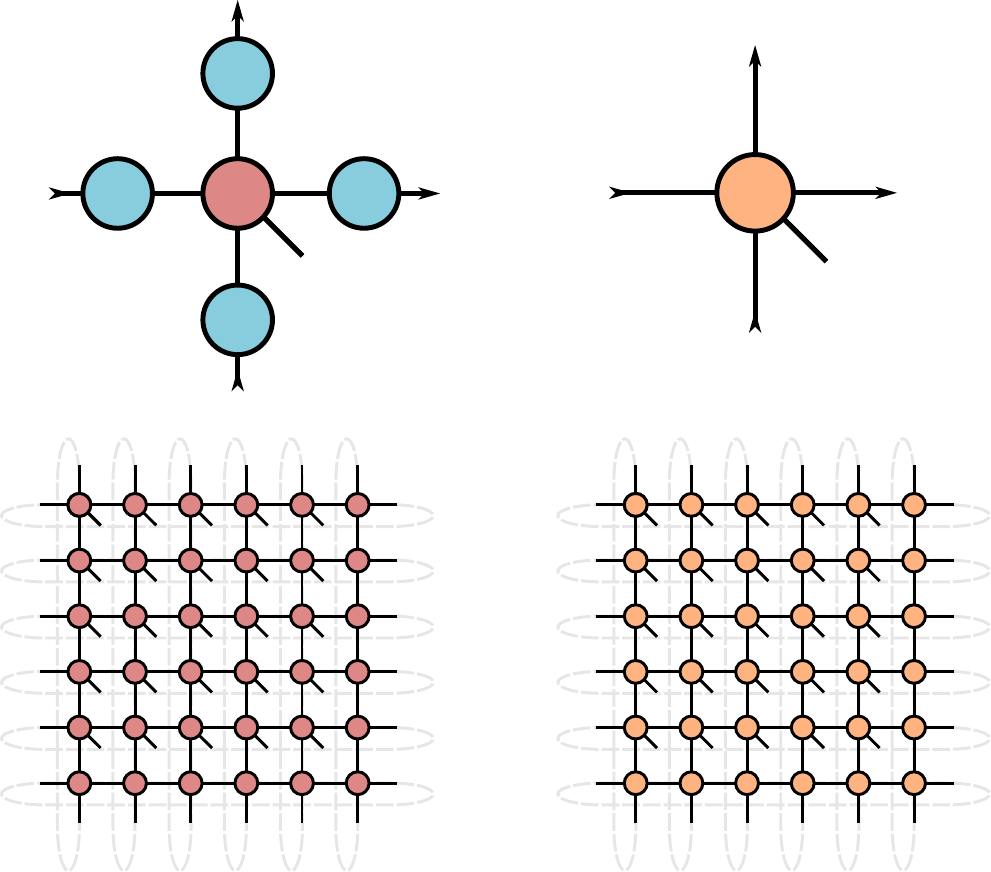}
\put(10,67){\footnotesize{$g_1$}} \put(34,67){\footnotesize{$g_1^{-1}$}}
\put(22.5,54.5){\footnotesize{$g_2$}} \put(21.5,79){\footnotesize{$g_2^{-1}$}}
\put(22.5,67){\footnotesize{$T$}}
\put(75,67){\footnotesize{$S$}}
\put(49,67){\Large{$=$}}

\put(48,47){\large{$\Downarrow$}}
\put(48,22){\Large{$=$}}
\end{overpic}
\vspace*{0.5cm}
\caption{\emph{Gauge invariance:} For $\vec g = (g_1, g_2) \in \GL(D_1) \times \GL(D_2)$, if one defines the tensor $S = \vec g \cdot T$ as in the figure, the corresponding states $\ket{T_{n,m}}$ and $\ket{S_{n,m}}$ are equal.}
\label{fig:gauge invariance}
\end{subfigure}
\hspace*{0.5cm}
\begin{subfigure}{0.45\textwidth}
\centering
\begin{overpic}[width=0.6\textwidth,grid=false]{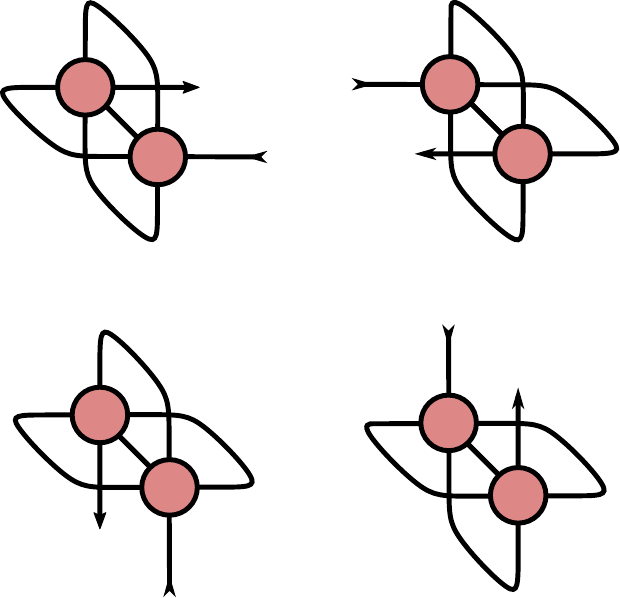}
\put(44,75){$=$} \put(45,21){$=$}
\end{overpic}
\vspace*{0.5cm}
\caption{\emph{Canonical form conditions:}
A tensor is in canonical form if the reduced density matrices of the tensor as a quantum state are equal up to transposition (corresponding to reversing the arrows in diagrammatic notation).}
\label{fig:mcf condition}
\end{subfigure}
\caption{}
\label{fig:mcf and gauge}
\end{figure}

We now define the new canonical form and highlight its main properties and the new fundamental theorem.
Here we only discuss uniform PEPS in $m$ spatial dimensions.
These are defined by a single tensor~$T$, with $2m+1$ legs, one associated to the physical Hilbert space, and two legs each for the spatial directions~$k \in \{1, \dots, m\}$, associated with virtual Hilbert spaces of bond dimension~$D_k$.
The gauge group $G = \GL(D_1) \times \dots \times \GL(D_m)$ acts on the virtual legs of the tensor as illustrated in \cref{fig:gauge invariance}.
We say $T_{\min}$ is a \textbf{minimal canonical form} of $T$ if it ``infimizes'' the $\ell_2$-norm
among all gauge equivalent tensors:
\begin{align}\label{eq:minimization}
   \fbox{ $T_{\min} = \argmin \left\{ \norm{S} : S \in \overline{G \cdot T} \right\}.$ }
\end{align}
Two important remarks are in order:
First, we consider the closure $\overline{G \cdot T}$ of the gauge group orbit of $T$, so that the minimum is attained.
Thus there need not be a single gauge transformation~$\vec g \in G$ such that~$\vec g \cdot T = T_{\min}$, but rather a sequence~$\vec g^{(k)} \in G$ such that $\vec g^{(k)} \cdot T \to T_{\min}$
(the same is true for the usual canonical forms of MPS when one has to set off-diagonal blocks to zero).
This is, however, natural, since the uniform PEPS determined by a tensor depend continuously on the tensor, hence remain unchanged even when taking limits.
Second, while any tensor clearly has a minimal canonical form, uniqueness up to unitaries is a priori unclear.
This is addressed by our first result, which justifies calling $T_{\min}$ a `canonical form'.

\begin{res}[\textbf{Canonical form}]\label{prp:minimal canonical form}
Any tensor has a minimal canonical form. It is unique up to unitary~gauge~symmetry.
Moreover, two tensors $T$, $T'$ have a common minimal canonical form if and only if $\overline{G \cdot T} \,\cap\, \overline{G \cdot T'} \neq \emptyset$.
\end{res}

The condition $\overline{G \cdot T} \,\cap\, \overline{G \cdot T'} \neq \emptyset$ is the natural definition of \emph{gauge equivalence}, since then~$T$, $T'$ determine the same PEPS as explained above.
\Cref{prp:minimal canonical form}, which we formally state as \cref{thm:oci mps} for MPS and \cref{thm:oci peps} for PEPS, states that this is captured by the minimal canonical form. It also guarantees the analogue of property~\ref{it:symmetry} for normal tensors, stated as \cref{cor:unitary-charact-symm-normal}.

We can characterize the minimal canonical form in terms of the reduced states of the virtual bonds.
To this end, interpret $T$ as a quantum state and denote by $\rho_{k,1}$ and $\rho_{k,2}$ the reduced states of the two virtual bonds in the $k$-th direction.
Then we have the following characterization, illustrated in \cref{fig:mcf condition}.

\begin{res}[\textbf{Characterization}]\label{prp:virtual rdm}
A tensor is in minimal canonical form if and only if $\rho_{k,1} = \rho_{k,2}^T$~for~$1\leq k\leq m$.
\end{res}

Interestingly, this shows our minimal canonical form does \emph{not} coincide with the usual ones for MPS~($m=1$); it also differs from previously proposed heuristics in higher dimensions.
We prove \cref{prp:virtual rdm} in \cref{thm:symmetric mps} for MPS and \cref{thm:symmetric virtual rdm} for PEPS.

This begs the question whether it can be computed effectively, even for MPS.
Our next result answers this in the affirmative.

\begin{res}[\textbf{Computation}]\label{prp:algorithm}
There is an algorithm which computes a minimal canonical form of a tensor~$T$ up to given $\ell_2$-error~$\delta>0$.
For fixed bond dimensions, it runs in time polynomial in $\log\frac1\delta$ and in~the~bitsize~of~$T$.
\end{res}

We prove this in \cref{cor:second order}.
The algorithm depends exponentially on the bond dimensions (for $m>1$).
We also give an algorithm whose runtime depends only polynomially on the bond dimension, but also on $\frac1\eps$, where $\eps$ measures the accuracy to which the condition in \cref{prp:virtual rdm} is fulfilled (see \cref{cor:peps scaling explicit guarantee}).
In \cref{sec:algos} we discuss these and another natural way of quantifying approximation error; we relate them in \cref{subsec:relerr}.

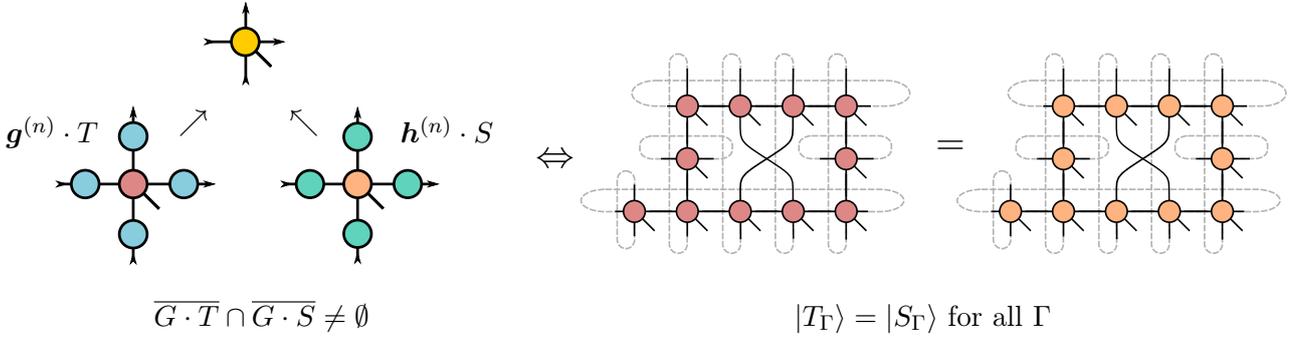
\begin{figure}
\centering
\begin{overpic}[width=0.9\textwidth,grid=false]{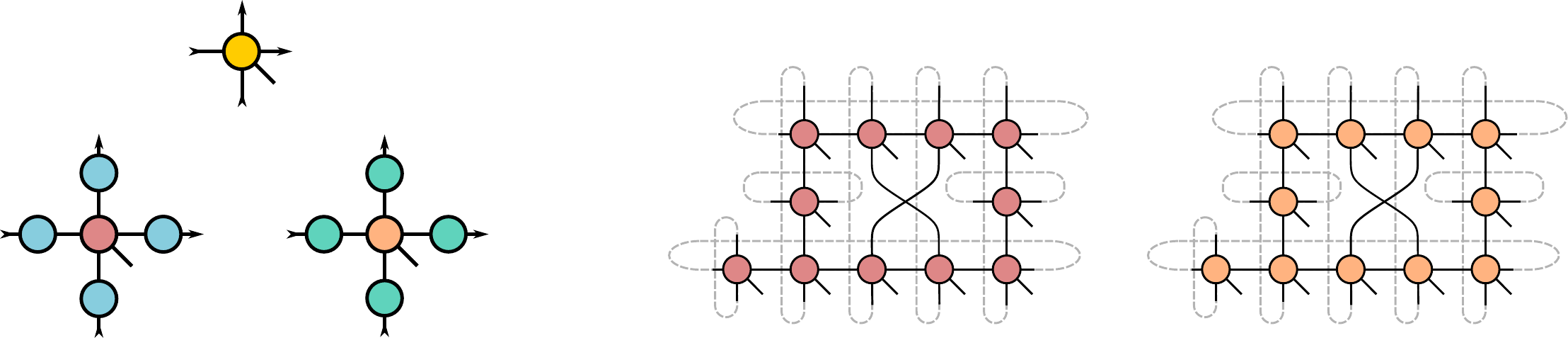}
\put(8,-5){$\overline{G \cdot T} \cap \overline{G \cdot S} \neq \emptyset$}
\put(60,-5){$\ket{T_{\Gamma}} = \ket{S_{\Gamma}}$ for all $\Gamma$}
\put(-4,10){$\vec g^{(n)} \cdot T$} \put(28,10){$\vec h^{(n)} \cdot S$}
\put(10,11){$\nearrow$} \put(19,11){$\nwarrow$}
\put(39,8){\Large{$\Leftrightarrow$}}
\put(71.5,9){\Large{$=$}}
\end{overpic}
\vspace*{0.8cm}
\caption{\emph{Fundamental theorem:} Two tensors $S$ and $T$ are gauge equivalent, meaning $\overline{G \cdot T}\cap \overline{G \cdot S}\neq\emptyset$ or that $\lim_{n\to\infty} \vec g^{(n)} \cdot T = \lim_{n\to\infty} \vec h^{(n)} \cdot S$ for certain $\vec g^{(n)}, \vec h^{(n)} \in G$ (equivalently, the two tensors have a common minimal canonical form), if and only if they contract to the same state on all contraction graphs.}
\label{fig:fundamental theorem}
\end{figure}

Finally, we discuss our fundamental theorem.
We start with the following observation (for simplicity in 2D):
If two tensors are gauge equivalent, they not only determine the same state~$\ket{T_{n,m}}$ on any $n\times m$ lattice, but also if we contract according to an arbitrary graph such that only left and right virtual legs, and only top and bottom virtual legs are connected.
We say~$\Gamma$ is a \emph{contraction graph} and write $\ket{T_\Gamma}$ for the corresponding uniform PEPS, see \cref{fig:fundamental theorem}.
Intuitively, this means we consider tensor networks on surfaces of \emph{arbitrary topology} rather than only on the torus.
Clearly, these notions generalize to any spatial dimension.
We find that this precisely captures gauge equivalence, in any spatial dimension!
Indeed, we have the following result which we formalize and prove as \cref{thm:peps fundamental thm}:

\begin{res}[\textbf{Fundamental theorem}]\label{thm:fundamental-theorem}
Two tensors $T$, $T'$ are gauge equivalent (meaning $\overline{G \cdot T}\cap \overline{G \cdot T'}\not=\emptyset$) if and only if $\ket{T_\Gamma} = \ket{T'_\Gamma}$ for all contraction graphs~$\Gamma$.
It suffices to consider to graphs on $e^{\tilde O(m D^2)}$ vertices.
\end{res}

We further show $e^{\Omega(m D)}$ vertices are necessary when~$m\geq2$, while for $m=1$ we find two MPS tensors to be gauge equivalent iff $\ket{T_n} = \ket{T'_n}$ for $1 \leq n \leq \tilde O(D)$, which is essentially tight~\cite{derksen2020algorithms}.
While we stress that our fundamental theorem is of independent interest, as it precisely characterizes when two tensors are gauge equivalent, we note that gauge equivalence is the same as having a common canonical form (by \cref{prp:minimal canonical form}).
Accordingly, our theorem proves a version of property~\ref{it:compare} for PEPS in any spatial dimension, and as we show in \cref{cor:u-invariance peps}, this also implies global symmetries of the states~$\ket{T_\Gamma}$ can be lifted to unitary gauge symmetries, as in property~\ref{it:symmetry}.
Strikingly, it shows that deciding whether two tensors generate the same uniform PEPS~$\ket{T_\Gamma}$ on arbitrary contraction graphs is \textbf{decidable} -- in stark contrast to the problem when we restrict to uniform PEPS~$\ket{T_{n,m}}$ on periodic rectangular lattices.
The undecidability of the latter is proved by relating it to the problem of deciding if a given set of tiles tiles a torus~\cite{scarpa2020projected}.
Our result implies that this problem becomes decidable if one allows for some arbitrary ``surface'' (contraction graph).

\smallskip

Given the practical and theoretical importance of canonical forms and fundamental theorems, we hope our results offer a useful new tool for the study and application of tensor networks.
From a theory perspective, our results may be helpful in studying virtual symmetries of tensor networks, which are crucial in understanding topological order.
From a practical perspective, it would be interesting to investigate if our canonical form can improve the numerical stability of variational optimization algorithms and other numerical methods~\cite{vanderstraeten2016gradient}, as it could be expected by the known close connection between gauge fixing and stability \cite{lubasch2014algorithms, phien2015fast}.
Our results also imply that one can sample uniformly from all PEPS tensors in minimal canonical form in the same orbit.
This has applications beyond quantum information, e.g., it allows to extend the technique of~\cite{pozas2022physics} for enhancing privacy in machine learning from MPS to PEPS.
Finally, we note that our approach generalizes naturally to other tensor network types and gauge groups; it would be exciting to explore this in followup work.
We discuss all these points further in \cref{sec:outlook}.

\subsection{Overview of methods: geometric invariant theory and geodesic convex optimization}
On a high level, our approach is to start with the desired gauge symmetry and explore its natural consequences (rather than with a specific class of networks, such as PEPS on a torus).
In our case this means starting with the action of the gauge group $G = \GL(D_1) \times \cdots \times \GL(D_m)$ on the vector space of PEPS tensors of a certain format, as above.
Geometric invariant theory (GIT) is a field of mathematics that studies group actions such as the above from the perspective of geometry and invariants~\cite{mumford1994geometric,wallach2017geometric}.
To find a canonical form, we wish to identify a special element in the orbit closure $\overline{G \cdot T}$.
Minimum norm tensors provide a natural candidate.
We prove \cref{prp:minimal canonical form,prp:virtual rdm} by relying on the Kempf-Ness theorem, a fundamental result in GIT that precisely studies minimum norm vectors.
Next, we observe that~\eqref{eq:minimization} is a \emph{non-commutative (group) optimization} problem of the kind that has recently been of great interest in TCS~\cite{garg2020operator,garg2017algorithmic,allenzhu2018operator,BGOWW18,burgisser2018efficient,kwok2018paulsen,burgisser2019theory}.
While non-convex in the usual sense, such problems are \emph{geodesically convex}, meaning the objective is convex along geodesics (shortest paths) of the domain.
To prove \cref{prp:algorithm}, we instantiate the general framework of~\cite{burgisser2019theory} (but give some improvements) and we relate the approximation guarantees provided by that framework to $\ell^2$-error (which is nontrivial).

So far, we have focused on geometry, but we now move to invariants to connect to tensor networks and sketch the proof of our fundamental theorem.
A theorem by Mumford implies that two tensors $T,T'$ are gauge equivalent (meaning $\overline{G \cdot T} \cap \overline{G \cdot T'} \neq \emptyset$) if and only if $P(T) = P(T')$ for any $G$-invariant polynomial~$P$.
Now, for any contraction graph~$\Gamma$, the tensor network state~$\ket{T_{\Gamma}}$ is unchanged by gauge symmetries, and therefore its coefficients are $G$-invariant polynomials in~$T$.
We use constructive invariant theory to prove that, conversely, \emph{any $G$-invariant polynomial can be obtained from coefficients of tensor network states}.
A theorem by Derksen~\cite{derksen2001polynomial} allows bounding the size of~$\Gamma$, which concludes the proof of \cref{thm:fundamental-theorem}.

\subsection{Organization of the paper}
We start in \cref{sec:git} by reviewing basic results from geometric invariant theory in a general setting.
Then we apply this to construct the minimal canonical form for MPS in \cref{sec:mps}.
Our main results are stated in \cref{sec:peps}, where we introduce the minimal canonical form for PEPS and prove its basic properties.
In \cref{sec:algos} we provide algorithms for computing the minimal canonical form and relate to recent work on non-commutative optimization.
We end with a brief outlook in \cref{sec:outlook}, suggesting applications for the minimal canonical form and avenues for future research.

\subsection{Notation}
We let $[k] := \{1,\dots,k\}$ and denote by $\Cstar$ the nonzero complex numbers.
We write $y = \argmin \{ f(x) : x \in X \}$ to denote that $y \in X$ and $f(y) = \min \{ f(x) : x \in X \}$; in general this will not uniquely determine~$y$.
Throughout we write $\braket{\cdot,\cdot}$ for inner products and $\norm{\cdot}$ for the corresponding Euclidean or $\ell^2$-norm.
We write $\Mat_{n,n'}$ for the complex vector space of complex $n\times n'$ matrices and $\Herm_n$ for the real vector space of Hermitian $n\times n$ matrices, which we will always endow with the Hilbert-Schmidt inner product $\braket{A,B} := \tr[A^\dagger B]$.
Thus, $\norm{A} := \sqrt{\braket{A,A}}$ denotes the Hilbert-Schmidt (or Frobenius or Schatten-2) norm of a matrix~$A\in\Mat_{n,n'}$.
We will (very rarely) also use the induced operator (or Schatten-$\infty$) norm, which we denote by $\opnorm{A}$.
We denote identity matrices by~$I$ and use subscripts to denote context when this increases clarity.
We write~$\GL(n)$ for the \emph{general linear group}, which consists of the invertible $n\times n$ matrices, and $\SL(n)$ for the \emph{special linear group}, which consists of the $n\times n$ matrices of unit determinant.
We will use boldface for $m$-tuples of matrices, e.g., $\vec X = (X_1,\dots,X_m)$ (as well as similarly in the case of open boundary conditions in \cref{sec:non ti mps}), but never for the $d$-tuples that make up uniform MPS or PEPS tensors.
Finally, we denote by~$\C[V]$ the algebra of polynomial functions on a vector space~$V$.

\section{Preliminaries in geometric invariant theory}\label{sec:git}
Geometric invariant theory (GIT) is a field of mathematics that studies orbits of group actions from a perspective that combines geometry and algebra.
In this section we give a gentle introduction to this theory and review some central results.
In subsequent sections we will apply it to define and analyze our new canonical form for tensor networks.
A good reference on GIT is the textbook by Wallach~\cite{wallach2017geometric} and we follow his concrete approach; for a more abstract account see the seminal monograph~\cite{mumford1994geometric}.

Throughout this section, we fix a subgroup $G \subseteq \GL(n)$ that is closed under taking adjoints, i.e.,~$g^\dagger \in G$ for every $g\in G$.
We furthermore assume that $G$ is defined by polynomial equations, i.e., $G = \{ g \in \GL(n) \;:\; P_i(g) = 0 \; \forall i \in [k] \}$ for certain polynomials $P_1,\dots,P_k$ in the matrix entries of~$g$.%
\footnote{That is, $G$ is a Zariski-closed subgroup of $\GL(n)$ that is closed under taking adjoints. Such groups can also be defined more abstractly and are known as \emph{complex reductive algebraic groups}.}
The unitary matrices in~$G$ form a maximally compact subgroup, which we denote by~$K = G \cap \U(n)$.

\begin{exa}
We will almost exclusively deal with groups of the form $G = \GL(D_1) \times \cdots \times \GL(D_m)$.
These can be realized as above as the subgroup of $\GL(n)$, $n=D_1+\dots+D_m$, consisting of block diagonal invertible matrices with blocks of size $D_k \times D_k$ for $k\in[m]$.
Then $K = \U(D_1) \times \cdots \times \U(D_m)$.
\end{exa}

Next, we fix a representation $\pi\colon G \to \GL(V)$ on a finite-dimensional Hilbert space~$V \cong \C^N$.
Recall that~$\pi$ is a representation if~$\pi(I_G) = I_V$ and $\pi(gh) = \pi(g)\pi(h)$ for all $g, h \in G$.
We will assume that $\pi$ is \emph{regular} or \emph{rational}, meaning that the matrix entries of~$\pi(g)$ with respect to any basis are polynomial functions of the matrix entries of~$g$ and of~$\det(g)^{-1}$.
Finally, we assume that $\pi(K) \subseteq \U(K)$, meaning that the unitary matrices in the group act by unitarily on the Hilbert space.

To emphasize that the group acts on vectors, we often write $g \cdot v := \pi(g)v$ for the \emph{action} of a group element $g\in G$ on a vector $v\in V$.
Then the \emph{orbit} of a vector $v\in V$ is the set of all vectors that can be obtained by the group action, denoted $G \cdot v := \{ g \cdot v : g \in G \}$.
Since the group~$G$ is never compact, orbits will in general not be closed; hence we will also be interested in the \emph{orbit closure} $\overline{G \cdot v}$.%
\footnote{The closure can be taken with respect to the standard topology induced by the norm.
For orbits, this coincides with the Zariski topology, and this is important for establishing the theory, but we will never need to use this explicitly.}
One of the central goals of GIT is to classify vectors under the group action, and it is natural to allow taking limits, as we explained in the introduction in the context of PEPS.
Thus, GIT is concerned with classifying orbit closures up to a natural notion of equivalence, where two vectors~$v, v'$ are called equivalent if
\begin{equation*}
  \overline{G \cdot v} \cap \overline{G \cdot v'} \neq \emptyset.
\end{equation*}
To this end, one would like to pick out special points in orbit closures.
Minimum norm vectors are natural candidates, generalizing \cref{eq:minimization}.
The following terminology is not standard, but natural:

\begin{dfn}[Minimum norm vectors]\label{dfn:minimum norm vector}
  For $v \in V$, we say that $\mcf{v}$ is a \emph{minimum norm vector} for $v$ if
  \begin{equation*}
    \mcf v = \argmin \{ \norm{w} : w \in \overline{G \cdot v} \}.
  \end{equation*}
  That is, $\mcf{v}$ is a minimum norm vector for $v$ if $\mcf v \in \overline{G \cdot v}$ and $\norm{\mcf v} = \min_{w \in \overline{G \cdot v}} \norm w = \inf_{g \in G} \norm{g \cdot v}$.
\end{dfn}

Clearly, any vector~$v$ has a minimum norm vector~$\mcf v$.
The latter is in general \emph{not} unique, since if $\mcf v$ is a minimum norm vector then so is $k \cdot \mcf v$ for any $k \in K$ (recall that $K$ preserves the norm).
Crucially, this is the only source of non-uniqueness.
Moreover, two orbit closures intersect if and only if they have a common minimum norm vector!
We summarize these fundamental results of GIT:

\begin{thm}\label{thm:git canonical}
Let $v\in V$.
Then minimum norm vectors for~$v$ exist and form a single $K$-orbit (meaning that any two minimum norm vectors $w,w'$ satisfy $K \cdot w = K \cdot w'$).
Moreover, if $v' \in V$, one has $\overline{G \cdot v} \cap \overline{G \cdot v'} \neq \emptyset$ if and only if $v$ and $v'$ have a common minimum norm vector.
\end{thm}

We now focus on the properties of the minimum norm vector itself and give some intuition why \cref{thm:git canonical} holds.
It is clear that if $w$ is a vector of minimal norm in an orbit closure, then it is in particular a vector of minimal norm in its own $G$-orbit, hence the derivatives of the norm (or norm squared) must vanish in any direction along the orbit.

What are these directions?
They are given by the Lie algebra $\Lie(G)$ of $G$, which is the complex vector space consisting of all matrices $X \in \Mat_{n,n}$ such that $e^{tX} \in G$ for all $t\in\R$.
Then $t \mapsto e^{tX} \cdot v$ is a smooth curve in the orbit of~$w$.
Accordingly, if $w$ is a vector of minimum norm in its orbit then $\norm{e^{tX} \cdot w}^2$ must have a minimum at $t=0$ and hence its derivative will vanish.
This motivates the following definition:

\begin{dfn}\label{dfn:critical}
  A vector $w \in V$ is called \emph{critical} if $\partial_{t=0} \norm{e^{tX} \cdot w}^2 = 0$ for every $X \in \Lie(G)$.
\end{dfn}

Since $K$ acts unitarily, the norm will always be preserved if we move in directions that keep us in the $K$-orbit.
The latter are given by the Lie algebra $\Lie(K)$ of~$K$, which is defined analogously.
One can show that \cref{dfn:critical} is equivalent to demanding that $\partial_{t=0} \norm{e^{tX} \cdot v}^2 = 0$ for all $X \in i \Lie(K)$; the latter are precisely the Hermitian matrices in~$\Lie(G)$.

Criticality is the natural first-order condition for a vector to have minimum norm in its orbit (``at a minimum, all derivatives vanish'').
Remarkably, this is also \emph{sufficient}!%
\footnote{As one might imagine the reason is a kind of convexity (in, as it turns out, a natural non-Euclidean geometry), and we will explain this in more detail in \cref{sec:algos}.}
This result is part of a key theorem by Kempf and Ness~\cite{kempf1979length}, which further characterizes the existence of minimum norm vectors:

\begin{thm}[Kempf--Ness]\label{thm:kempf ness}
  Let $v \in V$. Then:
  \begin{enumerate}
    \item $v$ is critical if and only if $\norm{g \cdot v} \geq \norm{v}$ for every $g \in G$ (i.e., $v$ has minimum norm in its orbit).
    \item If $v$ is critical and $w \in G \cdot v$ is such that $\norm{v} = \norm{w}$, then $w \in K \cdot v$.
    \item If $G \cdot v$ is closed then there exists a critical element $v' \in G \cdot v$.
    \item If $v$ is critical then $G \cdot v$ is closed.
  \end{enumerate}
In particular, $v$ is a minimum norm vector for itself (i.e., has minimum norm in $\overline{G \cdot v}$) if and only if it has minimum norm in its orbit (meaning $\norm{g \cdot v} \geq \norm v$ for all $g \in G$), which is the case if and only if $v$ critical.
\end{thm}

Thus, minimum norm vectors (or critical vectors) are unique up to the $K$-action, and they can be found precisely in closed $G$-orbits.
While $G$-orbits are not closed in general, it is well-known that any orbit closure contains a unique closed orbit.

\begin{lem}\label{thm:unique closed}
Every orbit closure $\overline{G \cdot v}$ contains a unique closed $G$-orbit.
\end{lem}

Accordingly, the minimum norm vectors $\mcf v$ for any vector $v\in V$ are precisely the vectors of minimal norm in the unique closed orbit inside $\overline{G \cdot v}$.
\Cref{thm:git canonical} follows from this and the Kempf-Ness theorem.
Indeed, the first claim in \cref{thm:git canonical} is immediate, and for the second claim we only need to argue that $\overline{G \cdot v} \cap \overline{G \cdot v'} \neq \emptyset$ implies that the two vectors have a common minimum norm vector.
To this end, take any $v'' \in \overline{G \cdot v} \cap \overline{G \cdot v'}$.
Then $\overline{G \cdot v} \cap \overline{G \cdot v'}$ contains $\overline{G \cdot v''}$, which in turn contains a closed orbit.
Thus both $\overline{G \cdot v}$ and $\overline{G \cdot v'}$ contain the same closed orbit, and hence~$v$ and~$v'$ have the same minimum norm vectors.

It is interesting to ask why \cref{thm:unique closed} is true.
Even though so far we only discussed geometry, to answer this question we have to turn towards invariant theory.

\begin{dfn}
  A \emph{($G$-)invariant polynomial} is a polynomial $P \in \C[V]$ such that, for every $g \in G$ and $v \in V$, $P(g \cdot v) = P(v)$.
  The \emph{invariant ring}, denoted $\C[V]^G$, is the algebra consisting of all $G$-invariant polynomials.
\end{dfn}

Then the point is that any two closed orbits can be separated by a $G$-invariant polynomial.

\begin{lem}\label{thm:mumford separation}
  Suppose that two orbits $G \cdot v$ and $G \cdot v'$ are closed and disjoint.
  Then there exists an invariant polynomial $P\in\C[V]^G$ such that $P(v) \neq P(v')$.
\end{lem}

This implies \cref{thm:unique closed} at once, since any $G$-invariant polynomial is a continuous function and hence constant not just on orbits but even on orbit \emph{closures}.
Furthermore, it implies the following important result, which connects geometry (orbit closures) and algebra (invariants):

\begin{thm}[Mumford]\label{thm:git fundamental}
Let $v,v'\in V$.
Then, $\overline{G\cdot v} \cap \overline{G \cdot v'} \neq \emptyset$ if and only if $P(v) = P(v')$ for all invariant polynomials $P \in \C[V]^G$.
\end{thm}

We end with a classical fact about invariant rings.

\begin{thm}[Hilbert finiteness]\label{thm:hilbert finiteness}
  The invariant ring $\C[V]^G$ is a finitely generated algebra.
\end{thm}

Moreover, there exist algorithms that compute generators $P_1,\dots,P_k \in \C[V]^G$~\cite{derksen2015computational}.
Accordingly, determining whether two vectors $v,v'$ are equivalent in the sense of GIT (i.e., $\overline{G \cdot v} \cap \overline{G \cdot v'} \neq \emptyset$) can in principle be decided by an algorithm -- simply check whether $P_i(v) = P_i(v')$ for all~$i\in[k]$.
However, this is impractical, since known algorithms for computing generators are inefficient (run in exponential time or worse) and in many situations one will have to deal with generators that have exponentially large degree (we will in fact see an example in \cref{sec:peps}) or are hard to evaluate in the sense of computational complexity~\cite{garg2019search}.
Moreover, it is not clear how such an algebraic approach could go beyond the decision problem to compute, e.g., minimum norm vectors.
This motivates the search for alternative algorithms.
We will return to this point in \cref{sec:algos}, but first we discuss in \cref{sec:mps,sec:peps} how the machinery of geometric invariant theory and in particular \cref{thm:git canonical,thm:git fundamental} allow defining new canonical forms for tensor networks that enjoy very good theoretical properties.

\section{Matrix product states}\label{sec:mps}
In this section, we discuss the setting of matrix product states (MPS).
While MPS are very well-understood theoretically, it is instructive to revisit this setting from our new perspective and contrast our minimal canonical form to the known ones, which also enjoy excellent theoretical properties.

We start by defining uniform (or translation-invariant) MPS and briefly reviewing existing canonical forms in \cref{sec:mps review}.
We then introduce the minimal canonical form in \cref{sec:mcf mps}.
Finally, in \cref{sec:non ti mps} we also discuss the case of non-uniform MPS with open boundary conditions.

\subsection{Gauge freedom and canonical forms for uniform MPS}\label{sec:mps review}
We denote by $\Mat_{D\times D}^d$ the vector space of $d$-tuples of $D \times D$-matrices.

\begin{dfn}[Uniform MPS]
For any matrix tuple $M = (M^{(i)})_{i=1}^d \in \Mat_{D\times D}^d$ and system size $n\in\N$, we define the \emph{uniform} (or \emph{translation-invariant}) \emph{matrix product state (MPS)} as the (not necessarily) quantum state $\ket{M_n} \in (\C^d)^{\ot n}$ whose coefficients are given by
\begin{align}\label{eq:mps}
  \braket{i_1,\dots,i_n | M_n} = \tr M^{(i_1)} \cdots M^{(i_n)}
\qquad (\forall i_1,\dots,i_n \in [d]).
\end{align}
We refer to $d$ as the \emph{physical dimension} and $D$ as the \emph{bond dimension}.
\end{dfn}

\begin{figure}
\centering
\begin{overpic}[width=0.7\textwidth,grid=false]{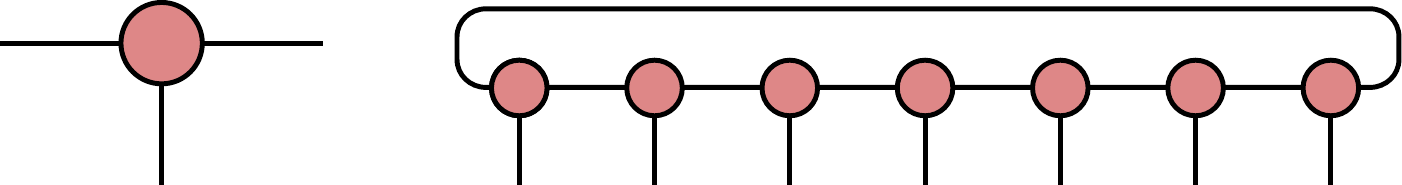}
\put(2,2){\Large{$M$}} \put(3,12){$\C^D$} \put(17,12){$\C^D$} \put(13,3){$\C^d$}
\put(102,9){$\ket{M_n}$}
\end{overpic}
\vspace*{0.4cm}
\caption{\emph{Matrix product state:} $M = (M^{(i)})_{i=1}^d \in \Mat_{D\times D}^d$ gives rise to a state $\ket{M_n}$ for any system size $n$.}
\label{fig:mps tensor}
\end{figure}
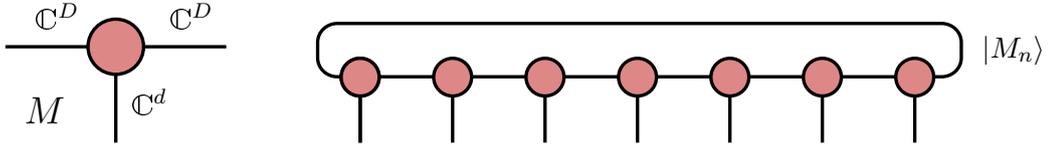

We will interchangeably refer to $M$ as a \emph{matrix tuple} or as an \emph{MPS tensor}.
Indeed, it is often useful to think of $M$ itself as a 3-tensor, or as an (unnormalized) quantum state $\ket M \in \HH_1 \ot \HH_2 \ot \HH_{\phys}$ on the tensor product of a physical Hilbert space $\HH_{\phys} = \C^d$ and two virtual Hilbert spaces $\HH_1 = \HH_2 = \C^D$, where $\HH_1$ is the `left' virtual Hilbert space and $\HH_2$ is the `right' virtual Hilbert space, see \cref{fig:mps tensor}.
Formally:
\begin{align*}
  \braket{a,b,i|M} = \braket{a|M^{(i)}|b}.
\end{align*}
We may then compute the reduced density matrices of $\rho = \proj{M}$ on either of the two virtual Hilbert spaces:
\begin{align}\label{eq:mps marginals}
  \rho_1 = \sum_{i=1}^d  M^{(i)}(M^{(i)})^\dagger \quad \text{ and } \quad \rho_2 = \sum_{i=1}^d (M^{(i)})^T \overline{M^{(i)}}.
\end{align}

An important property of MPS is that the states $\ket{M_n}$ are left invariant (for any $n$) if we conjugate each matrix $M^{(i)}$ in the tuple by the same invertible matrix. Formally:

\begin{dfn}[Gauge action]\label{dfn:gauge}
We define the \emph{gauge action} of $g\in\GL(D)$ on $M = (M^{(i)})_{i=1}^d$ by
\begin{align*}
  g \cdot M := ( g M^{(i)} g^{-1} )_{i=1}^d.
\end{align*}
If we think of $M$ as a quantum state $\ket{M}$ in $\HH_1 \ot \HH_2 \ot \HH_{\phys}$, the gauge action can be written as
\begin{align*}
  g \cdot \ket{M} := \ket{g \cdot M} = (g \ot g^{-T} \ot I) \ket{M}.
\end{align*}
\end{dfn}

\begin{lem}[Gauge symmetry]
For every $M \in \Mat_{D\times D}^d$, $g \in \GL(D)$, and $n \in N$, we have
\begin{align*}
  \ket{M_n} = \ket{(g \cdot M)_n}.
\end{align*}
\end{lem}

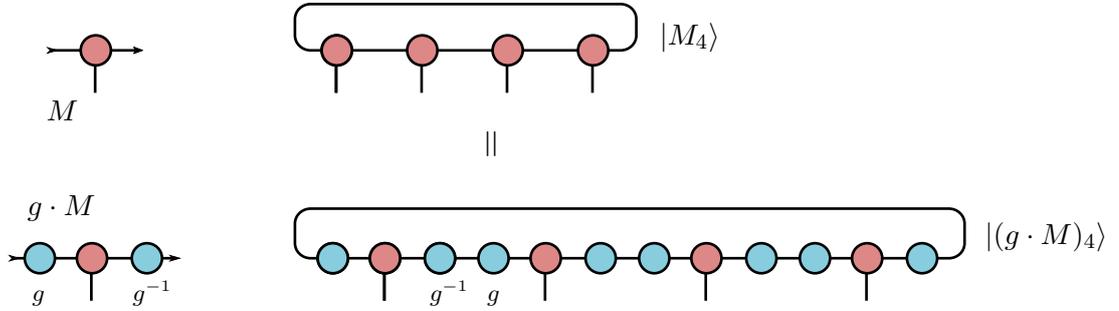
\begin{figure}
\centering
\begin{overpic}[width=0.7\textwidth,grid=false]{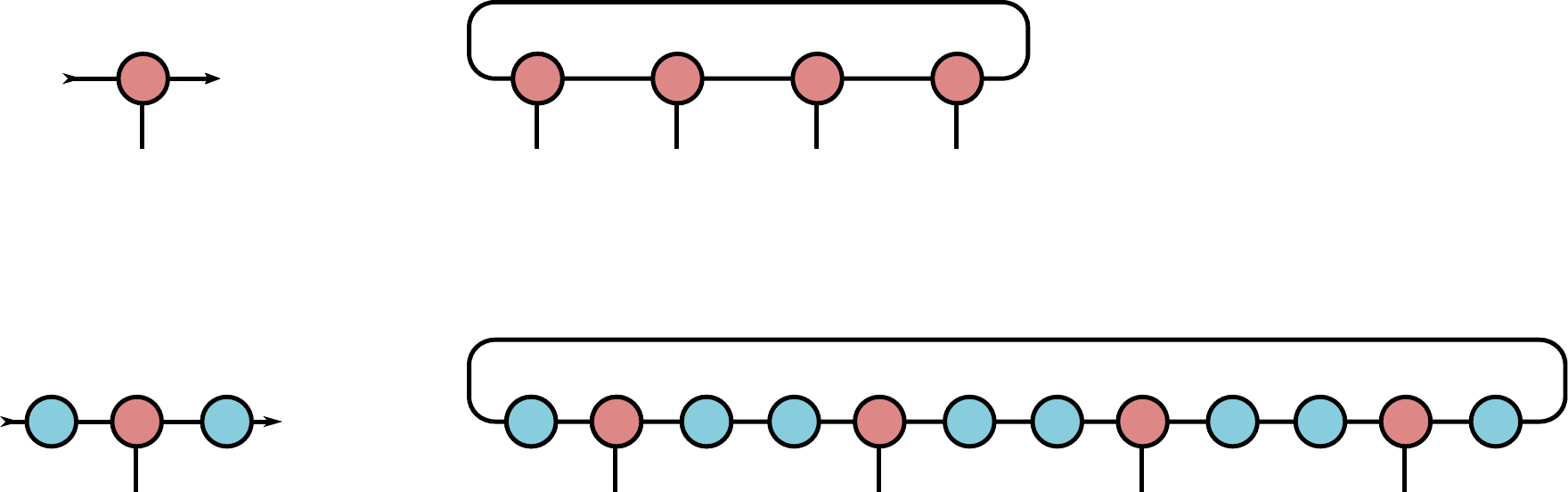}
\put(4,19){$M$} \put(2,9){$g \cdot M$}
\put(2.5,0){\footnotesize{$g$}} \put(13,0){\footnotesize{$g^{-1}$}}
\put(50,0){\footnotesize{$g$}} \put(44,0){\footnotesize{$g^{-1}$}}
\put(50,15){\rotatebox{90}{\Large{$=$}}}
\put(68,27){$\ket{M_4}$}
\put(102,6){$\ket{(g \cdot M)_4}$}
\end{overpic}
\vspace*{0.4cm}
\caption{\emph{MPS gauge invariance:} Tensors related by a gauge transformation give rise to the same MPS.}
\label{fig:mps gauge}
\end{figure}

This is shown in \cref{fig:mps gauge}.

It is then a natural question to ask whether this is the only freedom in the tensor $M$ to define the same state $\ket{M_n}$ for all~$n$.
The answer is \emph{no}, as is well-known and illustrated by the following example:

\begin{exa}\label{ex:ghz}
Let
\begin{align*}
  M^{(0)} = \begin{pmatrix} 1 & 1 \\ 0 & 0 \end{pmatrix} \qquad \text{and} \qquad M^{(1)} = \begin{pmatrix} 0 & 1 \\ 0 & 1 \end{pmatrix}.
\end{align*}
and
\begin{align*}
  \hat M^{(0)} = \begin{pmatrix} 1 & 0 \\ 0 & 0 \end{pmatrix} \qquad \text{and} \qquad \hat M^{(1)} = \begin{pmatrix} 0 & 0 \\ 0 & 1 \end{pmatrix}.
\end{align*}
Then both tensors define the same MPS, for any system size $n\in\N$, namely the \emph{GHZ states}
\begin{align*}
  \ket{M_n} =\ket{\hat M_n}= \ket{0}^{\ot n} + \ket{1}^{\ot n}.
\end{align*}
However, there is no $g \in \GL(D)$ so that $g\cdot M = \hat M$.
\end{exa}

The underlying problem is that when the matrices $M^{(i)}$ in a tuple are all in upper triangular form (with respect to some basis), the off-diagonal terms are totally irrelevant for the final state $\ket{M_n}$.
The standard way to deal with this is to \emph{remove} such off-diagonal terms in a structured manner.
Let us briefly sketch the procedure, but refer to \cite{cirac2021matrix} and \cite{sanz2010quantum} for details and nomenclature.

One starts looking for a minimal common invariant subspace of all $M^{(i)}$ and change $M^{(i)}$ by $PM^{(i)}P +QM^{(i)}Q$, with $P$ being the orthogonal projector onto such a subspace and $Q=I-P$.
It is not difficult to see that the new tensor defines the same original MPS.
Now one proceeds similarly with $QPM^{(i)}Q$ until one reaches a block diagonal form.
The minimality of the subspaces guarantees that, in each of the diagonal blocks $b$, the corresponding tensor, say $M_b$, fulfills the property that the associated completely positive (CP) map $\mathcal{E}_b$ given by $X_b\mapsto \sum_i M_b^{(i)}X_b (M_b{(i)})^\dagger$ is \emph{irreducible}.
Normalizing so that the spectral radius of the map is $1$, this implies that the eigenvalues of modulus $1$ are all non degenerate and they are exactly the $q$-th roots of unity with a $q$ dividing the size $D_b$ of the matrices $M_b^{(i)}$.
One can then distinguish two cases: $q=1$, in which case the map $\mathcal{E}_b$ is \emph{primitive}, or $q>1$, in which case one can ``block'' or group together~$q$ sites; then the resulting tensor in $\HH_1 \ot \HH_2 \ot \HH_{\phys}^{\otimes q}$ consists of block diagonal matrices whose associated CP maps are also primitive.

To make a long story short, starting with a matrix tuple~$M$, after projecting and blocking following the above procedure, one obtains a new matrix tuple $\tilde{M}$ such that each $\tilde M^{(i)}$ is block diagonal, $\tilde M^{(i)}=\oplus_b M_b^{(i)}$, and the CP maps $\mathcal{E}_b$ are all primitive.
It is now possible to act with a gauge $g \in \GL(D)$, which can also be taken to be block-diagonal, $g=\oplus_b g_b$, so that one obtains in each block $b$ of $\hat M := g\cdot \tilde{M}$ the \emph{canonical} condition. That is,
there exist constants $c_b \in \R_{+}$ such that
\begin{equation}\label{eq:left-canonical-MPS}
  \sum_{i=1}^d (\hat M_b^{(i)})^\dagger \hat M_b^{(i)} = c_b I_{D_b} \qquad (\forall b),
\end{equation}
meaning that, after normalization, the maps $\mathcal{E}_b:X_b\mapsto \sum_i \hat M_b^{(i)}X_b (\hat M_b^{(i)})^\dagger$ are trace preserving completely positive (TPCP) maps, i.e., quantum channels.
One could analogously have taken the \emph{dual} condition
\begin{equation}\label{eq:right-canonical-MPS}
  \sum_{i=1}^d  \hat M_b^{(i)} (\hat M_b^{(i)})^\dagger = c_b I_{D_b}  \qquad (\forall b),
\end{equation}
meaning the $\mathcal E_b$ are completely positive unital (CPU) maps.

For generic matrix tuples~$M$, the channel $X\mapsto \sum_i M^{(i)}X (M^{(i)})^\dagger$ is already primitive.
In this case, $M$ is called \emph{normal} and one can obtain a left or right canonical form~$\hat M$ by acting with a suitable gauge group element: $\hat M=g\cdot M$ for some $g\in \GL(D)$.

\begin{dfn}[Left and right canonical form]\label{dfn:lrcan}
A matrix tuple (MPS tensor) is said to be in \emph{left canonical form} if it is block diagonal, with each diagonal block a normal tensor fulfilling \cref{eq:left-canonical-MPS}.
The \emph{right canonical form} is defined analogously by imposing the dual condition in \cref{eq:right-canonical-MPS}.
\end{dfn}

The above procedure guarantees that, after discarding off-diagonal blocks and at the price of blocking, one can bring any MPS tensor into left or right canonical form.
For instance, in \cref{ex:ghz} the tensor $\hat M$ is block diagonal, its blocks are 1-dimensional and hence trivially primitive, and moreover $\hat\rho_1 = \hat\rho_2 = I_2$. Thus $\hat M$ is both in left and right canonical form.
\emph{For tensors in canonical form}, (unitary) gauge symmetry is the only freedom for two tensors to generate the same MPS:

\begin{thm}[Fundamental theorem of MPS, \cite{cirac2017matrix, cirac2021matrix}]\label{thm:old-fundamental}
Let $M$, $N$ be both in left (or right) canonical form and $\ket{M_n}=\ket{N_n}$ for all $n\in\N$.
Then there exists a unitary $u \in \U(D)$ such that $u \cdot M = N$.
\end{thm}

The name \emph{``fundamental theorem''} stems from its numerous applications, and we refer for instance to \cite{cirac2021matrix} or \cite{haegeman2017diagonalizing} for an accounting of several of these.

\subsection{The minimal canonical form for uniform MPS}\label{sec:mcf mps}
We now define a new canonical form for uniform MPS.
Its appeal is that it will naturally generalize to tensors with an arbitrary gauge symmetry and in particular to PEPS in higher dimensions, and that it can be analyzed using the powerful tools from geometric invariant theory.

Our starting point is the following simple but powerful observation:
For a given matrix tuple $M \in \Mat_{D \times D}^d$, we should not only consider gauge transformations $M \mapsto g \cdot M$ for some~$g \in \GL(D)$, but also limits of such.
Indeed, suppose we have a sequence of gauge group elements $g_k \in \GL(D)$ such that $g_k \cdot M$ converges to some~$\tilde M$.
Then, since the MPS $\ket{M_n}$ are continuous functions of the matrix tuple~$M$, we still have
\begin{align*}
  \ket{\tilde M_n} = \lim_{k\to\infty} \ket{(g_k \cdot M)_n} = \ket{M_n} \qquad (\forall n\in\N).
\end{align*}
In other words, all matrix tuples in the \emph{orbit closure} $\overline{\GL(D) \cdot M}$ determine the same MPS.
This naturally leads to the following definition:

\begin{dfn}[Gauge equivalence]\label{dfn:mps gauge eqv}
Let $M, N \in \Mat_{D\times D}^d$ be two matrix tuples.
We say that $M$ and $N$ are \emph{gauge equivalent} if and only if $\overline{\GL(D) \cdot M} \cap \overline{\GL(D) \cdot N} \neq \emptyset$.
\end{dfn}

This is the natural notion of gauge equivalence for MPS tensors, since if $M$ and $N$ are gauge equivalent in the sense just defined then
\begin{align*}
  \ket{M_n} = \ket{N_n} \qquad (\forall n \in \N).
\end{align*}
Indeed, it is the smallest equivalence relation generated by gauge transformations and taking limits.
In particular, to define a canonical form we should naturally look at orbit closures, not just at orbits.
How could we single out special elements in the orbit closure?
\Cref{sec:git} motivates the following definition:

\begin{dfn}[Minimal canonical form of MPS]\label{dfn:MPS canonical form}
We say $\mcf{M} \in \Mat_{D\times D}^d$ is a \emph{minimal canonical form} for a matrix tuple (MPS tensor)~$M \in \Mat_{D\times D}^d$ if it is an element of minimal norm in the orbit closure of~the~latter:
\begin{align*}
  \mcf{M} = \argmin \,\{ \norm{M'} : M' \in \overline{\GL(D) \cdot M} \},
\end{align*}
where we use the Euclidean norm of $M$ (or $\ket M$), that is,
\begin{align*}
  \norm M
= \sqrt{\braket{M|M}}
= \left( \sum_{i=1}^d \tr \left[ (M^{(i)})^\dagger M^{(i)} \right] \right)^{1/2}
= \left( \tr \left[ \sum_{i=1}^d (M^{(i)})^\dagger M^{(i)} \right] \right)^{1/2}.
\end{align*}
We say $M \in \Mat_{D\times D}^d$ is \emph{in minimal canonical form} if it is a minimal canonical form for itself.
\end{dfn}

Note that any MPS tensor has a minimal canonical form -- in contrast to the usual left or right canonical form of \cref{dfn:lrcan}, no explicit projecting and blocking is required.

Clearly, the minimal canonical form is a special case of the general notion of a minimum norm vector (\cref{dfn:minimum norm vector}) for the action of $G = \GL(D)$ on $V = \Mat_{D\times D}^d$ (\cref{dfn:gauge}).
We can now use the general theory of geometric invariant theory to understand the basic properties of this canonical form and we will see the usefulness of the general results of \cref{sec:git}.
First of all, while the minimal canonical form is not uniquely defined, it is uniquely defined up to unitary gauge transformations (the action of $K = \U(D)$), and it precisely characterizes gauge equivalence (\cref{dfn:mps gauge eqv}):

\begin{thm}[Minimal canonical form]\label{thm:oci mps}
Let $M, N \in \Mat_{D\times D}^d$.
Then the following are equivalent:
\begin{enumerate}
  \item $M$ and $N$ have a common minimal canonical form.
  \item If $\mcf M, \mcf N$ are minimal canonical forms of $M,N$ then $U(D) \cdot \mcf{M} = \U(D) \cdot \mcf{N}$.
  That is, minimal canonical forms of $M$ and $N$ are related by unitary gauge symmetries.
  \item $M$ and $N$ are gauge equivalent, i.e., $\overline{\GL(D) \cdot M} \cap \overline{\GL(D) \cdot N} \neq \emptyset$.
\end{enumerate}
\end{thm}
\begin{proof}
This is an immediate consequence of \cref{thm:git canonical}.
\end{proof}

\begin{figure}
\centering
\begin{overpic}[width=0.9\textwidth,grid=false]{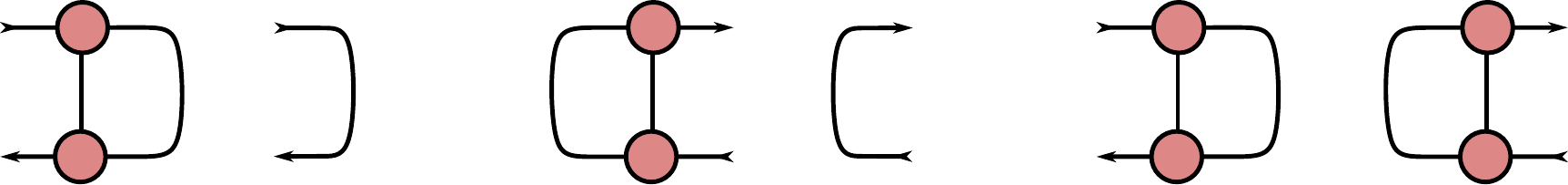}
\put(14,5){$=$} \put(49,5){$=$} \put(84,5){$=$}
\end{overpic}
\vspace*{0.2cm}
\caption{\emph{MPS canonical forms:} From left to right, the conditions for respectively right, left and minimal canonical forms for MPS.}
\label{fig:mps mcf}
\end{figure}

The characterization of minimum norm vectors as critical vectors (\cref{thm:kempf ness}) allows us to give an easy characterization for a matrix tiple to be in minimal canonical form.
To see this, we compute the condition for a matrix tuple $M \in \Mat_{D\times D}^d$ to be critical (\cref{dfn:critical}):
For $X \in \Herm_D = i\Lie(K)$, we have
\begin{align}
\nonumber
  \partial_{t=0} \norm{e^{tX} \cdot M}^2
&= \partial_{t=0} \sum_{i=1}^d \tr \left[ ( e^{tX} M^{(i)} e^{-tX} )^\dagger e^{tX} M^{(i)} e^{-tX} \right]
= \partial_{t=0} \sum_{i=1}^d \tr \left[ ( M^{(i)} )^\dagger e^{2tX} M^{(i)} e^{-2tX} \right] \\
\label{eq:mps deriv}
&= 2 \tr \left[ X \left( \sum_{i=1}^d M^{(i)} ( M^{(i)} )^\dagger - ( M^{(i)} )^\dagger M^{(i)} \right) \right].
\end{align}
Thus we arrive at the following (illustrated in \cref{fig:mps mcf}):

\begin{thm}[Characterization]\label{thm:symmetric mps}
Let $M \in \Mat_{D\times D}^d$.
Then $M$ is in minimal canonical form if and only if $\norm{g \cdot M} \geq \norm M$ for all $g\in\GL(D)$.
This is the case if and only if
\begin{align}\label{eq:symmetric condition 1}
  \sum_{i=1}^d M^{(i)} (M^{(i)})^\dagger = \sum_{i=1}^d (M^{(i)})^\dagger M^{(i)}.
\end{align}
Equivalently, the reduced density matrices of $\rho = \proj M$ on the virtual bonds are the same up to a transpose:
\begin{align}\label{eq:symmetric condition 2}
  \rho_1 = \rho_2^T.
\end{align}
\end{thm}
\begin{proof}
Note that $M$ is critical if and only if the derivative in \cref{eq:mps deriv} vanishes for all $X\in\Herm_D$.
Thus both statements follow from \cref{thm:kempf ness}.
\end{proof}

Given a tensor~$M$ it is perhaps at first glance surprising that there always exist gauge transformations $g_k \in \GL(D)$ such that $\lim_{k\to\infty} g_k \cdot M$ satisfies the condition in \cref{eq:symmetric condition 1,eq:symmetric condition 2} -- yet as we just saw this follows readily from geometric invariant theory.
We also note that \cref{thm:symmetric mps} also shows that the minimal canonical form for MPS will in general \emph{not} coincide with the usual left or right canonical form (\cref{dfn:lrcan}); there appears to be no obvious way to convert one into the other.
In \cref{sec:algos} we give a simple iterative algorithm that computes the minimal canonical form to arbitrary precision.

To get more intuition about the definition and the relevance of the orbit closure, we revisit \cref{ex:ghz}.

\begin{exa}\label{ex:ghz closure}
In \cref{ex:ghz} we saw that the following matrix tuples $M, \hat M \in \Mat_{2\times2}^2$ both define the GHZ states:
\begin{align*}
  M^{(0)} = \begin{pmatrix} 1 & 1 \\ 0 & 0 \end{pmatrix}, \quad M^{(1)} = \begin{pmatrix} 0 & 1 \\ 0 & 1 \end{pmatrix} \qquad \text{and} \qquad
  \hat{M}^{(0)} = \begin{pmatrix} 1 & 0 \\ 0 & 0 \end{pmatrix}, \qquad \hat{M}^{(1)} = \begin{pmatrix} 0 & 0 \\ 0 & 1 \end{pmatrix}.
\end{align*}
\Cref{thm:symmetric mps} shows that $\hat M$ is already in minimal canonical form, while $M$ is not.
Indeed, while $\hat\rho_1 = \hat\rho_2^T = I_2$ for~$\hat\rho = \ket{\hat M}\bra{\hat M}$, the reduced states of $\rho=\ket M\bra M$ satisfy
\begin{align*}
  \rho_1 &= M^{(0)}(M^{(0)})^\dagger + M^{(1)} (M^{(1)})^\dagger = \begin{pmatrix} 3 & 1 \\ 1 & 1 \end{pmatrix}, \\
  \rho_2^T &= (M^{(0)})^\dagger M^{(0)} + (M^{(1)})^\dagger M^{(1)} = \begin{pmatrix} 1 & 1 \\ 1 & 1 \end{pmatrix}.
\end{align*}
Moreover, in this example it is easy to see that there does \emph{not} exist a $g \in GL(2)$ such that $g \cdot M$ is in minimal canonical form.
However, if we let
\begin{align*}
  g_{\eps} = \begin{pmatrix} \eps & 0 \\ 0 & 1 \end{pmatrix}
\end{align*}
then we may verify that
\begin{align*}
   g_{\eps}  M^{(0)} g_{\eps}^{-1} = \begin{pmatrix} \eps & 0 \\ 0 & 1 \end{pmatrix} \begin{pmatrix} 1 & 1 \\ 0 & 0 \end{pmatrix} \begin{pmatrix} \eps^{-1} & 0 \\ 0 & 1 \end{pmatrix} = \begin{pmatrix} 1 & \eps \\ 0 & 0 \end{pmatrix} \\
  g_{\eps} M^{(1)} g_{\eps}^{-1} = \begin{pmatrix} \eps & 0 \\ 0 & 1 \end{pmatrix} \begin{pmatrix} 0 & 1 \\ 0 & 1 \end{pmatrix} \begin{pmatrix} \eps^{-1} & 0 \\ 0 & 1 \end{pmatrix} = \begin{pmatrix} 0 & \eps \\ 0 & 1 \end{pmatrix}
\end{align*}
so as we let $\eps \to 0$ we see that $g_\eps \cdot M \to \hat M$, which as just discussed is in minimal canonical form.
\end{exa}

\begin{exa}\label{ex:jordan}
An amusing special case is $d=1$, so we have a single matrix $M \in \Mat_{D \times D}$.
The minimal canonical form is given by the diagonal matrix with the same eigenvalues as $M$ (repeated according to their algebraic multiplicity).
Indeed, there are matrices $g_\eps$ such that $g_\eps \cdot M = g_\eps M g_\eps^{-1}$ is in Jordan normal form, but with $\eps$ instead of 1 as the offdiagonal entries.
Letting $\eps\to0$ we obtain the desired diagonal matrix.
\end{exa}

From \cref{ex:ghz closure,ex:jordan} it is clear that, by virtue of considering the orbit closure, the minimal canonical form \emph{automatically} sets off-diagonal blocks to zero, which is an additional step which needs to be manually taken in the usual approach to canonical forms for MPS (see \cref{sec:mps review}).
There, as already commented in \cref{sec:mps review}, it may also be necessary to block together multiple sites.
The geometric invariant theory approach makes these steps redundant.%
\footnote{As a side remark, there is actually no need to block in the usual canonical form for MPS. This is a consequence of Theorem 16 in \cite{delascuevas2017irreducible}, together with the overlooked observation that the matrix $Z$ appearing there can be absorbed in another gauge transformation.}

We will now prove a fundamental theorem for MPS where this will become explicit.
Before stating the result, we state the ingredient that will be used to prove it.
In invariant theory, the action of the gauge group on MPS tensors (\cref{dfn:gauge}) is known as the \emph{simultaneous conjugation action} of $\GL(D)$ on matrix tuples in $\Mat_{D\times D}^d$.
There, it is known that the ring of invariant polynomials is generated precisely by the coefficients~\eqref{eq:mps} of the corresponding matrix product states for system size $1 \leq n \leq D^2$, as stated in the following theorem:

\begin{thm}[Procesi-Razmyslov-Formanek~\cite{procesi1976invariant,razmyslov1974trace,formanek1986generating,deconcini2017invariant}]
\label{thm:mps inv}
The invariant ring for the simultaneous conjugation action, i.e., $\C[\Mat_{D\times D}^d]^{\GL(D)}$, is generated by the invariant polynomials $P_{\vec i}$, where
\begin{align*}
  P_{\vec i}(M) = \braket{i_1,\dots,i_n | M_n} = \tr M^{(i_1)} \cdots M^{(i_n)},
\end{align*}
for all $\vec i = (i_1,\dots,i_n) \in [d]^n$ and $n\in\N$. Moreover, it suffices to restrict to $n \in [D^2]$.
\end{thm}

Thus, geometric invariant theory implies that gauge equivalence of the tensors (which by \cref{thm:oci mps} is captured by the minimal canonical form) is precisely equivalent to equality of the corresponding matrix product states!
We summarize this in the following fundamental theorem for MPS (note that it works in full generality, without the need to block sites or remove off-diagonal terms):

\begin{thm}[Fundamental theorem for MPS]\label{thm:fundamental thm mps}
Let $M, N \in \Mat_{D\times D}^d$.
Then the following are equivalent:
\begin{enumerate}
  \item $M$ and $N$ are gauge equivalent, i.e., $\overline{\GL(D) \cdot M} \cap \overline{\GL(D) \cdot N} \neq \emptyset$.
  \item $\ket{M_n} = \ket{N_n}$ for all $n \in \N$.
  \item $\ket{M_n} = \ket{N_n}$ for $n=1,\dots,D^2$.
\end{enumerate}
\end{thm}
\begin{proof}
This follows from \cref{thm:git fundamental,thm:mps inv}.
\end{proof}

\begin{rem}\label{rem:improved mps}
It is also known that the invariant ring is not generated when restricting to~$n \leq D^2/8$~\cite{formanek1986generating}.
However, while a system of generators of the invariant ring always suffices to separate orbit closures, this is in fact \emph{not} necessary.
Theorem~1.14 in~\cite{derksen2020algorithms} shows that the third condition in \cref{thm:fundamental thm mps} can be improved almost quadratically to:
\begin{enumerate}
\item[3'.] $\ket{M_n} = \ket{N_n}$ for $n=1,\dots,4 D \log_2 D + 12 D - 4$,
\end{enumerate}
and it has been conjectured that $n = O(D)$ suffices~\cite{shitov2019improved}.
\Cref{exa:jordan revisited} shows that this is essentially tight.
\end{rem}

\begin{exa}
In \cref{ex:ghz} we saw two matrix tuples $M, \hat M \in \Mat_{2\times2}^2$ that defined the GHZ states, for all system sizes.
By our fundamental theorem, \cref{thm:fundamental thm mps}, this implies that they are gauge equivalent, meaning that
\begin{align*}
  \overline{\GL(D) \cdot M} \cap \overline{\GL(D) \cdot \hat M} \neq \emptyset.
\end{align*}
Now, in \cref{ex:ghz closure} we saw that $\hat M$ is already in minimal canonical form.
By the Kempf-Ness theorem (\cref{thm:kempf ness}) this means that the orbit of $\hat M$ is already closed.
It follows that
\begin{align*}
  \hat M \in \overline{\GL(D) \cdot M},
\end{align*}
which is in exact agreement with what we saw in \cref{ex:ghz closure}.
\end{exa}

\begin{exa}\label{exa:jordan revisited}
We also revisit \cref{ex:jordan}, the case of a single matrix.
For $M, N \in \Mat_D$, the equality of quantum states means that $\tr M^n = \tr N^n$ for all~$n$, which is the case if and only if $M,N$ have the same characteristic polynomial and hence the same eigenvalues with the same algebraic multiplicities -- in agreement with the discussion in \cref{ex:jordan}.
Thus we see that in this special case it suffices to have equality for all~$n=1,\dots,D$.
This is also necessary, since, e.g., for $M$ a $D\times D$-permutation matrix representing a $D$-cycle we have $\tr M^n = 0$ for $1\leq n<D$.
\end{exa}

Together, \cref{thm:oci mps,thm:fundamental thm mps} show that if $M$, $N$ are two matrix tuples in minimal canonical form that give rise to the same quantum states, then $M$ and $N$ are related by a \emph{unitary} gauge symmetry.
As a consequence, we can lift unitary symmetries to the virtual level.
Again, we do not need to make any assumptions about the tensor $M$.

\begin{cor}[Lifting symmetries]\label{cor:u-invariance}
Suppose that $M, N \in \Mat_{D\times D}^d$ are in minimal canonical form and $u\in \U(d)$ is a unitary such that $u^{\ot n} \ket{M_n} = \ket{N_n}$ for all $n \in \N$.
Then there exists a unitary $U \in \U(D)$ such that $(I \ot I \ot u) \ket M = (U \ot \bar U \ot I) \ket N$.
\end{cor}
\noindent
In other words, the action of $u$ on the physical degrees of~$M$ is implemented by the gauge action of $U$ on~$N$.
\begin{proof}
Let $M' \in \Mat_{D\times D}^d$ be the matrix tuple defined by
\begin{align*}
  \ket{M'} := (I \ot I \ot u) \ket{M}.
\end{align*}
Then $M'$ is also in minimal canonical form, since $u$ is unitary and hence we have $\norm{g \cdot M} = \norm{g \cdot M'}$ for all~$g\in\GL(D)$.
Moreover, by construction it holds that
\begin{align*}
  \ket{M'_n} = u^{\ot n} \ket{M_n} = \ket{N_n}
\end{align*}
for all $n\in\N$.
Thus \cref{thm:fundamental thm mps} shows that $M'$ and $N$ are gauge equivalent, and it follows from \cref{thm:oci mps} that there exists a unitary gauge transformation $U \in \GL(D)$ such that $U \cdot N = M'$.
\end{proof}

We note that $U$ need not be unique; for instance, $M$ itself may have a stabilizer, i.e., there may exist $U \in \U(D)$ such that $U \cdot M = M$.
Indeed, this is exactly the case in which the MPS given by $M$ has a global on-site symmetry, for which \cref{cor:u-invariance} reproduces, for the  minimal canonical form, the known local characterization of symmetries on MPS \cite{cirac2021matrix} usually obtained via the left or right canonical form and \cref{thm:old-fundamental}.

Such characterization is the key step in the classification of symmetry protected topological phases done in \cite{chen:1d-phases-rg, pollmann2012symmetry, schuch:mps-phases}.
The connection is as follows.
If a system is invariant under the action of an onsite (global) symmetry group~$u_g$, one gets $u_g^{\ot n} \ket{\Psi_n} =  \ket{\Psi_n}$ for its ground state $\ket{\Psi_n}$ (global phases do not play a relevant role here).
Since $\ket{\Psi_n}$ is known to be very well approximated by MPS one may want to solve equation $u_g^{\ot n} \ket{M_n} =  \ket{M_n}$ for the MPS generated by some tensor~$M$.
By \cref{cor:u-invariance}, this is characterized by the existence of $U_g \in \U(D)$ such that $(I \ot I \ot u_g) \ket M = (U_g \ot \bar U_g \ot I) \ket M$.
It is not difficult to see that $U_g$ must be a projective representation of the symmetry group.
The classification of SPT phases is given then by all non-equivalent projective representations, which is precisely described by the second cohomology group of the group cohomology of the symmetry group.
The general validity of this approach has been recently established by the groundbreaking results of Ogata \cite{ogata2020mathbb}.

The idea that the relevant topological content of a system lies in its boundary has also given rise to the study of a bulk-boundary correspondence, usually known in this context as ``entanglement spectra'' or ``entanglement Hamiltonian'' \cite{cirac2011entanglement}, in which one upgrades the boundary to a physical system and looks for a dictionary between bulk and boundary properties. This is precisely the reason that tensor networks have become rather popular in the context of AdS-CFT holography in quantum gravity.  For this program it is rather crucial that the boundary representations of the physical on-site symmetries are indeed given themselves by unitary representations, which is precisely what \cref{cor:u-invariance} guarantees for the MPS case.

\begin{rem}
As commented in \cref{sec:mps review}, a MPS state can also be interpreted as a CP map on the virtual Hilbert spaces, where $M \in \Mat_{D \times D}^d$ is interpreted such that the $M^{(i)}$ are Kraus operators of a CP map $\mathcal{E}$, usually called \emph{transfer operator}.
Equivalently, the reduced state~$\rho_{12}$ of the quantum state $\rho=\ket M\bra M$ on \emph{both} virtual Hilbert spaces is the Choi operator of $\mathcal{E}$.
As explained above \cref{dfn:lrcan}, the left and right canonical form conditions are equivalent to $\mathcal{E}$ either being completely positive trace-preserving (CPTP) or unital (CPU).
This perspective is particularly useful when dealing with contractions of large or infinite uniform MPS (the thermodynamic limit).

What is the interpretation of the minimal canonical form in this perspective?
It is not hard to see that a mixed quantum state $\rho_{12}$ with conjugate marginals (i.e., $\rho_1 = \rho_2^T$) that are full-rank contains exactly the same data as a CPTP map $\Phi$ along with a full-rank invariant density operator $\Omega$ (i.e., $\Phi(\Omega)=\Omega$).
The isomorphism
$\rho_{12} \mapsto (\Phi,\Omega)$
is defined by defining $\Phi = \Phi_{1\to2}$ as the CPTP map with Choi operator $\rho_1^{-1/2} \rho_{12} \rho_1^{-1/2}$ and $\Omega = \rho_2 = \rho_1^T$.
If the marginals do not have full rank we can restrict to its support.
By duality, this is in turn is the same as a CP unital map $\phi$ along with a \emph{faithful} invariant state $\omega$ in the algebraic sense:
We have an isomorphism
$(\Phi,\Omega) \mapsto (\phi,\omega)$,
defined by taking $\phi = \Phi^\dagger$ and $\omega(X) = \tr \Omega X$.
At this point we do not see a natural interpretation of these conditions for MPS contractions in the thermodynamic limit.
\end{rem}

\subsection{Canonical forms for MPS with open boundary conditions}\label{sec:non ti mps}
We will now consider open boundary conditions.
We use the invariant theory framework to define canonical forms, which in this case are closely related to well-known canonical forms.
Then it is natural to fix the system size $n$, and to consider the non-uniform setting.
Let
\begin{align*}
  V = \bigoplus_{k=0}^{n-1} \Mat^d_{D_k \times D_{k+1}}
\end{align*}
where $D_0=D_n=1$.
As usual, $d$ is the physical dimension and the $D_k$ are the bond dimensions (which may vary per bond).
Let $\vec M = (M_0, \dots, M_{n-1}) \in V$, then the associated MPS state $\ket{M}$ (note that now we have a fixed system size) is defined by
\begin{align*}
  \braket{i_0\dots i_{n-1} | M} = M_0^{(i_0)} M_1^{(i_1)} \dots M_{n-1}^{(i_{n-1})}.
\end{align*}
We let $G = \GL(D_1) \times \cdots \times \GL(D_{n-1})$ act on $V$ by gauge transformations.
To define this action, let $\vec g = (g_1, \dots, g_{n-1}) \in G$ and $\vec M = (M_0, \dots, M_{n-1}) \in V$. Then the action is given by
\begin{align*}
  \vec g \cdot \vec M = ((1,g_1) \cdot M_0, (g_1, g_2) \cdot M_1, \dots, (g_{n-2}, g_{n-1}) \cdot M_{n-2}, (g_{n-1},1) \cdot M_{n-1}).
\end{align*}
where for $M_i = (M_i^{(j)})_{j=1}^d$
\begin{align*}
  (g_i,g_{i+1}) \cdot M_i := (g_i M_i^{(j)}g_{i+1}^{-1})_{j=1}^d.
\end{align*}
It is clear that the resulting MPS state is invariant under the action of $G$.
For every `bond cut' $k \in \{1,\dots,n-1\}$, we let
\begin{align*}
  W_k := \Mat_{d^k \times D_k} \op \Mat_{D_k \times d^{n-k}}
\end{align*}
and we define a $G$-action on $W_k$ by
\begin{align*}
  \vec g \cdot (w_{\text{left}}, w_{\text{right}}) = (w_{\text{left}} g_k^{-1}, g_k w_{\text{right}}).
\end{align*}
Then we have a map $\iota_k \colon V \to W_k$, which maps the vector of MPS tensors $\vec M$ to a pair of `half-chain contractions' $(M_{k,\text{left}}, M_{k,\text{right}})$
\begin{align*}
  \bra{i_0 \dots i_{k-1}} M_{k,\text{left}} = M_0^{(i_0)}M_1^{(i_i)} \dots M_{k_1}^{(i_{k-1})} \qquad M_{k,\text{right}} = M_k^{(i_k)} \dots M_{n-1}^{(i_{n-1})} \ket{i_{k} \dots i_{n-1}}.
\end{align*}
This map is clearly $G$-equivariant.
We can patch the maps $\iota_k$ together to obtain a $G$-equivariant polynomial map
\begin{align*}
  \iota \colon V \to W := \bigoplus_{k=1}^{n-1} W_k.
\end{align*}
We can think of $M_{k,\text{left}}$ and $M_{k,\text{right}}$ as the states where we have contracted all the bonds except the $k$-th.
In this perspective the reduced density matrices on the left and right copies of $\C^{D_k}$ are given by
\begin{align*}
  \rho_{k,\text{left}} &= \sum_{\vec i} (M_{k-1}^{(i_{k-1})})^\dagger \dots (M_0^{(i_0)})^{\dagger} M_0^{(i_0)} \dots M_{k-1}^{(i_{k-1})} = M_{k,\text{left}}^\dagger M_{k,\text{left}}\\
  \rho_{k,\text{right}}^T &= \sum_{\vec i} M_{k}^{(i_{k})} \dots M_{n-1}^{(i_{n-1})}(M_{n-1}^{(i_{n-1})})^\dagger \dots ( M_{k}^{(i_{k})})^\dagger = M_{k,\text{right}} M_{k,\text{right}}^\dagger
\end{align*}
We claim that norm minimization in the image of $\iota$ leads to a canonical form where $\rho_{k,\text{left}} = \rho_{k,\text{right}}^T$, which we call the \emph{minimal canonical form} for non-uniform MPS:
\begin{dfn}
Let $\vec M \in V$.
Then $\mcf{\vec M}$ is a \emph{minimal canonical form} for $\vec M$ if $\iota(\mcf{\vec M})$ is an element of minimal norm with respect to the orbit closure~$\overline{G \cdot \vec M}$, i.e.,
\begin{align*}
  \mcf{\vec M} = \argmin \,\{ \norm{\iota(\vec{M'})} : M' \in \overline{\GL(D) \cdot \vec M} \}.
\end{align*}
\end{dfn}

The norm we are considering here is again the Euclidean one. Note also that $\vec g \cdot \iota_k(\vec M)$ only depends on $g_k$.
Therefore, we may also write $g_k \cdot \iota_k(\vec M)$.
In minimizing $\norm{\vec g \cdot \iota(\vec M)}$ we may minimize each $\norm{g_k \cdot \iota_k(\vec M)}$ separately.
By the same general theory as applied in \cref{sec:mcf mps} we deduce that the canonical form exists and is unique up to conjugation by unitary elements in $G$.
Moreover, as in \cref{thm:symmetric mps} we may set an appropriate derivative equal to zero to find a condition for when $\vec M$ is in minimal canonical form.

Letting $g_k(t) = e^{tX_k}$ for $X_k \in \Herm_{D_k}$ we see that
\begin{align*}
  \norm{g_k(t) \cdot \iota_k(\vec M)}^2 &= \tr\mleft[ (g_k(t)^{-1})^{\dagger} M_{k,\text{left}}^{\dagger} M_{k,\text{left}} g_k(t)^{-1} + g_k(t) M_{k,\text{right}} M_{k,\text{right}}^{\dagger} g_k(t)^\dagger \mright] \\
  &= \tr\mleft[ e^{-2tX_k} M_{k,\text{left}}^{\dagger} M_{k,\text{left}} + M_{k,\text{right}} M_{k,\text{right}}^{\dagger} e^{2tX_k} \mright]
\end{align*}
and hence, denoting by $\vec g(t) = (g_1(t),\dots, g_{n-1}(t))$ we have
\begin{align*}
  \partial_{t = 0} \norm{\iota(\vec g(t) \cdot \vec{M})}^2 &= \partial_{t=0} \sum_{k = 1}^{n-1} \norm{g_k(t) \cdot \iota_k(\vec M)}^2 \\
  &= 2\sum_{k = 1}^{n-1} \tr\mleft[X_k \left(M_{k,\text{right}} M_{k,\text{right}}^{\dagger} -  M_{k,\text{left}}^{\dagger} M_{k,\text{left}}\right)\mright].
\end{align*}
Setting this equal to zero is equivalent to $\rho_{k,\text{left}} = \rho_{k,\text{right}}^T$ for all $k$.

We may explicitly perform the minimization; it is closely related to Vidal's canonical form \cite{vidal2003efficient}.
We can perform a singular value decomposition
\begin{align*}
  M_{k,\text{left}} = V_1 \Sigma_1 U_1
\end{align*}
where $V_1 \in \Mat_{d^k \times D_k}$ is an isometry, $\Sigma_1 \in \Mat_{D_k \times D_k}$ is diagonal with nonnegative entries and $U_1 \in \Mat_{D_k \times D_k}$ is unitary.
Next, we perform a singular value decomposition on $\Sigma_1 U_1 M_{k,\text{right}}$ so
\begin{align*}
  \Sigma_1 U_1 M_{k,\text{right}} = U_2 \Sigma_2 V_2
\end{align*}
where $V_2 \in \Mat_{D_k \times d^{n-k}}$ is an isometry, $\Sigma_2 \in \Mat_{D_k \times D_k}$ is diagonal with nonnegative entries and $U_2 \in \Mat_{D_k \times D_k}$ is unitary.
Let $\Pi_i$ be the projection onto $\ker(\Sigma_i)$ and let $\tilde \Sigma_i = \Sigma_i + \Pi_i$.
Then let
\begin{align*}
  g_k = \sqrt{\tilde \Sigma_2^{-1}} U_2^{\dagger} \tilde \Sigma_1 U_1
\end{align*}
 and we let $\tilde M_{k,\text{left}} = M_{k,\text{left}} g^{-1}$ and $\tilde M_{k,\text{right}} = g M_{k,\text{right}}$.
Then we may verify that the associated reduced density matrices are
\begin{align*}
  \rho_{k,\text{left}} &= \tilde M_{k,\text{left}}^{\dagger} \tilde M_{k,\text{left}} \\
  &= \sqrt{\tilde\Sigma_2} U_2^\dagger \tilde\Sigma_1^{-1} U_1 M_{k,\text{left}}^{\dagger} M_{k,\text{left}}   U_1^\dagger \tilde\Sigma_1^{-1} U_2 \sqrt{\tilde\Sigma_2} \\
  &= \sqrt{\tilde\Sigma_2} U_2^\dagger \tilde \Sigma_1^{-1} U_1 U_1^\dagger \Sigma_1 V_1^\dagger V_1 \Sigma_1 U_1  U_1^\dagger \tilde \Sigma_1^{-1} U_2 \sqrt{\tilde \Sigma_2} \\
  &= \Sigma_2
\end{align*}
and
\begin{align*}
  \rho_{k,\text{right}}^T &= \tilde M_{k,\text{right}} \tilde M_{k,\text{right}}^{\dagger}\\
  &= \sqrt{\tilde\Sigma_2^{-1}} U_2^{\dagger} \tilde\Sigma_1 U_1 M_{k,\text{right}} M_{k,\text{right}}^{\dagger}   U_1^\dagger \tilde\Sigma_1 U_2 \sqrt{\tilde\Sigma_2^{-1}} \\
  &= \sqrt{\tilde \Sigma_2^{-1}} U_2^{\dagger} U_2 \Sigma_2 U_2 U_2^{\dagger} \Sigma_2 U_2^\dagger U_2 \sqrt{\tilde \Sigma_2^{-1}} \\
  &= \Sigma_2.
\end{align*}
Therefore, defining $g_k$ in this fashion for each $k$ gives $g \cdot M$ in minimal canonical form.
In this case it is not necessary to go to the closure to obtain the canonical form.

This canonical form coincides with the one of Vidal \cite{vidal2003efficient}, usually written in the form
\begin{equation}
    \sum_{i_0\ldots i_{n-1}} \Gamma_0^{(i_0)}\Lambda_1 \Gamma_1^{(i_1)}\Lambda2 \cdots \Lambda_{n-1}\Gamma_{n-1}^{(i_{n-1})}\ket{i_0,i_{n-1}}
\end{equation}
if one identifies $M_k^{(i_k)}$ with $\sqrt{\Lambda_k}\Gamma_k^{(i_k)} \sqrt{\Lambda_{k+1}}$. The reason is that, by the properties of Vidal's canonical form \cite{vidal2003efficient, schollwock2011density}, such choice fulfills the algebraic characterization of the minimal canonical form given by $\rho_{k,\text{left}} = \rho_{k,\text{right}}^T$ for all $k$. Since the positive diagonal matrices $\Lambda_k$ correspond to the Schmidt coefficients of the bipartition of the system in the cut $[0:k-1], [k_n-1]$, the minimal canonical form can be understood in this case as an even distribution of those weights. This particular distribution of weights has also appeared extensively in the standard MPS literature \cite{orus2008infinite}.

There are also left and right canonical forms \cite{schollwock2011density}.
These fit in the same framework, which we will now show for the left canonical form (with the right canonical form being completely analogous).
Let $V$ be as before, but now we consider the action of $G = \SL(D_1) \times \dots \times \SL(D_{n-1})$.
We let $W_k = \Mat_{d^k \times D_k}$ (which is only the left half chain) and we let $\iota_k : V \to W_k$ be given by $M_k = \iota_k(\vec M)$
\begin{align*}
  \bra{i_0 \dots i_{k-1}} M_{k} = M_0^{(i_0)}M_1^{(i_i)} \dots M_{k_1}^{(i_{k-1})}
\end{align*}
(so this is what previously was $M_{k,\text{left}}$).
The group action is given by the $M_k \mapsto M_k g_k^{-1}$.
We similarly define
\begin{align*}
  \iota \colon V \to := \bigoplus_{k=1}^{n-1} W_k.
\end{align*}
Computing the gradient as before, but now restricting to \emph{traceless} $X$ (as we are optimizing over $SL(D_k)$ we find that at the minimum of the norm $\norm{\vec g \cdot \iota(\vec M)}$ the reduced density matrix $\rho_{k,\text{left}}$ must be proportional to the identity for all $k$.
Again, we can explicitly realize the minimum, without going to the closure.
To this end we perform a singular value decomposition $M_k = V \Sigma U$.
Let $\Pi$ be the projection onto $\ker(\Sigma)$ and let $\tilde \Sigma = \Sigma + \Pi$.
Then taking $g_k = \det(\tilde \Sigma U)^{-1/D_k}\tilde\Sigma U \in \SL(D_k)$ yields a uniform reduced density matrix $\rho_{k,\text{left}}$.

\section{Projected entangled pair states}\label{sec:peps}
In this section we start by defining projected entangled pair states (PEPS), in particular uniform PEPS.
In \cref{sec:minimal-peps-basics} we introduce the minimal canonical form for PEPS.
We will see that by closely analogous arguments to the MPS case we may establish its basic properties.
In \cref{sec:tilings} we relate to two-dimensional tilings and explain how our results are compatible with earlier no-go results for the existence of canonical forms for PEPS.
In \cref{sec:closure} we study in more detail the role of the orbit closure and show that in many cases of interest the orbit is closed.

\subsection{Definition of uniform PEPS}
We will now define a generalization of MPS, known as Projected Entangled Pair States (PEPS).
We start by defining a rather general version, and then specialize to cases of interest.
As input we require a graph $\Gamma = (V,E)$ and dimensions $(D_e)_{e \in E}$ (the bond dimensions) and $(d_v)_{v \in V}$ (the physical dimensions).
Let $E(v)$ denote the set of edges incident to $v \in V$.
Then we let $\HH_v := \C^{d_v}$ and for each $e \in E(v)$ we let $\HH_{v,e} := \C^{D_e}$.
The PEPS will now be constructed from a collection of tensors $(T^{[v]})_{v \in V}$ where
\begin{align*}
  T^{[v]} \in \left( \bigotimes_{e \in E(v)} \HH_{v,e} \right) \ot \HH_v.
\end{align*}
The resulting PEPS is a state on $\bigotimes_{v \in V} \HH_v$ and is constructed by `contracting along the edges'.
If $e = (v,w)$ is an edge incident to $v$ and $w$, then the contraction map $\delta_e: \HH_{v,e} \ot \HH_{w,e} \rightarrow \C$ along $e$ is defined by
\begin{align*}
  \ket{ij} \mapsto \delta_{i,j}
\end{align*}
and extending by linearity.
We may apply these maps along each of the edges in $E$ and this yields a state $\ket{T_\Gamma}$ on $\bigotimes_{v \in V} \HH_v$.

A clean way of writing this contraction operation (and also explaining the nomenclature projected entangled pair states) is by the identity
\begin{align*}
  \ket{T_\Gamma} = \left(\bigotimes_{e = (v,w) \in E}\left( \sum_{i = 0}^{D_e - 1}\bra{ii}\right) \ot I_{V} \right) \bigotimes_{v \in V} T^{[v]}.
\end{align*}
where $I_V$ is the identity operator on $\bigotimes_{v \in V} \HH_v$.

\begin{figure}
\centering
\begin{overpic}[width=0.9\textwidth,grid=false]{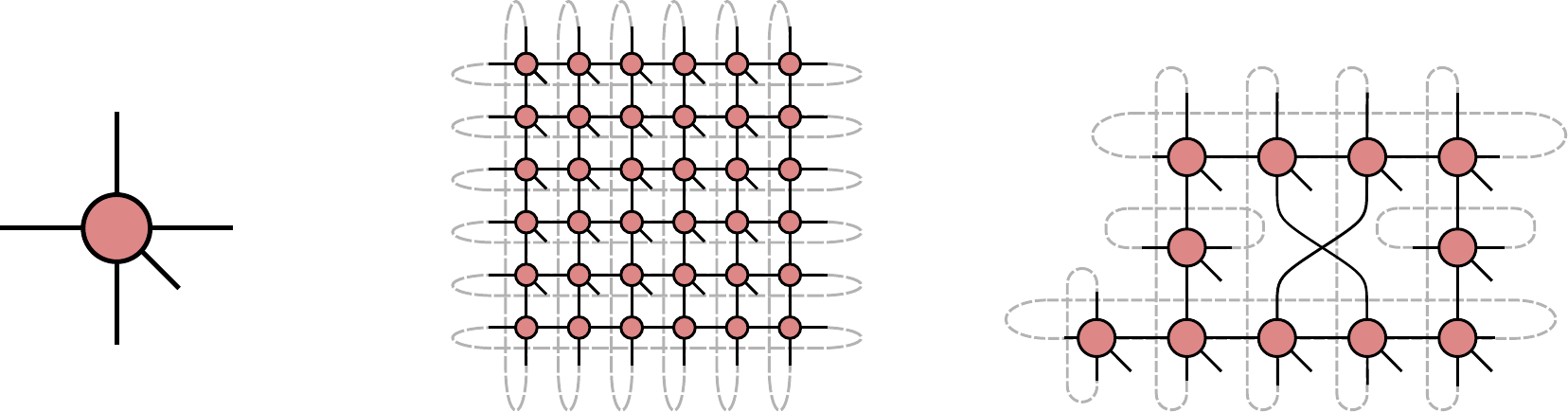}
\put(2,6){\Large{$T$}}
\put(39,-3){$\ket{T_{n,m}}$}
\put(80,-3){$\ket{T_{\vec \pi}}$}
\end{overpic}
\vspace*{0.8cm}
\caption{\emph{Projected entangled pair states:} Given a tensor $T$, here in two spatial dimensions, we may contract on a $n_1 \times n_2$ grid to obtain $\ket{T_{n_1,n_2}}$ or using arbitrary permutations $\vec \pi = (\pi_1,\pi_2)$ to get $\ket{T_{\vec \pi}}$}
\label{fig:peps definition}
\end{figure}

We will now specialize to the case of \emph{uniform PEPS}.
In this case we place the same tensor at each vertex.
It is natural to contract the tensors placed on periodic grids in $m$ spatial dimensions, but we will see that other graphs are also relevant.
We denote the physical dimension by $d$ and the associated physical Hilbert space by $\HH_{\phys} = \C^d$, and there are $m$ relevant bond dimensions in the different directions, which we will denote by $D_k$ for $k \in [m]$.
For each direction $k \in [m]$ we have two Hilbert spaces $\HH_{k,1} = \C^{D_k}$ and $\HH_{k,2} = \C^{D_k}$.
Similar to the MPS case, we may interpret the PEPS tensor $T$ either as a tensor
\begin{equation}\label{eq:ket-T-PEPS}
  \ket{T} \in \left(\bigotimes_{k=1}^m \HH_{k,1} \ot \HH_{k,2}\right) \ot \HH_{\phys}
\end{equation}
or as a matrix tuple
\begin{equation}\label{eq:mat-T-PEPS}
  T = (T^{(i)})_{i = 1}^d, \qquad T^{(i)} \in \bigotimes_{j=1}^m \Mat_{D_j \times D_j}
\end{equation}
and we will generally identify this space of matrix tuples as $\Mat_{D_1\dots D_m \times D_1\dots D_m}^d$.
Typically, one constructs corresponding quantum states by placing copies of the tensor on a grid and contracting along the bond dimensions, see \cref{fig:peps definition}.

\begin{dfn}[Uniform PEPS on a grid]
For any matrix tuple $T = (T^{(i)})_{i=1}^d \in \Mat_{D_1 \dots D_m \times D_1 \dots D_m}^d$ and system sizes $n_1, \dots, n_m \in\N$, we define the \emph{uniform} (or \emph{translation-invariant}) \emph{projected entangled pair state (PEPS)} as the (not necessarily) quantum state $\ket{T_{n_1, \dotsc, n_m}} \in (\C^d)^{\ot n}$, where $n = n_1 \dots n_m$ and which is given by contracting $n$ copies of $T$ on an $n_1 \times \dots \times n_m$ periodic grid.
\end{dfn}

We would like to allow a broader class of uniform PEPS, where one may use in principe any possible contraction graph.
In such a contraction graph we only demand that the directions are matched up, in the sense that we always contract $\HH_{k,1}$ with $\HH_{k,2}$.
A natural way to express such contractions is as follows.
Suppose that we have $n$ vertices, with at each vertex a copy of $T$, and we are given a contraction graph.
We will define permutations $\pi_k \in S_n$ for each direction $k \in [m]$.
Suppose that in direction $k$ $\alpha, \beta \in [n]$ are such that the Hilbert space $\HH_{k,2}$ of the $\alpha$-th copy of $T$ is contracted with the Hilbert space $\HH_{k,1}$ of the $\beta$-th copy of $T$, then we let $\pi_k$ map $\alpha$ to $\beta$.
Each contraction map (and ordering of the vertices) then uniquely determines permutations $\pi_k \in S_n$.
As permutations $\vec \pi = (\pi_1, \dots, \pi_m)$ completely determine the contraction of the $n$ copies of $T$ to a quantum state on $\HH_{\phys}^{\ot n} = (\C^d)^{\ot n}$ we denote this state by $\ket{T_{\vec \pi}}$.
For $k \in [m]$ let $R_{\pi_k}$ be the operator on $(\C^{D_k})^{\ot n}$ permuting the $n$ tensor factors.

\begin{dfn}[Uniform PEPS on arbitrary contraction graphs]
For any matrix tuple $T = (T^{(i)})_{i=1}^d \in \Mat_{D_1 \dots D_m \times D_1 \dots D_m}^d$, system size $n$ and for $\vec \pi = (\pi_1,\dots,\pi_m) \in S_n^{m}$ we define the associated \emph{uniform projected entangled pair state (PEPS)} as the (not necessarily) quantum state $\ket{T_{\vec \pi}} \in (\C^d)^{\ot n}$ which has coefficients defined by
\begin{align*}
  \braket{i_1, \dots, i_n| T_{\vec \pi}} = \tr\mleft[ (R_{\pi_1} \ot \dots \ot R_{\pi_m}) T^{(i_1)} \ot \dots \ot T^{(i_n)}\mright] \qquad \vec i = (i_1,\dots,i_n) \in [d]^n.
\end{align*}
\end{dfn}

We may use the coefficients of the contracted state $\ket{T_{\vec \pi}}$ to define functions $P_{\vec \pi, \vec i} \in \C[\Mat_{D_1 \dots D_m \times D_1 \dots D_m}^d]$ as
\begin{align}\label{eq:peps general permutations}
  P_{\vec \pi, \vec i}(T) = \braket{i_1, \dots, i_n| T_{\vec \pi}}.
\end{align}

For $m = 1$ we get back the usual notion of MPS.
Note that in this case, if we assume the contraction graph to be connected, there is a unique way to contract the tensors, corresponding to any full cycle in $S_n$.
Indeed, for $T \in \Mat_{D \times D}^d$ and $\pi = (1\, 2 \dots n) \in S_n$ we see that $\ket{T_{\pi}} = \ket{T_n}$ as defined in \cref{eq:mps}.

We also note that we recover the notion of uniform PEPS on a grid by choosing appropriate permutations.
For instance, for $m = 2$, and a grid of size $n_1 \times n_2$ this would correspond to using the permutations
\begin{align*}
  \pi_1 &= (1\; 2\dots n_1)(n_1+1\; n_1+2 \dots 2n_1) \dots ((n_2-1)n_1 + 1\; (n_2-1)n_1 + 2 \dots n_2n_1) \\
  \pi_2 &= (1 \; n_1+1 \dots (n_2-1)n_1 + 1) (2 \; n_1+2 \dots (n_2-1)n_1 + 2) (n_1 \; 2n_1 \dots n_2n_1).
\end{align*}
This yields (upon appropriately identifying the copies of $\HH_{\phys}$) an equivalence $\ket{T_{n_1,n_2}} = \ket{T_{(\pi_1,\pi_2)}}$.

As in the MPS case, we have a `gauge group' acting on the tensor.
We can now act with a different group element along each direction $k \in [m]$.
\begin{dfn}[Gauge action]\label{dfn:peps gauge action}
We define the \emph{gauge action} of $\vec g \in G = \GL(D_1) \times \dots \times \GL(D_m)$, where $\vec g = (g_1,\dots,g_m)$, on $T \in \Mat_{D_1 \dots D_m\times D_1 \dots D_m}^d$ as
\begin{align*}
  \vec g \cdot T = \left((g_1 \ot \dots \ot g_m) T^{(i)}(g_1^{-1} \ot \dots \ot g_m^{-1})\right)_{i=1}^d.
\end{align*}
If we think of $T$ as a quantum state $\ket{T}$ in $\left(\bigotimes_{k=1}^m \HH_{k,1} \ot \HH_{k,2}\right) \ot \HH_{\phys}$, the gauge action can be written as
\begin{align*}
  \vec g \cdot \ket{T} = \left(\left(\bigotimes_{k=1}^m g_k \ot g_k^{-T}\right) \ot I\right) \ket{T}.
\end{align*}
\end{dfn}

As in the MPS case, it is easy to see that this action keeps the associated PEPS invariant.
By continuity, this is also true after taking limits, giving rise to the following lemma.

\begin{lem}
For every $T \in \Mat_{D_1 \dots D_m\times D_1 \dots D_m}^d$, $G = \GL(D_1) \times \dots \times \GL(D_m)$, if $T' \in \overline{G \cdot T}$, then for all $\vec \pi \in S_n^{m}$
\begin{align*}
  \ket{T_{\vec \pi}} = \ket{T'_{\vec \pi}}.
\end{align*}
and in particular
\begin{align*}
  P_{\vec \pi, \vec i}(T) = P_{\vec \pi, \vec i}(T').
\end{align*}
\end{lem}

In other words, the coefficient functions $P_{\vec \pi, \vec i}$ are polynomials in the invariant ring $\C[\Mat_{D_1\dots D_m \times D_1 \dots D_m}^d]^{G}$.
We have a corresponding notion of gauge equivalence.

\begin{dfn}[Gauge equivalence]
Let $S, T \in \Mat_{D_1 \dots D_m\times D_1 \dots D_m}^d$ be two matrix tuples.
Let $G = \GL(D_1) \times \dots \times \GL(D_m)$.
We say that $S$ and $T$ are \emph{gauge equivalent} if and only if $\overline{G \cdot S} \cap \overline{G \cdot T} \neq \emptyset$.
\end{dfn}

\subsection{Minimal canonical form}\label{sec:minimal-peps-basics}
We consider uniform PEPS in $m$ spatial dimensions with bond dimensions $D_1, \dots, D_m$ and physical dimension $d$.
We denote the gauge group by $G = \GL(D_1) \times \dots \times \GL(D_m)$.
We denote by $K = \U(D_1) \times \dots \times \U(D_m) \subset G$ the unitary subgroup.
We can now follow exactly the same approach as in the MPS case to define the minimal canonical form, and the same general results from geometric invariant theory allow us to prove its basic properties.

\begin{dfn}[Minimal canonical form PEPS]
We say $\mcf{T} \in \Mat_{D_1 \dots D_m\times D_1 \dots D_m}^d$ is a minimal canonical form of~$T \in \Mat_{D_1 \dots D_m\times D_1 \dots D_m}^d$ if it is an element of minimal norm in the orbit closure~$\overline{G \cdot T}$, i.e.,
\begin{align*}
  \mcf{T} = \argmin \,\{ \norm{S} : S \in \overline{G \cdot T} \}.
\end{align*}
We say $T \in \Mat_{D_1 \dots D_m\times D_1 \dots D_m}^d$ is \emph{in canonical form} if it is a minimal canonical form for itself, i.e. an element of minimal norm in $\overline{G \cdot T}$.
\end{dfn}

The norm considered in the definition is, as in the MPS case, the Euclidean norm of $T$ (or $\ket T$):
\begin{align*}
  \norm{T} = \sqrt{\braket{T | T}} = \left(\sum_{i=1}^d \tr\mleft[(T{(i)})^\dagger T^{(i)}\mright]\right)^{1/2}.
\end{align*}

The minimal canonical form is not uniquely defined, but it is unique up to the action by the unitary group $K = \U(D_1) \times \dots \times \U(D_m)$:

\begin{thm}[Minimal canonical form]\label{thm:oci peps}
Let $S, T \in \Mat_{D_1 \dots D_m\times D_1 \dots D_m}^d$. Then the following are equivalent:
\begin{enumerate}
  \item\label{it:common canonical form} $S$ and $T$ have a common minimal canonical form.
  \item\label{it:related by unitary} If $\mcf{S}$ and $\mcf{T}$ are minimal canonical forms for $S$ and $T$, then $K \cdot \mcf{S} = K \cdot \mcf{T}$.
  \item $S$ and $T$ are gauge equivalent, i.e., $\overline{G \cdot S} \cap \overline{G \cdot T} \neq \emptyset$.
\end{enumerate}
\end{thm}
\begin{proof}
This is an immediate consequence of \cref{thm:git canonical}.
\end{proof}

Recall that if $T \in \Mat_{D_1 \dots D_m\times D_1 \dots D_m}^d$ is a PEPS tensor, we saw in \cref{eq:ket-T-PEPS} that we may consider it as a quantum state $\ket{T}$.
For each `direction' $k \in [m]$ we have two virtual Hilbert spaces $\HH_{k,1}$ and $\HH_{k,2}$ of dimension $D_k$ and there is the physical Hilbert space $\HH_{\phys}$ of dimension $d$.
We denote by $\rho_{k,j}$ the reduced state of $\rho = \proj{T}$ on $\HH_{k,j}$.

The characterization of minimum norm vectors as critical norm vectors in \cref{thm:kempf ness} can be used to give a condition for a tensor to be in minimal canonical form.
To find this condition we perform a computation similar to the MPS case.
We identify $i\Lie(K)$ with $\Herm_{D_1} \times \dots \times \Herm_{D_m}$ and compute for $\vec X = (X_1,\dots,X_m) \in \Herm_{D_1} \times \dots \times \Herm_{D_m}$
\begin{align}\label{eq:moment map peps}
\begin{split}
  &\partial_{t=0} \norm{(e^{tX_1}, \dots, e^{tX_m}) \cdot T}^2
  \\
  &\qquad = \partial_{t=0} \, \tr\mleft[\sum_{i=1}^d (e^{2tX_1} \ot \dots \ot e^{2tX_m}) T^{(i)} (e^{-2tX_1} \ot \dots \ot e^{-2tX_m})  (T^{(i)})^{\dagger} \mright]\\
  &\qquad = 2\sum_{k=1}^m \tr\mleft[I_{D_1} \ot \dots \ot X_k \ot \dots \ot I_{D_m}\left(\sum_{i=1}^d T^{(i)} (T^{(i)})^{\dagger} - (T^{(i)})^{\dagger}T^{(i)} \right)\mright]\\
  &\qquad = 2\sum_{k=1}^m \tr\mleft[X_k\left(\rho_{k,1} - \rho_{k,2}^T \right)\mright].
\end{split}
\end{align}

\begin{thm}[Characterization]\label{thm:symmetric virtual rdm}
Let $T  \in \Mat_{D_1 \dots D_m\times D_1 \dots D_m}^d$.
Then $T$ is in minimal canonical form if and only if $\norm{\vec g \cdot T} \geq \norm{T}$ for all $\vec g \in G$.
This is the case if and only if the reduced density matrices of $\rho = \proj{T}$ on the virtual bonds are the same in each direction, up to a transpose:
\begin{align}
  \rho_{k,1} = \rho_{k,2}^T \qquad (\forall k \in [m])
\end{align}
\end{thm}

\begin{proof}
By \cref{thm:kempf ness}, $T$ is in minimal canonical form if and only if it is critical, which means that the derivative in \cref{eq:moment map peps} should vanish for all $\vec X$.
This is equivalent to $\rho_{k,1} = \rho_{k,2}^T$ for all $k \in [m]$.
\end{proof}

\begin{figure}
\centering
\begin{overpic}[width=0.7\textwidth,grid=false]{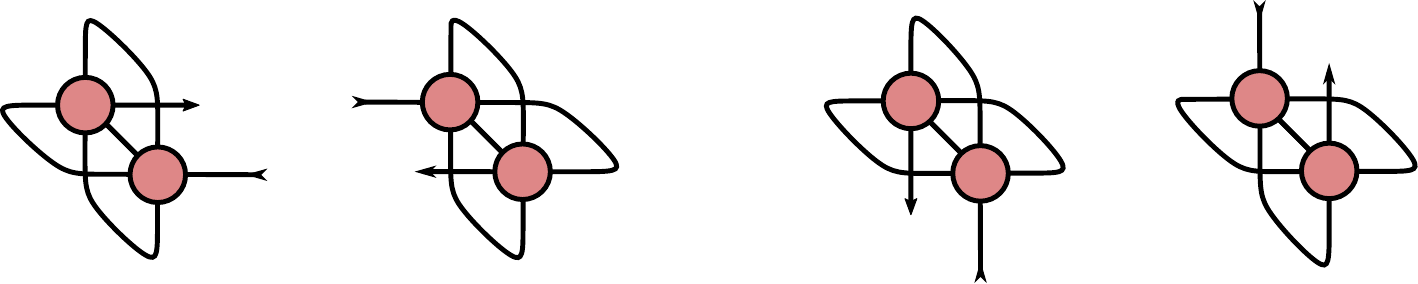}
\put(21,9.5){$=$} \put(78.5,9.5){$=$}
\end{overpic}
\caption{\emph{Minimal canonical form PEPS:} Graphical version of the conditions $\rho_{1,1} = \rho_{1,2}^T$ and $\rho_{2,1} = \rho_{2,2}^T$ from \cref{thm:symmetric virtual rdm}}
\label{fig:peps mcf condition}
\end{figure}

These conditions are illustrated in \cref{fig:peps mcf condition} for $m = 2$.
Without the framework of invariant theory, it is not clear that one can indeed transform any tensor by gauge transformations to satisfy the conditions in \cref{thm:symmetric virtual rdm}.
This is an important difference with earlier proposals for canonical forms for PEPS.
For instance, \cite{phien2015fast} proposes a canonical form based on a similar (but different) condition.
However, in that case, it is not clear that such a canonical form indeed exists for any tensor.

Both \cref{thm:oci peps} and \cref{thm:symmetric virtual rdm}, giving the ``uniqueness'' of the canonical form and its algebraic characterization respectively, only require situations in which one is already interested in analyzing tensors related by gauge transformations. Reducing to such a situation is the goal of the \emph{Fundamental Theorems}.
For MPS we already saw such fundamental theorems, in particular \cref{thm:fundamental thm mps}, which apply to general MPS.

For PEPS the situation is more complicated, but for important special cases, fundamental theorems are known.
In particular, fundamental theorems are known for the family of \emph{normal} tensors \cite{cirac2017matrix}, proven for the uniform 2D case in \cite{perez2010characterizing}, and extended to the general case in \cite{molnar2018normal}.

To define normal tensors,  we first recall the notion of an \emph{injective} PEPS tensor.
A tensor $T \in \Mat_{D_1 \dots D_m \times D_1 \dots D_m}^d$ is injective if it is injective as a map from the virtual legs to the physical legs, i.e. if it is injective as a $d \times D_1^2\dots D_m^2$ matrix.
The tensor $T$ is \emph{normal} if it is injective after blocking together a number of copies to a single new tensor.
Let us explain what we mean by `blocking'.
Given $T \in \Mat_{D_1 \dots D_m \times D_1 \dots D_m}^d$ we can contract $n = n_1 \dots n_m$ copies of $T$ on a rectangular lattice of size $n_1 \times \dots \times n_m$ sites to obtain a new tensor $\tilde T$ with physical dimension $d^n$ and bond dimensions $D_1^{n_2 \dots n_m}, D_2^{n_1 n_3 \dots n_m}, \dots, D_m^{n_1 \dots n_{m-1}}$.
The tensor $T$ is normal if there exists some blocking such that the resulting tensor $\tilde T$ is injective.

Hence in the normal case, which is a generic condition, \cref{thm:oci peps} and \cref{thm:symmetric virtual rdm} already apply to show the following statement (for simplicity we only write down the two-dimensional case):

\begin{cor}\label{cor:unitary-charact-symm-normal}
  Two normal tensors $T$ and $S$ in $\Mat_{D_1D_2 \times D_1 D_2}^d$ define the same state in all $n_1\times n_2 $ grids, i.e. $\ket{T_{n_1,n_2}}= \ket{S_{n_1,n_2}}$ for all $n_1, n_2 \in \N$, if and only if their corresponding minimal canonical forms $\mcf{S}$ and $\mcf{T}$ are related by local unitary gauges: $\mcf{S} = \vec U \cdot \mcf{T}$ for a suitable unitary $\vec U \in \U(D_1) \times \U(D_2)$.
\end{cor}

Moreover, we will see below in \cref{lem:injective} that the orbit of a normal tensor is always closed.
However, this is not the end of the story.
There are other (non-normal) tensors which define the same state in all $n_1\times n_2$ grids, but are nevertheless \emph{not} related by a gauge transformation.
An explicit example appears in \cite{molnar2018generalization}, in the context of 2D SPT phases.
We provide the example here:

\begin{exa}\label{ex:Andras-SPT}
The idea of the example is simple but ingenious.
Take pairs of MPS normal tensors $A$ and $B$ so that $\ket{A_4}=\ket{B_4}$ but $\ket{A_j}\not = \ket{B_j}$ for all $j>4$.%
\footnote{It is only proven in \cite{molnar2018generalization} that $\ket{A_5}\not = \ket{B_5}$, but since it is also shown that both $A$ and $B$ become injective when blocking two sites, known bounds for the fundamental theorem \cite{molnar2018normal} imply already that if $\ket{A_j} = \ket{B_j}$ for any $j\ge 6$, then $A$ and $B$ would be gauge-related and then  $\ket{A_5} = \ket{B_5}$.}
The explicit examples of \cite{molnar2018generalization} have physical dimension $2$ and are given by the matrices:
$$
  A^{(1)} = \begin{pmatrix} 1 & 0 \\ 0 & 2 \end{pmatrix} \quad , \quad
  A^{(2)} = \begin{pmatrix} 24 & -10 \\ 17 & -3 \end{pmatrix}.
$$
where $B^{(1)}=A^{(1)}$ and $B^{(2)}=-A^{(2)}$.

Now, in each vertex of a two dimensional grid, place four qubits and, by joining each one of those qubits with the closest one in each of the nearest neighbor sites, fill in the lattice with a set of non-overlapping plaquettes $\mathcal{P}$. The states we are interested are $\ket{M_A}=\bigotimes_{p\in\mathcal{P}}  \ket{A_4}_p$ and $\ket{M_B}=\bigotimes_{p\in\mathcal{P}}  \ket{B_4}_p$. It is now obvious how to define the associated PEPS tensors $M_A$ and $M_B$ for the vertices. Just take, with the appropriate identification of indices, $M_A=A^{\otimes 4}$, $M_B= B^{\otimes 4}$ (recall that each vertex contains four qubits and therefore the physical dimension is $16$). It is shown in \cite{molnar2018generalization} that tensors $M_A$, $M_B$ are not in the same $\GL_4\times \GL_4$ orbit. One can indeed show that the closure of their orbits do not intersect. One possibility is just to realize that, because of the symmetry of the tensors  $M_A$ and $M_B$, they are already in minimal canonical form, and therefore their orbits are already closed. The other possibility is to compare $M_A$ and $M_B$ in different contraction graphs $\Gamma$. It is easy to find some $\Gamma$ for which the length of some of the plaquettes are larger than $4$ and then the fact that $\ket{A_j}\not = \ket{B_j}$ for $j>4$ implies that the associated states $\ket{M_{A,\Gamma}}$ and $\ket{M_{B,\Gamma}}$ are different, which in turn implies that the orbits of $M_A$ and $M_B$ cannot intersect.
\end{exa}

\subsection{Fundamental theorem and invariant theory of uniform PEPS}\label{subsec:peps fund thm}
This example makes clear that we have to change perspective to derive a Fundamental Theorem which is an analog to the MPS one (\cref{thm:fundamental thm mps}).
Instead of starting with the condition $\ket{S_{n_1,n_2}} = \ket{T_{n_1,n_2}}$ for all $n_1, n_2$, and asking how the tensors $S$ and $T$ are related, we start with the condition that $S$ and $T$ are gauge equivalent, and we ask how we can characterize this based on the corresponding tensor network states.
It turns out that we need to compare the states not just on grids, but on arbitrary contraction graphs.
That is, the appropriate conditions is $\ket{S_{\vec \pi}} = \ket{T_{\vec \pi}}$ for tuples of permutations $\vec \pi$.

Additionally, for MPS we found that it suffices to consider systems of size at most $D^2$ (\cref{thm:fundamental thm mps}) or even~$\tilde O(D)$ (\cref{rem:improved mps}).
For $m \geq 2$ we prove a similar bound, but now we need a system size exponential in~$D$ (and we show below, in \cref{prp:degree lower bound}, that this exponential dependence cannot be avoided).
Formally, we have the following weak version of a Fundamental Theorem, illustrated in \cref{fig:fundamental theorem}.

\begin{thm}[Fundamental Theorem for PEPS]\label{thm:peps fundamental thm}
Let $S, T \in \Mat_{D_1 \dots D_m\times D_1 \dots D_m}^d$. Then the following are equivalent:
\begin{enumerate}
  \item\label{it:orbit closures intersect} The $G$-orbit closures of $S$ and $T$ intersect, i.e., $\overline{G \cdot S} \cap \overline{G \cdot T} \neq \emptyset$.
  \item\label{it:same state} $\ket{S_{\pi_1,\dots,\pi_m}} = \ket{T_{\pi_1,\dots,\pi_m}}$ for all $\pi_k \in S_r$ for all $r \in \N$.
  \item\label{it:same state bounded degree} $\ket{S_{\pi_1,\dots,\pi_m}} = \ket{T_{\pi_1,\dots,\pi_m}}$ for all $\pi_k \in S_r$  for $r \leq \exp(c m D^2 \log D)$ where $D = \max\{D_1,\dots,D_m\}$ and  $c$ is a constant.
\end{enumerate}
\end{thm}

To prove this result, we start with the following lemma (which is a basic result in invariant theory \cite[\S 4.6]{kraft-procesi-invariant}), which allows us to reduce the study of invariant polynomials $\C[\Mat_{D\times D}^d]^{G}$ to the study of multilinear invariant polynomials. While the result is a basic one, it is a key component in proving a number of \emph{first fundamental theorems} in invariant theory, see \cite{kraft-procesi-invariant} for more details.

\begin{lem}\label{lem:reduce to multilinear}
For any subgroup $G \subset GL(D)$, any polynomial $P$ in the ring of invariant polynomials $\C[\Mat_{D\times D}^d]^{G}$ can be written as a linear combination of multihomogeneous invariant polynomials $P_{\vec n}$ of some multidegree $\vec n=(n_1,\dots,n_d)$, each of which can be written as
\begin{align}\label{eq:P to Q}
  P_{\vec n}(M^{(1)},\dots,M^{(d)}) = Q(\underbrace{M^{(1)},\dots,M^{(1)}}_{n_1 \text{ times}},\dots,\underbrace{M^{(d)},\dots,M^{(d)}}_{n_d \text{ times}}),
\end{align}
where $Q$ is a multilinear $G$-invariant polynomial in $n = \sum_{i=1}^d n_i$ matrix variables.
\end{lem}

\begin{proof}
Let $P = P(M^{(1)},\dots,M^{(d)}) \in \C[\Mat_{D\times D}^d]^{G}$.
First we show that we may assume that $P$ is multihomogeneous, i.e., homogeneous of some degree~$n_i$ in each matrix variable $M^{(i)}$.
Indeed, we can write
\begin{align*}
  P(M^{(1)},\dots,M^{(d)}) = \sum_{\vec n=(n_1, \dots, n_d)} P_{\vec n}(M^{(1)},\dots,M^{(d)}),
\end{align*}
where $P_{\vec n}$ is homogeneous of degree $n_i$ in the matrix variable $M^{(i)}$.
Since the space of homogeneous polynomials of multidegree $\vec n$ is invariant under $GL(D)$, and spaces of different multidegree are linearly independent, each $P_{\vec n}$ is $G$-invariant.
Thus we may without loss of generality assume that $P = P_{\vec n}$.
Next, we reduce to \emph{multilinear} invariants of some possibly larger number of matrices, as follows.
Consider $P(M^{(1,1)} + \dots + M^{(1,n_1)}, \dots, M^{(d,1)} + \dots + M^{(d,n_d)})$, a polynomial in formal matrix variables $M^{(i,j)}$ for $i\in[d]$ and $j\in[n_i]$, and write
\begin{align*}
  P(M^{(1,1)} + \dots + M^{(1,n_1)}, \dots, M^{(d,1)} + \dots + M^{(d,n_d)})
= \sum_{\vec h = (h_{1,1}, \dots, h_{d,r})} P_{\vec h}(M^{(1,1)}, \dots, M^{(d,n_d)}),
\end{align*}
where $P_{\vec h}$ is homogeneous of degree $h_{i,j}$ in each matrix variable $M^{(i,j)}$.
Now note that for all $t_{1,1},\dots,t_{d,n_d}$,
\begin{align}\label{eq:define P_h}
\begin{split}
&\qquad P(t_{1,1} M^{(1,1)} + \dots + t_{1,n_1} M^{(1,n_1)}, \dots, t_{d,1} M^{(d,1)} + \dots + t_{d,n_d} M^{(d,n_d)}) \\
&= \sum_{\vec h = (h_{1,1}, \dots, h_{d,n_d})} \vec t^{\vec h} P_{\vec h}(M^{(1,1)}, \dots, M^{(d,n_d)}),
\end{split}
\end{align}
so if we take $M^{(i,j)} \equiv M^{(i)}$ for all $i\in[d]$ and $j\in[n_i]$ we have
\begin{align*}
&\qquad P(t_{1,1} M^{(1)} + \dots + t_{1,n_1} M^{(1)}, \dots, t_{d,1} M^{(d)} + \dots + t_{d,n_d} M^{(d)}) \\
&= \sum_{\vec h = (h_{1,1}, \dots, h_{d,n_d})} \vec t^{\vec h} P_{\vec h}(\underbrace{M^{(1)},\dots,M^{(1)}}_{n_1 \text{ times}},\dots,\underbrace{M^{(d)},\dots,M^{(d)}}_{n_d \text{ times}}).
\end{align*}
On the other hand, by multihomogeneity,
\begin{align*}
&\qquad P(t_{1,1} M^{(1)} + \dots + t_{1,n_1} M^{(1)}, \dots, t_{d,1} M^{(d)} + \dots + t_{d,n_d} M^{(d)}) \\
&= (t_{1,1} + \dots + t_{1,n_1})^{n_1} \cdots (t_{d,1} + \dots + t_{d,n_d})^{n_d} P(M^{(1)}, \dots, M^{(d)}) \\
&= \sum_{\vec h = (h_{1,1}, \dots, h_{d,n_d})} \binom{n_1}{h_{1,1} \ \ldots \ h_{1,n_1}} \cdots \binom{n_d}{h_{1,1} \ \ldots \ h_{1,n_d}} \vec t^{\vec h} P(M^{(1)}, \dots, M^{(d)}).
\end{align*}
Comparing coefficients and specializing to $\vec h=(1,\dots,1)$, we find that
\begin{align*}
  P(M^{(1)}, \dots, M^{(d)}) = \frac1{n_1! \cdots n_d!} P_{1,\dots,1}(\underbrace{M^{(1)},\dots,M^{(1)}}_{n_1 \text{ times}},\dots,\underbrace{M^{(d)},\dots,M^{(d)}}_{n_d \text{ times}}).
\end{align*}
Note that $P_{1,\dots,1}$ is a multilinear polynomial in $\sum_{i=1}^d n_i$ matrix variables.
Since the left-hand side of \cref{eq:define P_h} is $G$-invariant, we may also assume that $P_{1,\dots,1}$ is $G$-invariant.
\end{proof}

We now return to our setting, where $G = \GL(D_1) \times \dots \times \GL(D_m)$, and use this lemma to prove.

\begin{prp}\label{prp:peps inv}
The ring of invariant polynomials $\C[\Mat_{D_1 \dots D_m\times D_1 \dots D_m}^d]^{G}$ is generated by functions $P_{\vec \pi,\vec i}$ as in \cref{eq:peps general permutations} for $n \leq \exp(c m D^2 \log(mD))$ where $D = \max\{D_1,\dots,D_m\}$ and $c > 0$ is a universal constant.
\end{prp}

\begin{proof}
Let $P = P(T^{(1)}, \dots, T^{(d)}) \in \C[\Mat_{D_1 \dots D_m\times D_1 \dots D_m}^d]^{G}$.
By \cref{lem:reduce to multilinear} with $D = D_1 \dots D_m$ we may reduce to the case where $P =  P_{\vec n}$ for some $\vec n = (n_1, \dots, n_d)$, and we can write
\begin{align*}
  P(T^{(1)}, \dots, T^{(d)}) = \braket{R, \underbrace{T^{(1)} \ot \dots \ot T^{(1)}}_{n_1 \text{ times}} \ot \underbrace{T^{(2)} \ot \dots \ot T^{(2)}}_{n_2 \text{ times}} \ot \dots \ot \underbrace{T^{(d)} \ot \dots, T^{(d)}}_{n_d \text{ times}}},
\end{align*}
where $\braket{\cdot,\cdot}$ is the trace inner product and where
\begin{equation*}
  R \in \left( \End(\C^{D_1} \ot \dots \ot \C^{D_m})^{\ot r} \right)^{G}.
  \vspace{-1mm}
\end{equation*}
The total degree is given by $n = \sum_{i=1}^d n_i$.
Now note that
\begin{align*}
  \Bigl( \End(\C^{D_1} \ot \dots \ot \C^{D_m})^{\ot n} \Bigr)^{G}
&\cong \End((\C^{D_1})^{\ot n})^{\GL(D_1)} \ot \dots \ot \End((\C^{D_m})^{\ot n})^{\GL(D_m)} \\
&\cong \C[R_{\pi_1} : \pi_1 \in S_n] \ot \dots \ot \C[R_{\pi_m} : \pi \in S_n]\\
&\cong \C[R_{\pi_1} \ot \dots \ot R_{\pi_m} : \pi_1, \dots, \pi_m \in S_n]
\end{align*}
where we denote by $R_{\pi_k}$ the operator acting on $(\C^{D_k})^{\ot n}$ permuting the $n$ copies of $\C^{D_k}$ according to $\pi_k$.
Thus, $R$ is a linear combination of elements of the form $R_{\vec \pi} = R_{\pi_1} \ot \dots \ot R_{\pi_m}$, for $\vec \pi = (\pi_1, \dots, \pi_m)$.
We conclude that the ring of invariant polynomials $\C[\Mat_{D_1 \dots D_m\times D_1 \dots D_m}^d]^{G}$ is generated \emph{as a vector space} by the polynomial functions~$P_{\vec{\pi},\vec i}$ as in \cref{eq:peps general permutations} for $\vec \pi \in S_n^{m}$ and $n\in\N$.
In particular, the invariant polynomials of degree at most~$r$ are spanned by the~$P_{\vec \pi, \vec i}$ for~$\vec \pi \in S_n^m$ and~$\vec i \in [d]^n$ for $n \leq r$.

We now use general results in invariant theory to bound the degree necessary to generate the invariant ring \emph{as an algebra}.
For convenience, we write $V := \Mat_{D_1 \dotsb D_m \times D_1 \dotsb D_m}$, so we are interested in degree bounds for the action of $G = \GL(D_1) \times \dotsb \times \GL(D_m)$ on $V^d = \Mat_{D_1 \dotsb D_m \times D_1 \dotsb D_m}^d$.
We first appeal to a classical theorem by Weyl \cite[II.5~Thm.~2.5.A]{weyl1946classical} which states that if $d > \dim(V)$, a generating set of invariants for~$V^d$ can be obtained by acting with $\GL(d)$ on a generating set for $\C[V^{\dim(V)}]^G \hookrightarrow \C[V^d]^G$ (cf.~\cite[\S7.1]{kraft-procesi-invariant}).
In particular, any degree bound for $d = \dim(V)$ also applies to $d > \dim(V)$.
Accordingly, we may assume without loss of generality that $d \leq \dim(V)$.
Next, we observe that since we act by simultaneous conjugation, the invariants for the action of~$G$ are the same as for~$G' := \SL(D_1) \times \dotsb \times \SL(D_m)$, so we can restrict to the latter.
By results of Derksen~\cite{derksen2001polynomial} the ring of invariants is generated by invariant polynomials of degree at most
\begin{equation}
  \label{eq:peps degree abstract}
  r \leq \frac{3}{8} \dim(V^d) (H^{t-\dim(G')} A^{\dim(G')})^2
\end{equation}
where $t$, $H$, $A$ are integers computed as follows.
We think of $G'$ as being embedded in $\oplus_{k=1}^m \Mat_{D_k \times D_k} \cong \C^t$, with $t = \sum_{k=1}^m D_k^2$.
Then $G'$ is defined as the common zero set of the polynomials $\det(g_k) - 1$ for $k \in [m]$.
The integer~$H$ is the maximal degree of these polynomials, i.e., $H = \max_k D_k$.
If one fixes an arbitrary basis of~$V^d$, the matrix entries of the representation of~$G'$ are polynomial functions of the coordinates of $\C^t$ (that is, the entries of the $g_k$).
The integer~$A$ is the maximal degree of these polynomials.
To compute it, note that~$(g_1,\dots,g_m) \in G'$ acts on a matrix tuple $T = (T^{(i)})_{i=1}^d \in V^d$ by simultaneous conjugation by~$g_1 \otimes \cdots \otimes g_m$.
Thus, we left multiply each matrix $T^{(i)}$ with $g_1 \otimes \dotsb \otimes g_m$, the entries of which are polynomials of degree~$m$ in the entries of the~$g_k$, and we right multiply each~$T^{(i)}$ with
\begin{equation}\label{eq:inv}
  g_1^{-1} \otimes \dotsb \otimes g_m^{-1} = \adj(g_1) \otimes \dotsb \otimes \adj(g_m),
\end{equation}
where $\adj(g_k)$ is the adjugate matrix of $g_k$ (here we used that $g_k \in \SL(D_k)$, so that we did not have to divide by the determinant when computing the inverse); since the entries of the adjugate matrix are given by cofactors of~$g_k$ and hence have degree $D_k - 1$, the entries of~\eqref{eq:inv} are polynomials of degree $\sum_{k=1}^m (D_k - 1)$.
Therefore, each matrix entry of the representation of~$G'$ is a polynomial of degree~$A = m + \sum_{k=1}^m (D_k - 1) = \sum_{k=1}^m D_k$.

Evaluating \cref{eq:peps degree abstract} with $\dim(V^d) = d \prod_{k=1}^m D_k^2$, $d \leq \dim(V)$, $\dim(G') = \sum_{k=1}^m (D_k^2 - 1)$, $H = \max_k D_k$, $t = \sum_{k=1}^m D_k^2$ and $A = \sum_{k=1}^m D_k$ shows that we can bound the required degree by
\begin{align*}
  n
\leq \frac{3}{8} \left(d \prod_{k=1}^m D_k^2\right) \left(\left(\max_k D_k\right)^{m} \left( \sum_{k=1}^m D_k \right)^{\sum_{k=1}^m (D_k^2 - 1)}\right)^2
\leq \exp(c mD^2 \log(mD))
\end{align*}
for some universal constant $c \geq 0$.
\end{proof}

\begin{cor}[Lifting symmetries]\label{cor:u-invariance peps}
Suppose that $S, T \in \Mat_{D_1\dots D_m\times D_1 \dots D_m}^d$ are in minimal canonical form and $u\in \U(d)$ is a unitary such that $u^{\ot n} \ket{S_{\vec \pi}} = \ket{T_{\vec \pi}}$ for all $\vec \pi \in S_{n}^{m}$ and $n \in \N$.
Then there exist unitaries $U_k \in \U(D_k)$ such that $(I \ot u) \ket S = ((\bigotimes_{k=1}^m U_k \ot \bar U_k) \ot I) \ket T$.
\end{cor}
\begin{proof}
Let $S' \in \Mat_{D\times D}^d$ be the matrix tuple defined by
\begin{align*}
  \ket{S'} := (I \ot u) \ket{S}.
\end{align*}
Then $S'$ is also in minimal canonical form, since $u$ is unitary and hence we have $\norm{\vec g \cdot S} = \norm{\vec g \cdot S'}$ for all~$\vec g \in G$.
Moreover, by construction it holds that
\begin{align*}
  \ket{S'_{\vec \pi}} = u^{\ot n} \ket{S_{\vec \pi}} = \ket{T_{\vec \pi}}
\end{align*}
for all $\vec \pi \in S_{n}^{m}$ and $n\in\N$.
Thus \cref{thm:peps fundamental thm} shows that $S'$ and $T$ are gauge equivalent, and it follows from \cref{thm:oci peps} that there exist unitary gauge transformations $U_k \in \U(D_k)$ such that $(I \ot u) \ket S = ((\bigotimes_{k=1}^m U_k \ot \bar U_k) \ot I) \ket T$.
\end{proof}

The degree bounds in \cref{prp:peps inv} are a direct consequence of deep and completely general results in invariant theory.
These bounds are in general not necessarily sharp.
As an example, the degree bounds obtained in this way for the MPS case are still exponential, while we know from \cref{thm:mps inv} that in this special case we have a degree bound of~$D^2$.
Moreover, we know from \cref{rem:improved mps} that in this case invariants of degree~$\tilde O(D)$ already suffice to determine whether two MPS tensors are gauge equivalent.

However, this is quite special for one spatial dimension.
For PEPS with spatial dimension~$m \geq 2$, we now show that one in general needs to consider invariants of degree exponential in the bond dimension in order to decide whether two PEPS tensors are gauge equivalent (even if one is the zero tensor).
For convenience we take~$m = 2$, $D_1 = D_2 = D$, and $d=1$ (that is, the tensor networks defined by the PEPS tensors are scalars).

\begin{prp}[Degree lower bound]\label{prp:degree lower bound}
There exists a function $n_{\min}(D) = \exp(\Omega(D))$ and, for every $D$, a tensor $T \in \Mat_{D^2 \times D^2}$ with the following properties:
\begin{enumerate}
\item For any invariant polynomial $P \in \C[\Mat_{D^2 \times D^2}]^{\GL(D) \times \GL(D)}$ of degree less than $n_{\min}(D)$, we have $P(T) = P(0)$.
\item There exists an invariant polynomial~$P$ of degree $n_{\min}(D)$ such that $P(T) \neq P(0)$. In particular, we have~$0 \not\in \overline{G \cdot T}$, meaning that $T$ is \emph{not} gauge equivalent to the zero tensor.
\end{enumerate}
In particular, the ring of invariant polynomials $\C[\Mat_{D^2 \times D^2}^d]^{\GL(D) \times \GL(D)}$ for any $d \geq 1$ is not generated by the polynomials of degree $n < n_{\min}(D)$.
\end{prp}

\begin{proof}
The last statement of the proposition is an immediate consequence of the described properties of $T$.
Indeed, if the ring of invariants were generated by invariant polynomials of degree smaller than $n_{\min}(D)$, then $P(T) = P(0)$ for all such polynomials~$P$ would imply that $P(T) = P(0)$ for all invariant polynomials~$P$ -- but we know that $P(T) \neq P(0)$ for at least one invariant polynomial of degree $n_{\min}(D)$.

We will explicitly construct a tensor $T \in \Mat_{D^2 \times D^2}$.
For $d = 1$ and for $\vec \pi \in S_n^{m}$ and $\vec 1 = (1,\dots,1)$ we abbreviate $P_{\vec \pi,\vec 1} = P_{\vec \pi}$.
Since the $P_{\vec \pi}$ for $\vec \pi \in S_n^{\times 2}$ for $n < r$ are homogeneous and span the degree $r$ polynomials in $\C[\Mat_{D^2 \times D^2}]^{\GL(D) \times \GL(D)}$ it suffices to show that $P_{\vec \pi}(T) = 0$ for $\vec \pi \in S_n^{\times 2}$ for $n < n_{\min}$, while there exists some $\vec \pi \in S_n^{\times 2}$ for $n = n_{\min}$ such that $P_{\vec \pi}(T) \neq 0$.
We will take $n_{\min} = 2^{D} + 2^{D-1} - 2$.

To explain the construction and the argument we start with a construction where we allow the physical dimension $d$ to grow with $D$, and we construct a tensor $S \in \Mat_{D^2 \times D^2}^{2D-1}$ with certain properties.
Then, we will use a trick to reduce the physical dimension.
Let $\{\ket{j}\}_{j=0}^{D-1}$ denote the standard basis of $\C^D$.
We choose the tensor $S$ as follows:
\begin{align*}
  S^{(1)} = \ketbra{0}{1} \ot \ketbra{0}{1}, \qquad S^{(2j)} = \ketbra{j}{0} \ot \ketbra{0}{j}, \qquad S^{(2j+1)} = \ketbra{0}{j+1} \ot \ketbra{j}{j+1}
\end{align*}
for $j = 1,\dots,D-1$ and where the index $j$ should be read modulo $D$ (so $\ket{D} = \ket{0}$).
We will now argue that one the one hand, for all $\vec i = (i_1,\dots,i_n) \in [2D-1]^n$ and $n < 2^{D} + 2^{D-1} - 2$ we have $P_{\vec \pi,\vec i}(S) = 0$ for all $\vec \pi$, while on the other hand for $n = 2^{D} + 2^{D-1} - 2$ there is some $\vec \pi$ and $\vec i = (i_1,\dots,i_n)$ with $P_{\vec \pi,\vec i}(S) \neq 0$.

We start by showing that if $\vec i = (i_1,\dots,i_n)$ with $n < 2^{D} + 2^{D-1} - 2$ then we have $P_{\vec \pi,\vec i}(S) = 0$.
To conveniently reason about contractions in the tensor network picture we will name the four virtual legs of the tensors as follows:
\begin{align*}
  \ketbra{\text{left}}{\text{right}} \ot \ketbra{\text{down}}{\text{up}}
\end{align*}
and call the two directions `horizontal' and `vertical'.
In the tensor network picture, we observe that for each even $i = 2j$ one can only contract the upper leg of $S^{(2j)}$ along the vertical direction with a copy of $S^{(2j+1)}$ in order for the result to be nonzero.
That is, if we have $i_k = i$ even, then $\pi_2$ must map $k$ to $l$ where $i_l = i+1$.
Similarly, for $i = 2j+1 < 2D-1$ odd we need to contract the right leg of $S^{(2j+1)}$ with the left leg of a copy of $S^{(2j + 2)}$ in the horizontal direction and its upper leg with a copy of $S^{(2j + 3)}$ in the vertical direction.
Together these conditions imply that if $n_i$ denotes the number of copies of $S^{(i)}$ one requires in order for the contraction to be nonzero, we have $n_{i+2} \geq n_{i+1} + n_{i}$ for $i < 2D-1$ odd and $n_{i+1} \geq n_{i}$ for $i \leq 2D$ even.
By similar reasoning, for even $i = 2j$, the left leg of a copy of $S^{(2j)}$ needs to be contracted in the horizontal direction with a copy of $S^{(2j - 1)}$, and for odd $i = 2j+1 > 1$, the down leg of a copy of $S^{(2j+1)}$ needs to be contracted in the vertical direction with a copy of $S^{(2j)}$ or $S^{(2j-1)}$.
This implies that if $n_i \neq 0$ for $i \geq 2$ we also need either $n_{i-1}$ or $n_{i-2}$ to be nonzero and in particular $n_1 \geq 1$.

Solving the recursion with $n_1 \geq 1$ gives $n_{2i+1} \geq 2^i$ and $n_{2i} \geq 2^{i-1}$ for $i = 1,\dots,D-1$.
We then have
\begin{align*}
  n = \sum_{i=1}^{2D - 1} n_i
  \geq 2^{D} + 2^{D-1} - 2.
\end{align*}
On the other hand, it is easy to see that if we take $n_i$ copies of $S^{(i)}$ with $n_1 = 1$, $n_{2i} = 2^{i-1}$ and $n_{2i+1} = 2^{i}$ we can indeed contract to something nonzero.

Now, to prove the proposition, we adapt the previous construction to $d = 1$.
We construct $T \in \Mat_{D^2 \times D^2}$ as
\begin{align*}
  T = T^{(1)} = \sum_{i=1}^{2D-1} S^{(i)}.
\end{align*}
Consider some arbitrary $\vec \pi \in S_n^{m}$.
We may expand $T = \sum_{i} S^{(i)}$ for each copy of $T$ to find
\begin{align*}
  P_{\vec \pi}(T) = \sum_{\vec i \in [d]^n} P_{\vec \pi,\vec i}(S).
\end{align*}
By construction of $S$, each $P_{\vec \pi,\vec i}(S)$ is either zero or one, proving that $P_{\vec \pi}(T) \neq 0$ if and only if there is some $\vec i = (i_1,\dots,i_n)$ such that $P_{\vec \pi,\vec i}(S) \neq 0$.

By our previous arguments for $S$ this implies that for all $n < 2^{D} + 2^{D-1} - 2$ and $\vec \pi \in S_n^{m}$ we have $P_{\vec \pi}(T) = 0$, but that for $n = 2^{D} + 2^{D-1} - 2$ we can find some  $\vec \pi \in S_n^{m}$ such that $P_{\vec \pi}(T) \neq 0$.
\end{proof}

\begin{rem}
The argument of \cref{prp:degree lower bound} can be extended to $m > 2$.
We define a generalization of $S \in \Mat_{D^m \times D^m}^{m(D-1) + 1}$ as follows: for $i = 1, \dotsc, D-1$ and $j = 1, \dotsc, m-1$ set
\begin{equation*}
  S^{(1)} = (\ket{0}\bra{1})^{\otimes m}, \quad
  S^{(m (i - 1) + j + 1)} = (\ket{0}\bra{0})^{\otimes (j - 1)} \otimes \ket{i}\bra{0} \otimes (\ket{0} \bra{i})^{\otimes (m-j)}.
\end{equation*}
and
\begin{align*}
  S^{(m i + 1)} = (\ket{0}\bra{i+1})^{\otimes (m-1)} \otimes \ket{i}\bra{i+1}.
\end{align*}
Note that as before we interpret the basis states modulo $D$, i.e., $\ket{D} = \ket{0}$.
Then again define $T \in \Mat_{D^m \times D^m}$ by
\begin{align*}
  T = T^{(1)} = \sum_{i=1}^{m(D - 1) + 1} S^{(i)}
\end{align*}
Essentially the same argument yields
\begin{align*}
  n_{\min} = 1 + \sum_{i=1}^{D-1} 2^{i(m-1)} \sum_{j=0}^{m-1} 2^j = \exp(\Omega(mD))
\end{align*}
so the degree lower bound also scales exponentially in $m$.
\end{rem}

We note here that proving degree lower bounds is not often an easy task, and in literature often has to employ rather involved and indirect techniques to get exponential lower bounds even in very familiar cases, see e.g., \cite{DERKSEN2020107136, derksen2021polystability}. The technique we use above is far more straightforward and explicit even though the setting we study here is somewhat similar to some of the cases handled in the aforementioned papers.

\subsection{Two-dimensional tensor networks, tilings and topology}\label{sec:tilings}
Consider the following question: given a PEPS tensor $T$ in two spatial dimensions, determine whether there exist $n_1, n_2$ such that the associated state $\ket{T_{n_1,n_2}}$ on a rectangular periodic lattice of size $n_1 \times n_2$ is nonzero.
This problem is undecidable, see \cite{scarpa2020projected}.
The proof of the undecidability given in \cite{scarpa2020projected} is by reducing to the problem of the existence of a periodic tiling given some set of tiles.
Given a set of square tiles where each edge of the tile is associated one of $D$ boundary colors, the question is whether there exists a tiling (meaning that the boundary colors of adjacent tiles match) which is periodic.
Equivalently, this gives a tiling of the two-dimensional torus.
It is known that the existence of such tilings, given a set of tiles, is undecidable in general \cite{gurevich1972remarks}, and in \cite{scarpa2020projected} it was shown how to embed this problem into a PEPS tensor $T$ of bond dimensions $D_1 = D_2 = D$ such that the associated state $\ket{T_{n_1,n_2}}$ on a $n_1 \times n_2$ periodic rectangular lattice is nonzero if and only if there exists a $n_1 \times n_2$ periodic tiling.
The construction of such a tensor $T$ is as follows.
Let $d$ be the number of tiles, label the tiles with an index $i \in [d]$, and similarly label the colors with an index $j \in [D]$.
Then if the tile $i$ has colors $j_1, j_2, j_3, j_4$ on respectively the left, right, upper and lower sides, define $T^{(i)} := \ketbra{j_1}{j_2} \ot \ketbra{j_3}{j_4}$.
It is not very hard to see that under this construction the resulting PEPS state $\ket{T_{n_1,n_2}}$ is nonzero if and only if there exists a $n_1 \times n_2$ periodic tiling.
In fact, the argument in \cite{scarpa2020projected} is for PEPS tensors with boundary conditions, but the undecidability of the existence of periodic tilings \cite{gurevich1972remarks} yields the same result for PEPS with periodic boundary conditions.

Interestingly, \cref{prp:peps inv} shows that if one relaxes the problem to asking whether a PEPS tensor yields the zero state on any contraction graph, the problem is decidable, as we only have to check all graphs of size at most $\exp(\mathcal O(D^2\log D))$.
Alternatively, the PEPS tensor yields the zero state on any contraction graph if and only if its minimal canonical form is the zero tensor.
In the language of invariant theory, the PEPS tensor yields the zero state on any contraction graph if and only if it is in the null cone.

\begin{exa}
The following is the smallest set of tiles that only gives aperiodic tilings, meaning that if we take any rectangle with periodic boundary conditions, the associated PEPS equals zero \cite{jeandel2021aperiodic}.
\begin{center}
\includegraphics[width=5cm]{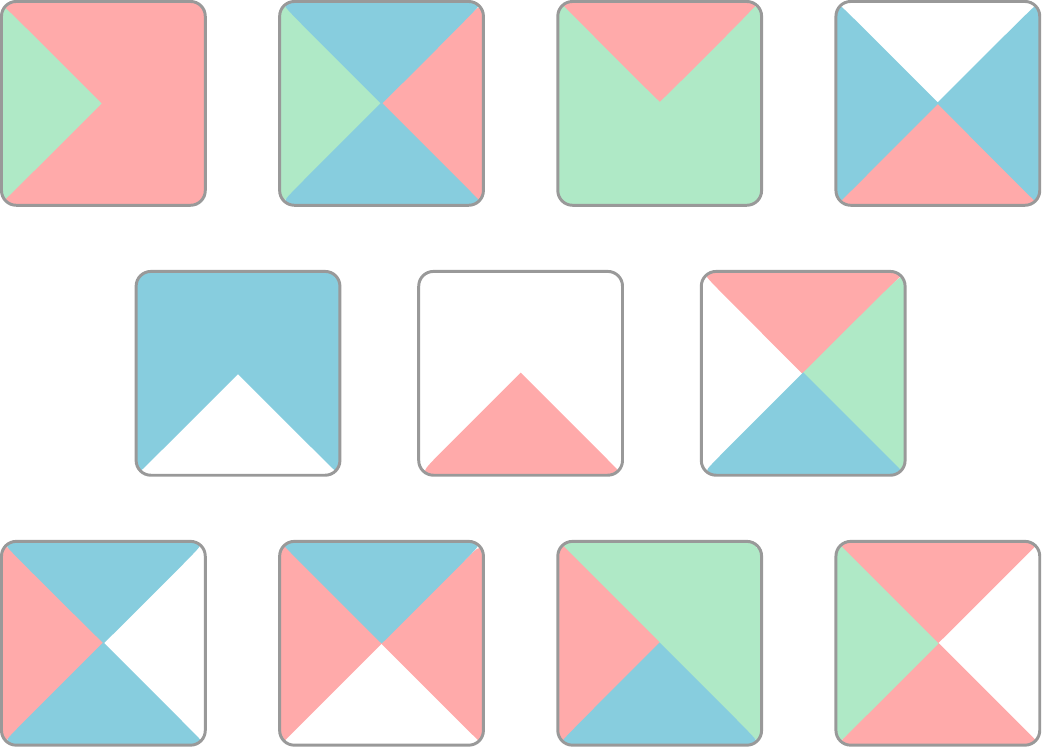}
\end{center}
On the other hand it is easy to construct a geometry for which the associated PEPS is nonzero:
\begin{center}
\includegraphics[width=4cm]{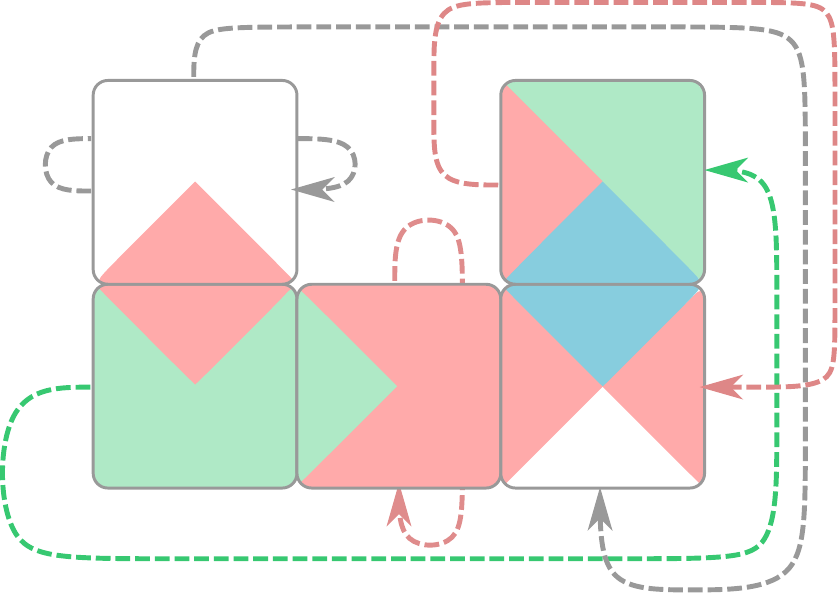}
\end{center}
In general, \cref{prp:peps inv} together with the reduction in \cite{scarpa2020projected} shows that given a set of tiles with $D$ colors, then there exists a `generalized tiling' (i.e. an arbitrary way to glue together the edges of the tiles) on some closed (possibly non-orientable) surface if and only if such a generalized tiling exists using at most $\exp(\mathcal O(D^2 \log D)$ tiles.
The problem of deciding, given a set of tiles, whether there exists some generalized tiling is thus a decidable problem.
The construction in \cref{prp:degree lower bound} in fact used a PEPS corresponding to a tiling problem, showing that there are indeed situations where the smallest possible generalized tilings are of size at least $\exp(\Omega(D))$.
\end{exa}

As argued in \cite{scarpa2020projected} their undecidability result excludes the possibility of a computable canonical form for two-dimensional PEPS which is such that two tensors $T, S$ yield the same state on all periodic lattices (so $\ket{T_{n_1,n_2}} = \ket{S_{n_1,n_2}}$ for all $n_1, n_2$) if and only if they have the same canonical form.
On the other hand, we saw in \cref{thm:old-fundamental} that any two \emph{normal} tensors which yield the same state on a periodic lattice are related by a local gauge transformation.
However, even if generic tensors are normal, in two spatial dimensions many interesting tensors describing physical systems are not normal, in particular those associated to topological order, either conventional or symmetry-protected \cite{cirac2021matrix}.
One way to interpret our Fundamental Theorem (\cref{thm:peps fundamental thm}) is that for some tensors it does not suffice to place them on periodic lattices and that the state they describe has a type of topological order which is only revealed by placing the states on a (possibly non-orientable) two-dimensional manifold other than a torus. This is an idea which is worth exploring in the future, and it is reminiscent of the well-known fact that different topological sectors can be detected by imposing different boundary conditions \cite{cirac2021matrix}.

\subsection{When does one need the orbit closure?}\label{sec:closure}
In general, finding the minimal canonical form requires one to go to the closure of the orbit of the action by the gauge group.
In other words, if $T \in \Mat_{D_1 \dots D_m \times D_1 \dots D_m}^d$ is a PEPS tensor in $m$ spatial dimensions, then there may not exist a minimal canonical form $\mcf{T}$ of the form $(g_1,\dots,g_m) \cdot T$, but only one that can be written as a limit of such tensors: $\mcf{T} = \lim_{j \to \infty}(g_1^{(j)},\dots,g_m^{(j)}) \cdot T$.
When is this really necessary?
In this section we will discuss conditions under which one does not need to go to the closure and give an example where it is required.
We consider PEPS tensors in $m$ spatial dimensions, and fix bond dimensions $D_1, \dots, D_m$ and physical dimension $d$.
We denote by $G = \GL(D_1) \times \dots \times \GL(D_m)$.

We will now argue that given a tensor $S \in \Mat_{D_1 \dots D_m \times D_1 \dots D_m}^d$ in minimal canonical form, if there exists a $T$ which has $S$ as a canonical form and which requires taking an orbit closure, then the tensor $S$ must have a continuous symmetry.
We formalize the notion of a continuous symmetry by a \emph{multiplicative one-parameter subgroup} of $G$, which is a homomorphism of Lie groups $\phi\colon \Cstar \rightarrow G$.
Given such a homomorphism we will write $\vec g(z)$ for $\phi(z)$ and we will say that $\vec g(z)$ is nontrivial if $\vec g(z)$ is not proportional to the identity for all $z \in \Cstar$.

The result we are aiming for is a consequence of the Hilbert-Mumford criterion in geometric invariant theory.
If $T \in \Mat_{D_1 \dots D_m \times D_1 \dots D_m}^d$ is any tensor, and $\mcf{T}$ is an associated minimal canonical form, then $G \cdot \mcf{T}$ is a closed orbit (by the Kempf--Ness Theorem, see \cref{thm:kempf ness}).
The Hilbert-Mumford criterion (see for instance Theorem 3.24 in \cite{wallach2017geometric}) then implies that there exists a one-parameter subgroup $\vec g(z) \in G$ such that
\begin{align*}
  \lim_{z \to 0} \vec g(z) \cdot T = S
\end{align*}
where $S \in G \cdot \mcf{T}$.

\begin{prp}[Non-closed implies symmetry]\label{prp:closure symmetry}
Suppose $S \in \Mat_{D_1 \dots D_m \times D_1 \dots D_m}^d$ is such that $G \cdot S$ is closed (in particular this is valid if $S$ is in minimal canonical form).
Suppose that there exists $T$ such that $S \in \overline{G \cdot T}$ but $S \notin G \cdot T$, then there exists a nontrivial one-parameter subgroup $\vec g(z) \subset G, z \in \Cstar$ such that $\vec g(z) \cdot S = S$ for all $z \in \Cstar$.
\end{prp}

\begin{proof}
By the Hilbert-Mumford criterion there exists $\vec g \in G$ and a one-parameter subgroup $\vec h(z) \in G$ such that
\begin{align*}
  \lim_{z \to 0} \vec h(z) \cdot T = \vec g \cdot S.
\end{align*}
This one-parameter subgroup must be nontrivial since $S \notin G \cdot T$.
Let $\vec g(z) = \vec g^{-1}\vec h(z)\vec g$.
Then
\begin{align*}
  \vec g(z) \cdot S = \vec g^{-1}\vec h(z)\vec g \cdot S = \lim_{w \to 0} \vec g^{-1} \vec h(z) \vec h(w) \cdot T = \lim_{w \to 0} \vec g^{-1} \vec h(zw) \cdot T = \vec g^{-1} \cdot (\vec g \cdot S) = S
\end{align*}
confirming that $\vec g(z)$ is a symmetry for $S$.
\end{proof}

\begin{exa}
Returning to the GHZ state in \cref{ex:ghz}, we note that it indeed has a one-parameter subgroup symmetry, for instance for
\begin{align*}
  g(z) = \begin{pmatrix} 1 & 0 \\ 0 & z \end{pmatrix}
\end{align*}
it holds that $g(z) \cdot M = M$.
\end{exa}

An important class of examples of PEPS tensors which lead to closed orbits are injective and normal tensors, already defined in \cref{sec:minimal-peps-basics}. For those tensors (in particular for normal MPS) one does not need to take closures to construct the minimal canonical form.
In fact we show that if there is any normal tensor in $\overline{G \cdot T}$, then $G \cdot T$ is closed (and in particular contains a minimal canonical form for $T$).

\begin{prp}[Canonical form normal PEPS]\label{lem:injective}
Suppose $T \in \Mat_{D_1 \dots D_m \times D_1 \dots D_m}^d$ is such that its orbit closure $\overline{G \cdot T}$ contains a normal tensor.
Then $\overline{G \cdot T} = G \cdot T$.
\end{prp}

\begin{proof}
By \cref{prp:closure symmetry} it suffices to show that if $T$ is normal, then $G \cdot T$ is closed and there is no nontrivial one-parameter subgroup $\vec g(z)$ such that $\vec g(z) \cdot T = T$ for all $z \in \Cstar$.

Let $\tilde T$ be the $n = n_1 \times \dots \times n_m$ blocking of $T$ such that $\tilde T$ is injective.
So, if we let $\tilde D_i = D_i^{n_1\dots n_{i-1}n_{k+1} \dots n_m}$ and $\tilde d = d^n$,
\begin{align*}
  \tilde T \in \Mat_{\tilde D_1 \dots \tilde D_m \times \tilde D_1 \dots \tilde D_m}^{\tilde d}.
\end{align*}

Let $S$ be any tensor in $\overline{G \cdot T}$ and let $\tilde S$ be the $n_1 \times \dots \times n_m$ blocking of $S$.
Since $S \in \overline{G \cdot T}$ there must be a sequence $\vec g^{(j)} = (g_1^{(j)},\dots,g_m^{(j)}) \in G$ for $j \in \N$ such that
\begin{align*}
  \lim_{j \to \infty} \vec g^{(j)} \cdot T = S.
\end{align*}
Since $\vec g \cdot T$ is invariant under rescaling the $g_k$ by a constant, we may assume that $\opnorm{g_k^{(j)}} = 1$ for all $k$ and $j$.
If we let $\tilde g_k^{(j)} = (g_k^{(j)})^{\ot n_1 \dots n_{i-1} n_{i+1} \dots n_m}$ and $\tilde{\vec g}^{(j)} = (\tilde g_1^{(j)}, \dots, \tilde g_m^{(j)})$ then
\begin{align*}
  \lim_{j \to \infty} \tilde{\vec g}^{(j)} \cdot \tilde T = \tilde S.
\end{align*}
Now, interpret $\tilde T$ as an element of $(\C^{\tilde D} \ot \C^{\tilde D})^{\tilde d}$ where $\tilde D = \tilde D_1 \dots \tilde D_m$, so
\begin{align*}
  \tilde T = (\tilde T^{(i)})_{i = 1}^{\tilde d}, \qquad \tilde T^{(i)} \in \C^{\tilde D} \ot \C^{\tilde D}.
\end{align*}
Then the fact that $\tilde T$ is injective implies that there exists a tensor $\tilde M \in (\C^{\tilde D} \ot \C^{\tilde D})^{\tilde d}$ which is an inverse to $\tilde T$ in the sense that
\begin{align*}
  \sum_{i = 1}^{\tilde d} T^{(i)} (M^{(i)})^{\dagger} = I_{\tilde D^2}
\end{align*}
is the identity map.
Let $\tilde N^{(j)}$ be the contraction of $\tilde{\vec g}^{(j)} \cdot \tilde T$ with $\tilde M$:
\begin{align*}
  \tilde N^{(j)} = \sum_{i=1}^{\tilde d}  \left(\tilde{\vec g}^{(j)} \ot (\tilde{\vec g}^{(j)})^{-T} \tilde T^{(i)}\right) (M^{(i)})^{\dagger}
\end{align*}
(writing $\tilde{\vec g}^{(j)} = \tilde g_1^{(j)} \ot \dots \ot \tilde g_m^{(j)}$ in a slight abuse of notation).
Then, $\tilde N^{(j)}$ must be a converging sequence (since $\tilde{\vec g}^{(j)} \cdot \tilde T$ is so).
On the other hand, since $\tilde M$ is the inverse to $\tilde T$,
\begin{align*}
  \tilde N^{(j)} = \tilde{\vec g}^{(j)} \ot (\tilde{\vec g}^{(j)})^{-T}.
\end{align*}
The fact that this sequence converges implies that $\opnorm{(\tilde{\vec g}^{(j)})^{-T}} = \opnorm{(\tilde{\vec g}^{(j)})^{-1}}$ is bounded and hence there is some constant $C$ such that for all $k \in [m]$ and $j \in \N$ we may bound $\opnorm{(g_k^{(j)})^{-1}} \leq C$.
However, this implies that $\vec g^{(j)}$ is contained in a compact subset of $G$ and therefore has a converging subsequence, which in turn implies that
\begin{align*}
  S = \lim_{j \to \infty} \vec g^{(j)} \cdot T \in G \cdot T.
\end{align*}
So, we conclude that $G \cdot T$ is closed.
Secondly, suppose that there exists a nontrivial one-parameter subgroup $\vec g(z)$ such that $\vec g(z) \cdot T = T$ for all $z \in \Cstar$.
Using the same notation as before, this implies that there exists a one-parameter subgroup $\tilde{\vec g}(z)$ such that $\tilde{\vec g}(z) \cdot \tilde T = \tilde T$.
However, applying the inverse $\tilde M$, this implies
\begin{align*}
  \tilde{\vec g}(z) \ot \tilde{\vec g}(z)^{-T} = I
\end{align*}
which implies that $\vec g(z)$ must be proportional to the identity for all $z \in \Cstar$.
\end{proof}

Beyond normal PEPS states there are also other states of interest where \cref{prp:closure symmetry} implies that one never needs to go to the closure to obtain the minimal canonical form.

\begin{exa}\label{ex:toric code}
In two spatial dimensions an important example of a PEPS state which is not normal is the toric code.
This is a state usually defined on a qubit lattice.
To write it as a PEPS state one may group together four physical sites into a single site of four qubits.
The toric code PEPS tensor is then given, as a map from the bond legs to the physical legs, by $T = \frac12 I^{\ot 4} + \frac12 Z^{\ot 4}$.
Alternatively, for $i,j,k,l \in \{0,1\}$
\begin{align*}
  T^{(i,j,k,l)} = \begin{cases}
    \ketbra{i}{j} \ot \ketbra{l}{k} & \text{if $i + j + k + l$ is even,}\\
    0 & \text{if $i + j + k + l$ is odd.}
  \end{cases}
\end{align*}
This tensor is in minimal canonical form, since all virtual marginals are maximally mixed.
We will now verify that this tensor has a finite symmetry group, and hence (as opposed to the GHZ state) there are no tensors for which $T$ is in their orbit closure while not in the orbit itself.
Suppose that $\vec g \cdot T = T$ for $\vec g = (g_1, g_2)$ with $g_k \in \GL(2)$ for $k = 1,2$.
This is equivalent to
\begin{align*}
  g_1 \ot g_1^{-T} \ot g_2 \ot g_2^{-T} \ket{i} \ket{j} \ket{k} \ket{l} = \ket{i} \ket{j} \ket{k} \ket{l}.
\end{align*}
for all $i + j + k + l = 0 \mod 2$.
We can choose $i$ and $j$ arbitrary, so $g_1$ must be diagonal.
By the same reasoning, $g_2$ must be diagonal as well.
If we let
\begin{align*}
  g_i = \begin{pmatrix} g_{i,0} & 0 \\ 0 & g_{i,1} \end{pmatrix}
\end{align*}
then we find $g_{1,i} g_{2,k} = g_{1,j} g_{2,l}$ for all $i + j + k + l = 0 \mod 2$.
By choosing $i \neq j$ and $k \neq l$ it is easy to see that this implies that after scaling by a global constant (which is irrelevant) $g_{i,j} \in {\pm 1}$ so we cannot have a nontrivial one-parameter subgroup symmetry.
\end{exa}

\begin{exa}\label{ex:abelian quantum double}
The previous example can be generalized to arbitrary quantum double models for abelian groups $G$.
For an arbitrary finite group $G$ we may construct a PEPS tensor (also known as a $G$-isometric PEPS tensor) as follows.
The Hilbert space along each of the bond legs consists of the group algebra $\C[G]$ with basis $\{\ket{g}\}_{g \in G}$, so the bond dimension is $D = \abs{G}$.
The group $G$ acts by the regular representation on $\C[G]$ as $g \ket{h} = \ket{gh}$.
The physical Hilbert space is given by $\C[G]^{\ot 4}$.
Then the PEPS tensor is given, as a map from the bond Hilbert spaces to the physical Hilbert space as
\begin{align*}
  T = \frac{1}{\abs{G}} \sum_{g \in G} g \ot \bar g \ot g \ot \bar g
\end{align*}
The toric code tensor is a special case of this construction for $G = \Z_2$.
Essentially the same argument as for the toric code shows that (up to a global constant) the symmetries of this tensor form a discrete set if the group $G$ is abelian and hence $\C[G]$ decomposes into one-dimensional irreducible representations.
Therefore, $\GL(D) \times \GL(D) \cdot T = \overline{\GL(D) \times \GL(D) \cdot T}$.
\end{exa}

\begin{exa}\label{ex:continuous virtual symmetry}
To give a nontrivial example where we do have a continuous symmetry, and we have non-closed orbits, we use a construction inspired by \cite{dreyer2018projected}, which investigates PEPS with continuous virtual symmetries.
Consider a 2-dimensional PEPS tensor $T$ with physical and bond dimensions all equal to two, given by
\begin{align*}
  T^{(0)} &= \sum_{i,j \in \{0,1\}} \ketbra{i}{j} \ot \ketbra{j}{i} \\
  T^{(1)} &= \sum_{i,j \in \{0,1\}} \ketbra{i}{j} \ot X\ketbra{i}{j}X.
\end{align*}
In the standard basis we may write this out as
\begin{align*}
  T^{(0)} = \begin{pmatrix} 1 & 0 & 0 & 0 \\  0 & 0 & 1 & 0 \\  0 & 1 & 0 & 0 \\  0 & 0 & 0 & 1 \end{pmatrix}
  \qquad \text{ and } \qquad T^{(1)} = \begin{pmatrix} 0 & 0 & 0 & 0 \\  0 & 1 & 1 & 0 \\  0 & 1 & 1 & 0 \\  0 & 0 & 0 & 0 \end{pmatrix}.
\end{align*}
See \cite{dreyer2018projected} for a graphical notation, expressing contractions as loop diagrams.
All the virtual marginals of $T$ are maximally mixed, so $T$ is in minimal canonical form.
It is now easy to see that $\vec g(z) = (h(z), h(z))$ is a one-parameter subgroup symmetry for
\begin{align*}
  h(z) = \begin{pmatrix} 1 & 0 \\ 0 & z \end{pmatrix}.
\end{align*}
Indeed, since $h(z) \ketbra{i}{j} h(z)^{-1} = z^{i - j} \ketbra{i}{j}$ and $h(z) X\ketbra{i}{j}X h(z)^{-1} = z^{j-i} X\ketbra{i}{j}X$
\begin{align*}
  \left(h(z) \ot h(z)\right) T^{(0)} \left(h(z)^{-1} \ot h(z)^{-1}\right) &= \sum_{i,j \in \{0,1\}} z^{i-j}\ketbra{i}{j} \ot z^{j - i} \ketbra{j}{i} = T^{(0)} \\
  \left(h(z) \ot h(z)\right) T^{(1)} \left(h(z)^{-1} \ot h(z)^{-1}\right) &= \sum_{i,j \in \{0,1\}} z^{i-j}\ketbra{i}{j} \ot z^{j - i} X\ketbra{i}{j}X = T^{(1)}.
\end{align*}
Let us construct an explicit example where we need the closure to reach the minimal canonical form.
Let $N = \ketbra{1}{0} \ot \ketbra{1}{0}$ and let
\begin{align*}
  S^{(0)} = T^{(0)} + N \qquad \text{ and } \qquad S^{(1)} = T^{(1)} + N.
\end{align*}
In the standard basis
\begin{align*}
  S^{(0)} = \begin{pmatrix} 1 & 0 & 0 & 0 \\  0 & 0 & 1 & 0 \\  0 & 1 & 0 & 0 \\  1 & 0 & 0 & 1 \end{pmatrix}
  \qquad \text{ and } \qquad S^{(1)} = \begin{pmatrix} 0 & 0 & 0 & 0 \\  0 & 1 & 1 & 0 \\  0 & 1 & 1 & 0 \\  1 & 0 & 0 & 0 \end{pmatrix}.
\end{align*}
Now, since $h(z) \ketbra{1}{0} h(z)^{-1} = z \ketbra{1}{0}$ and $T$ is invariant under $\vec g(z)$,
\begin{align*}
  \left(h(z) \ot h(z)\right) S^{(i)} \left(h(z)^{-1} \ot h(z)^{-1}\right) = T^{(i)} + z^2 N
\end{align*}
so
\begin{align*}
  \lim_{z \to 0} \vec g(z) \cdot S = T.
\end{align*}
On the other hand, since $1 = \rank(T^{(1)}) \neq \rank(S^{(1)}) = 2$ we see  that $S$ is not in the orbit of $T$.
\end{exa}

\section{Algorithms for computing minimal canonical forms}\label{sec:algos}
In this section we address the question of how to compute minimal canonical forms algorithmically.
We will discuss two algorithms (and sketch potential applications in \cref{sec:outlook}).
The first one is eminently practical and stated explicitly in \cref{algo:PEPS grad descent}.
The second one has a better runtime dependence in theory, but is less practical.
Both algorithms have their origin in a series of recent works on norm minimization and scaling problems, in increasing generality, including matrix, operator and tensor scaling (see \cite{garg2020operator,garg2017algorithmic,allenzhu2018operator,BGOWW18,burgisser2018efficient,kwok2018paulsen,burgisser2019theory} and references therein).
We follow and apply the general framework of~\cite{burgisser2019theory} but give some tighter bounds in our setting.

Before discussing our results and presenting our algorithm in more detail, we discuss what it means to compute a minimal canonical form.
In general, minimal canonical forms cannot be represented exactly in finite precision, so one is naturally led to look for approximations.
Then there are at least three natural choices of what it might mean to \emph{approximately compute a minimal canonical form} of a given PEPS tensor~$T$:
\begin{itemize}
\item \emph{$\ell^2$-error in the space of tensors:}
Given $\delta>0$, find a tensor $S \in G \cdot T$ that is $\delta$-close in $\ell^2$-norm to a minimal canonical form $\mcf T$ of~$T$.
It is natural consider relative error (but see \cref{rem:abserr}):
\begin{align}\label{eq:delta error}
  \frac {\norm{S - \mcf T}} {\norm S} \leq \delta.
\end{align}

\item \emph{$\ell^2$-error in the first-order characterization:}
Given $\eps>0$, find a tensor $S \in G \cdot T$ such that
\begin{align}\label{eq:eps error}
  \frac{1}{\tr\sigma} \sqrt{ \sum_{k=1}^m \norm*{\sigma_{k,1} - \sigma_{k,2}^T}^2 } \leq \eps
  \quad\text{where}\quad
  \sigma = \ket S\bra S.
\end{align}

\item \emph{error in the norm of the tensor:}
Given $\zeta>0$, find a tensor $S \in G \cdot T$ whose norm is almost minimal:
\begin{align}\label{eq:zeta error}
  \frac {\norm{\mcf T}}{\norm S} \geq 1 - \zeta.
\end{align}
\end{itemize}
We already know that \cref{eq:delta error} holds with $\delta=0$ if and only if \cref{eq:eps error} holds with $\eps=0$ if and only if \cref{eq:zeta error} holds with $\zeta=0$ (by \cref{thm:symmetric virtual rdm} and the definition of the minimal canonical form).
In \cref{subsec:relerr} we will show that the three error measures can be related in a precise way.
Accordingly, we may target either, and we will see that \cref{eq:eps error,eq:delta error} arise naturally when designing approximation algorithms.

\subsection{First-order algorithm}\label{subsec:first order}
We start by motivating our first algorithm, which we present explicitly in \cref{algo:PEPS grad descent}.
Suppose we are given a tensor~$0 \neq T = (T^{(i)})_{i=1}^d  \in \Mat_{D_1 \dots D_m \times D_1 \dots D_m}^d$ and we would like to approximately compute a minimal canonical form~$\mcf T$.
Since the latter is defined as a minimum norm tensor in the orbit closure, a natural way to address this is by minimizing or ``infimizing'' the norm or, equivalently, one half the norm square $\frac12\norm{\vec g \cdot T}^2$ over $\vec g \in G = \GL(D_1) \times \cdots \times \GL(D_m)$.
Because the norm is invariant under the action of $K = \U(D_1) \times \cdots \times \U(D_m)$, the objective function
\begin{align*}
  f_T(\vec g)
:= \frac12\norm{\vec g \cdot T}^2
= \frac12\braket{T, (g_1^\dagger g_1 \ot \cdots \ot g_m^\dagger g_m) \cdot T}
\end{align*}
only depends on the tuple $\vec p = (g_1^\dagger g_1,\dots,g_m^\dagger g_m)$ of positive definite matrices in $P = \PD(D_1) \times \cdots \times \PD(D_m)$.
However, $P$ is \emph{not} a convex subset of $H := \Herm(D_1) \op \cdots \op \Herm(D_m)$ and accordingly $\inf_{\vec p \in P} f_T(\vec p)$ is \emph{not} a convex optimization problem that can be addressed by standard methods (e.g., by semidefinite programming)!

Instead, we proceed differently.
Since $f_T(\vec g) = f_T(\vec k\vec g)$ for all $k\in K$ and $g\in G$, the objective function $f_T$ can be defined on the space $K\backslash G := \{ K \vec g : \vec g \in G \}$ of right $K$-cosets in the gauge group~$G$.
This space may be endowed with a natural Riemannian metric, yielding a simply-connected complete Riemannian manifold with \emph{non-positive curvature}~\cite{bridson2013metric,bhatia2009positive}.
In particular, between any two points there exist unique \emph{geodesics} (here: shortest paths).
Explicitly, the geodesics through $\vec g=(g_1,\dots,g_m)\in G$ take the form $K (e^{tX_1} g_1,\dots,e^{tX_m} g_m)$ for $\vec X = (X_1,\dots,X_m) \in H$.%
\footnote{We can also identify $K\backslash G$ with $P = \PD(D_1) \times \cdots \times \PD(D_m)$ by the map $K\vec g \mapsto \vec g^\dagger \vec g$.
Then the geodesics can be written as $(\sqrt{p_1} e^{tY_1} \sqrt{p_1}, \dots, \sqrt{p_m} e^{tY_m} \sqrt{p_m})$, where $p_k = g_k^\dagger g_k$ and the $Y_k$ are certain Hermitian matrices. These are tuples of the familiar geodesics of $\PD(D_k)$, see, e.g., \cite{bhatia2009positive}.}

The point then is the following:
While not convex in the ordinary sense, the function $f_T(p)$ is \emph{geodesically convex}, that is, convex along these geodesics.
This means for any $(g_1,\dots,g_m) \in G$ and $(X_1,\dots,X_m) \in H$,
\begin{align*}
  \partial_{t=0}^2 f_T\mleft( e^{tX_1} g_1, \dots, e^{tX_m} g_m \mright) \geq 0.
\end{align*}
Therefore, a reasonable approach to minimizing $f_T$ is to use a gradient descent.
What is the gradient in this setting at, say, $\vec g=\vec I=(I_{D_1},\dots,I_{D_m})$?
The computation done in \cref{eq:moment map peps} shows that
\begin{align*}
  \partial_{t=0} f_T\mleft( e^{tX_1}, \dots, e^{tX_m} \mright)
= \frac12 \partial_{t=0} \norm*{ (e^{tX_1}, \dots, e^{tX_m}) \cdot T}^2
= \sum_{k=1}^m \tr\mleft[X_k\left(\rho_{k,1} - \rho_{k,2}^T \right)\mright].
\end{align*}
where $\rho = \ket T \bra T$, and hence we should think of
\begin{align}\label{eq:grad norm sqr}
  \nabla f_T(\vec I) = \left( \rho_{k,1} - \rho_{k,2}^T \right)_{k=1}^m \in H=\Herm(D_1) \op \dots \op \Herm(D_m)
\end{align}
as the \emph{gradient} at $\vec g=\vec I$!
Accordingly, starting at $\vec g=\vec I$ and moving along the geodesic with this direction, we should take a gradient step of the form
\begin{align*}
   T \mapsto \vec g \cdot T, \quad\text{where}\quad \vec g := \left( e^{-\eta \left( \rho_{1,1} - \rho_{1,2}^T \right)}, \dots, e^{-\eta \left( \rho_{m,1} - \rho_{m,2}^T \right)} \right).
\end{align*}
for some suitable step size $\eta>0$.
Note that, crucially, this amounts to acting by the gauge group, i.e., will automatically remain in the $G$-orbit!

\begin{algorithm}[t]
  \SetAlgoLined
  \Input{A uniform PEPS tensor $T \in \Mat_{D_1 \dotsb D_m \times D_1 \dotsb D_m}^d$ and $\eps>0$.}
  \Output{A gauge transformation $\vec g \in \GL(D_1) \times \dotsb \times \GL(D_m)$.}
  \caption{Computing PEPS normal forms}
  \label{algo:PEPS grad descent}

  \BlankLine

  $\vec g^{(0)} \leftarrow (I_{D_1}, \dotsc, I_{D_m})$\;

  \For{$t=0,1,\dots$}{
    $T^{(t)} \leftarrow \vec g^{(t)} \cdot T$\;
    $\rho^{(t)} \leftarrow \ket{T^{(t)}}\bra{T^{(t)}}$\;
    \If{$\frac1{\tr\rho^{(t)}} \sum_{k=1}^m \norm{\rho_{k,1}^{(t)} - (\rho_{k,2}^{(t)})^T}^2 \leq \eps^2$} {\Return $\vec g^{(t)}$\;}
    \For{$k = 1, \dotsc, m$}{
      $g^{(t+1)}_k \leftarrow e^{- \frac{1}{4m} \frac1{\tr\rho^{(t)}} (\rho_{k,1}^{(t)} - (\rho_{k,2}^{(t)})^T) } g^{(t)}_k$\;
    }
  }
\end{algorithm}

Now we have almost derived \cref{algo:PEPS grad descent}, except for one observation:
The function $f_T$ is not only convex along geodesics, but even log-convex!
This gives stronger guarantees, so we consider
\begin{align}\label{eq:concrete kempf ness}
  F_T(\vec g) = \frac12 \log \mleft( 2 f_T(\vec g) \mright) = \log \norm{\vec g \cdot T},
\end{align}
with gradient
\begin{align}\label{eq:log grad}
  \nabla F_T(\vec I)
= \frac1{2 f_T(\vec I)} \left( \rho_{k,1} - \rho_{k,2}^T \right)_{k=1}^m
= \frac1{\tr\rho} \left( \rho_{k,1} - \rho_{k,2}^T \right)_{k=1}^m
\end{align}
and updates
\begin{align*}
   T \mapsto \vec g \cdot T, \quad\text{where}\quad \vec g := \left( e^{-\eta \frac1{\tr\rho} \left( \rho_{1,1} - \rho_{1,2}^T \right)}, \dots, e^{-\eta \frac1{\tr\rho} \left( \rho_{m,1} - \rho_{m,2}^T \right)} \right).
\end{align*}
Similarly to the ordinary gradient descent in Euclidean space, under suitable hypotheses on a geodesically convex objective one can provide a ``safe'' choice for the step size~$\eta$.
In the present case, the objective $F_T$ is $4m$-smooth along geodesics:
for every $\vec g=(g_1,\dots,g_m) \in G$ and $\vec X=(X_1,\dots,X_m) \in H$, one has
\begin{equation*}
  \partial_{t=0}^2 F_T(e^{tX_1} g_1,\dots,e^{tX_m} g_m) \leq 4m \norm{\vec X}^2,
\end{equation*}
where $\norm{\vec X}^2 = \sum_{k=1}^m \norm{X_k}^2$.
For such functions, $\eta = \frac1{4m}$ is a suitable step size and this is what we use in \cref{algo:PEPS grad descent}.
Below, we give formal guarantees for the performance of the algorithm.
We remark that \cref{thm:peps mcf algo guarantee} is a special case of~\cite[Thm.~4.2]{burgisser2019theory}.

\begin{thm}\label{thm:peps mcf algo guarantee}
  Let $T \in \Mat_{D_1 \dotsb D_m \times D_1 \dotsb D_m}^d$ be such that $\mcf T \neq 0$ (for some and hence for any minimal canonical form), and let $\eps > 0$.
  Then \cref{algo:PEPS grad descent} outputs a group element $\vec g \in \GL(D_1) \times \dotsb \times \GL(D_m)$ such that the tensor $S := \vec g \cdot T$ satisfies
  \begin{equation*}
    \frac1{\tr\sigma} \sqrt{ \sum_{k=1}^m \norm*{\sigma_{k,1} - \sigma_{k,2}^T}^2 } \leq \eps,
    \quad\text{where}\quad
    \sigma = \ket S\bra S,
  \end{equation*}
  within $O(\frac{m}{\eps^2} \log \frac {\norm T}{\norm{\mcf T}})$ iterations.
\end{thm}
\begin{proof}
  We analyze \cref{algo:PEPS grad descent}.
  For $t=0,1,2,\dots$ and $g^{(t)}$ the group elements produced by the algorithm.
  If the algorithm does not terminate in the $t$-th iteration, then, using \cref{eq:log grad},
  \begin{align*}
    F_T(\vec g^{(t+1)}) - F_T(\vec g^{(t)})
  &= F_{T^{(t)}}(e^{- \frac{1}{4m} \frac1{\tr\rho^{(t)}} (\rho_{1,1}^{(t)} - (\rho_{1,2}^{(t)})^T) }, \dots, e^{- \frac{1}{4m} \frac1{\tr\rho^{(t)}} (\rho_{m,1}^{(t)} - (\rho_{m,2}^{(t)})^T) }) - F_{T^{(t)}}(\vec I) \\
  &= F_{T^{(t)}}(e^{-\frac1{4m} \nabla F_{T^{(t)}}(\vec I)}) - F_{T^{(t)}}(\vec I) \\
  &\leq \tr\mleft[\nabla F_{T^{(t)}}(\vec I) \cdot \left( -\frac1{4m} \nabla F_{T^{(t)}}(\vec I) \right)\mright] + \frac m8 \norm*{-\frac1m \nabla F_{T^{(t)}}(\vec I)}^2 \\
  &= -\frac1{8m} \norm*{\nabla F_{T^{(t)}}(\vec I)}^2
  < -\frac{\eps^2}{8m},
  \end{align*}
  where the first inequality follows since $F_T$ is a convex and $4m$-smooth function \cite[Lemma~3.8]{burgisser2019theory}.
  Accordingly, if the algorithm has not terminated up to and including the $t$-th iteration, then
  \begin{align*}
    \log \frac{\norm{\mcf T}}{\norm T}
  \leq \log\,\norm{\vec g^{(t)} \cdot T} - \log\, \norm{T}
  = F_T(\vec g^{(t)}) - F_T(\vec g^{(0)})
  < -t \,  \frac{\eps^2}{8m},
  \end{align*}
  or
  \begin{equation*}
    t < \frac{8m}{\eps^2}\log \frac{\norm T}{\norm{\mcf T}}. \qedhere
  \end{equation*}
\end{proof}

The iteration bound of \cref{thm:peps mcf algo guarantee} involves~$\norm{\mcf T}$.
If the entries of~$T$ are given by some finite number of bits then this quantity can be estimated in an \emph{a priori} fashion, by first rescaling $T$ such that its entries are given by Gaussian integers, i.e., are in $\Z[i]$, and then using the following result.

\begin{prp}\label{prp:peps mcf integral norm bound}
  Let $T \in \Mat_{D_1 \dotsb D_m \times D_1 \dotsb D_m}^d$ with $\mcf T \neq 0$, and assume that all entries of $T$ are in $\Z[i]$.
  Then,
  \begin{equation*}
    \norm{\mcf{T}} \geq \frac{1}{\prod_{j=1}^m D_j}.
  \end{equation*}
\end{prp}
\begin{proof}
  We use the fact that the invariant ring is generated by the functions $P_{\vec \pi,\vec i}$ defined in~\cref{eq:peps general permutations}.
  Since $\mcf{T} \neq 0$, there exist $n \geq 1$, $\vec \pi \in S_n^{m}$ and $\vec i \in [d]^n$ such that $P_{\vec \pi, \vec i}(T) \neq 0$.
  But $P_{\vec \pi, \vec i}$ is a polynomial with integer coefficients in the entries of $T$; therefore, evaluating it on $T$ with entries in $\Z[i]$ must yield $\abs{P_{\vec\pi, \vec i}(T)} \geq 1$.
  Furthermore, it is an invariant under the PEPS action, so we deduce for any $g \in G$:
  \begin{align*}
    1 & \leq \abs{P_{\vec\pi, \vec i}(T)} = \abs{P_{\vec\pi, \vec i}(\vec g \cdot T)} = \abs*{\tr \left[ (R_{\pi_1} \otimes \dotsb \otimes R_{\pi_m}) ((\vec g \cdot T^{(i_1)}) \otimes \dotsb \otimes (\vec g \cdot T^{(i_n)})) \right]} \\
      & \leq \norm{R_{\pi_1} \otimes \dotsb \otimes R_{\pi_m}} \cdot \norm{(\vec g \cdot T^{(i_1)}) \otimes \dotsb \otimes (\vec g \cdot T^{(i_n)})}.
  \end{align*}
  Since each $R_{\pi_j}$ is unitary, the same is true of their tensor product.
  As it acts on a space of dimension $(\prod_{j=1}^m D_j^2)^n$, one obtains
  \begin{equation*}
    \norm{R_{\pi_1} \otimes \dotsb \otimes R_{\pi_m}} = \left( \prod_{j=1}^m D_j \right)^n.
  \end{equation*}
  Furthermore,
  \begin{equation*}
    \norm{(\vec g \cdot T^{(i_1)}) \otimes \dotsb \otimes (\vec g \cdot T^{(i_n)})} \leq (\max_i \norm{\vec g \cdot T^{(i)}})^n \leq \norm{\vec g \cdot T}^n.
  \end{equation*}
  Combining the two estimates, taking $n$-th roots and the infimum over $\vec g \in G$ yields the desired estimate.
\end{proof}

The above approach of evaluating an invariant to prove norm lower bounds is used in other settings as well, e.g., for tensor scaling in \cite[Thm.~7.12]{BGOWW18}, and for much more general actions in \cite[Cor.~7.19]{burgisser2019theory}; but appealing to the latter result would result in a worse bound.

We obtain the following corollary, which implies an $\poly(\frac{1}{\eps}, \text{input size})$-time algorithm, cf.\ \cite[Rem.~8.1]{burgisser2019theory}:

\begin{cor}\label{cor:peps scaling explicit guarantee}
  Let $T \in \Mat_{D_1 \dotsb D_m \times D_1 \dotsb D_m}^d$ be a tensor such that $\mcf T \neq 0$ (for some and hence for any minimal canonical form).
  Assume that the entries of $T$ are in $\Q[i]$ and given by storing the numerators and denominators in binary.
  Let $\eps>0$.
  Then \cref{algo:PEPS grad descent} outputs a group element $\vec g \in \GL(D_1) \times \dotsb \times \GL(D_m)$ such that the tensor $S := \vec g \cdot T$ satisfies
  \begin{equation*}
    \frac1{\tr\sigma} \sqrt{ \sum_{k=1}^m \norm*{\sigma_{k,1} - \sigma_{k,2}^T}^2 } \leq \eps,
    \quad\text{where}\quad
    \sigma = \ket S\bra S.
  \end{equation*}
  within $O(\frac1{\eps^2} \cdot \poly(\inpsize T))$ iterations, where $\inpsize T$ denotes the total number of bits used to represent~$T$.
\end{cor}

\subsection{Relation between approximation errors}\label{subsec:relerr}
In \cref{subsec:first order}, we discussed three natural notions of approximation error in \cref{eq:eps error,eq:delta error,eq:zeta error}, and we gave an algorithm targeting \cref{eq:eps error}, i.e., given a tensor~$T$ and $\eps>0$, we discussed how to obtain a tensor $S \in G \cdot T$ such that
\begin{align*}
  \frac{1}{\tr\sigma} \sqrt{ \sum_{k=1}^m \norm*{\sigma_{k,1} - \sigma_{k,2}^T}^2 } \leq \eps
  \quad\text{where}\quad
  \sigma = \ket S\bra S.
\end{align*}
We will now see that there is a precise quantitative relationship between these notions.
As we will see, the following quantity will play a crucial role.

\begin{dfn}\label{def:gamma}
  Given bond dimensions $D_1,\dots,D_m$, define
  \begin{equation*}
    \gamma
  := \gamma(D_1, \dotsc, D_m)
  :=
    \begin{cases}
      \displaystyle\frac1{D_1^{3/2}}, & \text{if } m = 1, \\
      \displaystyle\frac1{\sum_{i=1}^m D_i} \cdot \frac 1 {(2m)^{(\sum_{i=1}^m D_i - 1)/2} } & \text{if } m \geq 2.
    \end{cases}
  \end{equation*}
\end{dfn}

Note that $\gamma$ is only inverse polynomially small in the bond dimension for $m=1$, while it is exponentially small for $m\geq2$.
Then we have the following relation between \cref{eq:eps error,eq:zeta error}.

\begin{thm}\label{thm:peps error conversion}
  Let $0\neq T \in \Mat_{D_1 \dotsb D_m \times D_1 \dotsb D_m}^d$ and $S \in G \cdot T$.
  Then:
  \begin{align*}
    1 - \frac{\eps}{\gamma} \leq \frac{\norm{\mcf T}^2}{{\norm S}^2} \leq 1 - \frac{\eps^2}{8m}
  \qquad\text{for}\qquad \eps := \frac{1}{\tr\sigma} \sqrt{ \sum_{k=1}^m \norm*{\sigma_{k,1} - \sigma_{k,2}^T}^2 },
  \end{align*}
  where $\sigma=\ket S\bra S$ and $\gamma$ is the constant defined in \cref{def:gamma}.
  In particular, if $\eps < \gamma$, then $\mcf T \neq 0$.
\end{thm}

We will prove \cref{thm:peps error conversion} by appealing to a non-commutative duality theorem given in \cite[Thm.~1.17]{burgisser2019theory}.
To explain how this theorem applies in our setting, we must define a complexity measure defined by combinatorial data associated with representations known as the \emph{weight margin}.
The parameter $\gamma$ which appears in \cref{def:gamma} is a lower bound on this weight margin.
We shall do this using the language from \cref{sec:git} to make it easier to bridge the gap, and show how the definitions specialize for our representation.

Let $\pi\colon G \to \GL(V)$ be a representation of a group $G \subseteq \GL(n)$, where we make the same assumptions as in \cref{sec:git}.
It is known that such $G$ contains a \emph{maximal algebraic torus}, denote by $T_G$, which is a maximal connected abelian subgroup, and any two maximal algebraic tori in $G$ are conjugate to one another.
For $\GL(D)$, a canonical choice is the subgroup $T(D)$ of invertible \emph{diagonal} matrices, and for our $G = \GL(D_1) \times \cdots \times \GL(D_m)$, a canonical choice is given by the subgroup $T(D_1) \times \cdots \times T(D_m)$ consisting of all tuples of such matrices.
Then, viewing $\pi$ as a representation of $T_G$, we may simultaneously diagonalise the action.
The simultaneous eigenvalues are captured by the concept of weights of the representation:
\begin{dfn}
  Let $T_G \subseteq G$ be a maximal algebraic torus.
  Then there exists a unique finite set of \emph{weights} $\Omega(\pi) \subseteq \Lie(T_G)^*$ of the representation $\pi$, such that
  \begin{equation*}
    V = \bigoplus_{\omega \in \Omega(\pi)} V_\omega
  \end{equation*}
  is an orthogonal decomposition into \emph{weight spaces} $V_\omega$, where
  \begin{equation*}
    \pi(e^Y) v_\omega = e^{\omega(Y)} v_\omega
  \end{equation*}
  for all $Y \in \Lie(T_G)$ and $v_\omega \in V_\omega$.
\end{dfn}

\begin{exa}\label{exa:peps weights}
  Let $\GL(D)$ act on $\Mat_{D \times D}$ by conjugation.
  As said before, a maximal subtorus of $\GL(D)$ is given by the set $T(D)$ consisting of invertible diagonal $D \times D$ matrices, and its Lie algebra $\Lie(T(D))$ consists of all diagonal matrices, which may be identified with $\C^D$.
  Then for $Y \in \C^D$, we have
  \begin{equation*}
    e^{\diag(Y)} E_{ij} e^{-\diag(Y)} = e^{Y_i - Y_j} E_{ij},
  \end{equation*}
  where $E_{ij}$ are the elementary matrices.
  Therefore the weights are given by the functionals $\omega_{ij}(Y) = Y_i - Y_j$, with corresponding weight spaces $V_{\omega_{ij}} = \C E_{ij}$.
  Note that $\omega_{ij}$ can be identified with $e_i - e_j \in \C^D$.
  The action of $\GL(D)$ on $\Mat_{D \times D}^d$ has the same weights, but now each weight space is $d$-dimensional.

  Now consider the action of the gauge group $G = \GL(D_1) \times \dotsb \times \GL(D_m)$ on $V = \Mat_{D_1 \dotsb D_m \times D_1 \dotsb D_m}^d$,  the space of PEPS tensors, as defined in \cref{dfn:peps gauge action}.
  As mentioned, a maximal torus for $G$ is given by $T_G = T(D_1) \times \dotsb T(D_m)$, and the Lie algebra of $T_G$ may be identified with $\C^{D_1} \op \cdots \op \C^{D_m}$.
  Then it is easy to show that the weights are just tuples of weights as above, i.e.,
  \begin{equation*}
    (e_{i_1} - e_{j_1}, \dotsc, e_{i_m} - e_{j_m})
  \end{equation*}
  with $i_k, j_k \in [D_k]$ for $k \in [m]$.
\end{exa}

Given the general setting as above, we can now define the following two parameters:

\begin{dfn}
  The \emph{weight margin} $\gamma(\pi)$ of the representation $\pi$ is defined as
  \begin{equation*}
    \gamma(\pi) = \min\{d(0, \conv \Gamma) : \Gamma \subseteq \Omega(\pi), \, 0 \not\in \conv \Gamma\}.
  \end{equation*}
  Here, $\conv \Gamma$ refers to the convex hull of $\Gamma \subseteq \Lie(T_G)^*$.
  The \emph{weight norm} $N(\pi)$ is defined by
  \begin{equation*}
    N(\pi) = \max \{ \norm{\omega} : \omega \in \Omega(\pi) \}.
  \end{equation*}
  The distance $d(\cdot,\cdot)$ and $\norm{\cdot}$ are defined in terms of the Hilbert-Schmidt inner product and identifying $\Lie(T_G)^* \cong \Lie(T_G) \subseteq \Mat_{n\times n}$.
\end{dfn}

While these parameters are somewhat abstract, we give a short justification for their appearance being natural.
As in \cref{eq:concrete kempf ness}, for a vector $0\neq v\in V$ consider the function
\begin{align*}
  F_v(g) := \log \norm{g \cdot v}.
\end{align*}
Considered as a function on the space of right cosets $K\backslash G$, this is known as the \emph{Kempf-Ness function} and it plays an important role in the general theory.
Its gradient at the identity coset generalizes \cref{eq:log grad} and goes by the following name:

\begin{dfn}\label{eq:moma}
  The \emph{moment map} $\mu\colon V \setminus \{0\} \to i \Lie(K)$ is defined by
  \begin{equation*}
    \mu(v) = \grad_{X = 0} F_v(e^X) = \grad_{X=0} \log \norm{e^X \cdot v}.
  \end{equation*}
  where $X \in i \Lie(K)$.
\end{dfn}

This is also a moment map for the $K$-action on the projective space $\PP(V)$ in the sense of symplectic geometry, which serves as a ``collective Hamiltonian'' for the action.
If we restrict to the case where $G$ is commutative, i.e., $G = T_G$, then observe that for $Y \in \Lie(G)$ and $v = \sum_{\omega \in \Omega(\pi)} v_\omega$ one has
\begin{equation*}
  e^Y \cdot v = \sum_{\omega \in \Omega(\pi)} e^{\omega(Y)} v_\omega,
\end{equation*}
and since the decomposition into weight spaces is orthogonal, we get
\begin{equation*}
  \norm{e^Y \cdot v}^2 = \sum_{\omega \in \Omega(\pi)} e^{2 \omega(Y)} \norm{v_\omega}^2.
\end{equation*}
From this expression, one can already see that if $G$ is commutative, then $F_v(e^Y) = \log \norm{e^Y \cdot v}$ is convex in $Y \in i \Lie(K)$,
and moreover that it is $2 N(\pi)^2$-smooth.
(This also holds for non-commutative $G$, as can in fact be deduced from the preceding.)
With the expression for $\norm{e^Y \cdot v}^2$ we can compute the moment map by
\begin{equation*}
\braket{\mu(v), Y} = \partial_{t=0} \log \norm{\exp{t Y} \cdot v} = \frac{1}{\norm{v}^2} \sum_{\omega \in \Omega(\pi)} \norm{v_\omega}^2 \omega(Y),
\end{equation*}
and we deduce that $\mu(v)$ is a \emph{convex} combination of the weights $\omega$ for which $v_\omega \neq 0$.
We observe now that the support $\supp v$, i.e., the set of $\omega$ such that $v_\omega \neq 0$, does not change when one acts with $G = T_G$.
This implies that if $v \in V \setminus \{0\}$ is such that the convex hull of its support does not contain $0$, then $\norm{\mu(g \cdot v)} \geq \gamma(\pi)$ for all $g \in G$.
Note that this also implies that $\mcf{v} = 0$: one can use a separating hyperplane between $0$ and $\conv(\supp v)$ to find a direction $Y \in i \Lie(K)$ such that $e^{t Y} \cdot v \to 0$ as $t \to \infty$.
Similar statements hold for non-commutative $G$, and we refer the interested reader to \cite{burgisser2019theory}.

We are now in a position to state a quantitative relationship between the norm of the moment map and the approximation ratio $\norm{\mcf{v}}/\norm{v}$ for $v \in V \setminus \{0\}$:

\begin{thm}[Non-commutative duality, {\cite[Thm.~1.17]{burgisser2019theory}}]\label{thm:nc duality}
  For $v \in V \setminus \{0\}$ with minimum norm vector $\mcf v$ (\cref{dfn:minimum norm vector}), we have
  \begin{equation*}
    1 - \frac{\norm{\mu(v)}}{\gamma(\pi)} \leq \frac{\norm{\mcf{v}}^2}{\norm{v}^2} \leq 1 - \frac{\norm{\mu(v)}^2}{4 N(\pi)^2}.
  \end{equation*}
\end{thm}

To prove \cref{thm:peps error conversion}, we still need to bound the parameters $\gamma(\pi)$ and $N(\pi)$ for our specific representations.

\begin{lem}
  \label{lem:weight margin bounds}
  For the action of $G=\GL(D_1) \times \dotsb \times \GL(D_m)$ on $V=\Mat_{D_1 \dotsb D_m \times D_1 \dotsb D_m}^d$, the weight norm~$N(\pi)$ is given by
  \[ N(\pi) = \sqrt{2 m}, \]
  and the weight margin~$\gamma(\pi)$ is lower bounded as
  \begin{equation*}
    \gamma(\pi) \geq \gamma,
  \end{equation*}
  where $\gamma$ is the constant defined in \cref{def:gamma}.
\end{lem}
\begin{proof}
  The expression for the weight norm follows directly from \cref{exa:peps weights}.

  For $m=1$, the lower bound on the weight margin follows from \cite[Thm.~6.21]{burgisser2019theory}: the representation is a quiver representation, where the quiver is given by one vertex with $d$ self-loops.
  For $m\geq2$, the lower bound on the weight margin follows from \cite[Thm.~6.10]{burgisser2019theory}.
\end{proof}

\begin{proof}[Proof of \cref{thm:peps error conversion}]
  This follows by combining \cref{thm:nc duality,lem:weight margin bounds}.
\end{proof}

Now that we know that \cref{eq:zeta error,eq:eps error} can be related to each other, we will relate these to \cref{eq:delta error}.
In the one direction, it is clear that \cref{eq:delta error} implies a small error in the sense of \cref{eq:zeta error}:
\begin{align*}
  \frac {\norm{S - \mcf T}} {\norm S} \leq \delta
\quad\Rightarrow\quad
  \frac {\norm{\mcf T}}{\norm S}
\geq 1 - \frac {\norm{\mcf T - S}}{\norm S}
\geq 1 - \delta
\end{align*}

In the remainder of this section we show that \cref{eq:eps error} implies a small error in the sense of \cref{eq:delta error}, closing the circle.
It is useful to make the following abbreviation for \cref{eq:grad norm sqr}, the gradient of the norm square function at the identity:
\begin{align*}
  \tilde\mu(S) := \nabla f_S(\vec I) = \left( \sigma_{k,1} - \sigma_{k,2}^T \right)_{k=1}^m
\in \Herm(D_1) \op \cdots \op \Herm(D_m),
\quad\text{where}\quad
\sigma := \ket S \bra S.
\end{align*}
We write $\tilde\mu$ and not $\mu$ to distinguish it from the gradient of the log-norm, as in \cref{eq:log grad,eq:moma}, but note that
\begin{align}\label{eq:mu tilde norm vs eps}
  \norm{\tilde\mu(S)}
= \eps \tr(\sigma)
= \eps \norm S^2,
\quad\text{where}\quad
\eps = \frac{1}{\tr\sigma} \sqrt{ \sum_{k=1}^m \norm*{\sigma_{k,1} - \sigma_{k,2}^T}^2 }
\end{align}
Then we will consider the gradient flow of $\norm{\tilde\mu(S)}^2 := \sum_{k=1}^m \norm{\rho_{k,1} - \rho_{k,2}^T}^2$:
\begin{align}\label{eq:gradflow}
  \begin{cases}
  S'(t)
  = - \nabla\norm{\tilde\mu}^2 \mleft(S(t)\mright) \\
  S(0) = S
  \end{cases}
\end{align}
We will see that the solution $S(t)$ to this ODE remains in the gauge orbit of $S$ and that it converges to a minimal canonical form~$\mcf S$ whose distance to $S$ in the sense of \cref{eq:delta error} can be controlled using \cref{eq:eps error}.

The study of the gradient flow for the norm square of the moment map was pioneered in seminal work by Kirwan~\cite{kirwan1984cohomology}, and it found widespread use in mathematics.
It was first proposed as an algorithmic tool in~\cite{walter2013entanglement,walter2014multipartite} in the context of the quantum marginal problem, and analyzed quantitatively in~\cite{kwok2018paulsen} (to resolve the Paulsen problem) and \cite{allenzhu2018operator} for the operator scaling action and then in~\cite{burgisser2019theory} for general reductive group actions.
While the following arguments work in complete generality, here we restrict to the gauge action of $G=\GL(D_1)\times\cdots\times\GL(D_m)$ since this is all we need.

We start by analyzing \cref{eq:gradflow}.
Existence and uniqueness of the solution~$S(t)$ of this ordinary differential equation on some maximal (possibly infinite) interval of definition $[0,t_{\max})$, where~$t_{\max} \in (0,\infty]$, follows from Picard--Lindel\"of theory.
Then one can prove the following lemma, cf.~\cite[Prop.~3.27 and its proof]{burgisser2019theory}:

\begin{lem}\label{lem:flow}
Let $S(t)$ be the solution to the dynamical system~\eqref{eq:gradflow}.
Then, for all $t \in [0,t_{\max})$, we have
\begin{enumerate}
  \item\label{it:grad1} $\partial_t \norm{\tilde\mu\bigl(S(t)\bigr)}^2 = -\norm{S'(t)}^2$.
  \item\label{it:grad2} $\partial_t \norm{S(t)}^2 = -8 \norm{ \tilde\mu(S(t)) }^2$.
  \item\label{it:grad3} $S(t) \in G \cdot S$, i.e., the solution remains in the $G$-orbit of $S$ at all times.
\end{enumerate}
\end{lem}
\begin{proof}
The first claim holds for any gradient flow.

Next, we note that, for all $\vec Y \in \Herm(D_1) \op \cdots \op \Herm(D_m)$,
\begin{align}\label{eq:mu via lie}
  \braket{\tilde\mu(S), \vec Y}
= \braket{\nabla f_S(\vec I), \vec Y}
= \frac12 \partial_{t=0} \norm*{ ( e^{tY_1},\dots,e^{tY_m} ) \cdot S }^2
= \braket{S, \Pi(\vec Y) S},
\end{align}
where $\braket{\vec X,\vec Y} = \sum_{k=1}^m \tr[X_k Y_k]$ and we denote by $\Pi(\vec Y)$ the Lie algebra action of $\vec Y$, which is defined by
\begin{align*}
  \Pi(\vec Y) S := \partial_{t=0} \left( ( e^{tY_1},\dots,e^{tY_m} ) \cdot S \right).
\end{align*}
By differentiating \cref{eq:mu via lie} with respect to~$S$ in some direction~$W\in V$ (an operation we denote by $D_W$),
\begin{align*}
  \braket{D_W \tilde\mu(S), \vec Y}
= \braket{W, \Pi(\vec Y) S} + \braket{S, \Pi(\vec Y) W}
= 2 \Re \braket{W, \Pi(\vec Y) S}.
\end{align*}
Accordingly, for all $W\in V$,
\begin{align*}
  D_W \norm{\tilde\mu(S)}^2
= 2 \braket{D_W \tilde\mu(S), \tilde\mu(S)}
= 4 \Re \braket{W, \Pi(\tilde\mu(S)) S}.
\end{align*}
Thus we have proved that the gradient of~$\norm{\tilde\mu}^2$ is given by the following clean formula:
\begin{align}\label{eq:inorbit}
  \nabla\norm{\tilde\mu}^2(S) = 4 \Pi(\tilde\mu(S)) S.
\end{align}
The second item follows from this and \cref{eq:mu via lie},
\begin{align*}
  \partial_t \norm{S(t)}^2
= 2 \braket{S(t), S'(t)}
= -2 \braket{S(t), \nabla\norm{\tilde\mu}^2(S(t))}
= -8 \braket{S(t), \Pi(\tilde\mu(S(t))) S(t)}
= -8 \norm{ \tilde\mu(S(t)) }^2.
\end{align*}
As \cref{eq:inorbit} states that $S'(t)$ is a tangent vector of the $G$-orbit through $S(t)$, the third item also follows.
\end{proof}

Using the preceding, the following key lemma shows that if $\mcf S\neq0$ then $\tilde\mu(S(t))\to0$ sufficiently quickly, without $S(t)$ moving too much.
Our argument follows~\cite{kwok2018paulsen}, which treats the case $m=1$.

\begin{lem}\label{key lemma}
Let $S(t)$ denote the solution of \cref{eq:gradflow} for a tensor $S(0) = S$ with $\mcf S\neq0$ (for some and hence for any minimal canonical form).
Consider any $\tau$ such that $\tilde\mu(S(\tau))\neq0$.
Then there exists
\begin{align*}
  \tau' \leq \tau + \frac{1}{4 \gamma \norm{\tilde\mu(S(\tau))}}.
\end{align*}
such that
\begin{align}
  \norm{\tilde\mu(S(\tau'))}^2 = \frac {\norm{\tilde\mu(S(\tau))}^2} 2
\end{align}
(in fact $\tau'$ is the first time such that this is true) and, moreover,
\begin{align}\label{eq:meh}
  \norm{S(\tau') - S(\tau)} 
  \leq \frac{1}{2\sqrt{2}} \sqrt{ \frac {\norm{\tilde\mu\bigl(S(\tau)\bigr)}} {\gamma} },
\end{align}
where $\gamma$ is the constant from \cref{def:gamma}.
\end{lem}
\begin{proof}
Suppose that $\tau'>\tau$ is such that
\begin{align}\label{eq:mu large}
  \norm{\tilde\mu(S(\tau'))}^2 > \frac{\norm{\tilde\mu(S(\tau))}^2}2.
\end{align}
By \cref{it:grad1} of \cref{lem:flow},
\begin{align*}
  \norm{\tilde\mu(S(t))}^2 > \frac{\norm{\tilde\mu(S(\tau))}^2}2 \qquad \forall t \in [\tau,\tau']
\end{align*}
and hence, by \cref{it:grad2} of the same lemma,
\begin{align*}
  \partial_t \norm{S(t)}^2 = -8 \norm{ \tilde\mu(S(t)) }^2 < -4 \norm{\tilde\mu(S(\tau))}^2 \qquad \forall t \in [\tau,\tau'].
\end{align*}
Accordingly,
\begin{align*}
  \norm{S(\tau')}^2 - \norm{S(\tau)}^2 < -4 (\tau' - \tau) \norm{\tilde\mu(S(\tau))}^2.
\end{align*}
On the other hand, using the lower bound in \cref{thm:peps error conversion} and \cref{eq:mu tilde norm vs eps},
\begin{align*}
  \norm{S(\tau')}^2 - \norm{S(\tau)}^2
&\geq \norm{S(\tau')_{\min}}^2 - \norm{S(\tau)}^2
= \norm{S(\tau)_{\min}}^2 - \norm{S(\tau)}^2 \\
&= \norm{S(\tau)}^2 \left( \frac{\norm{S(\tau)_{\min}}^2}{\norm{S(\tau)}^2} - 1 \right)
\geq - \frac{\norm{\tilde\mu(S(\tau))}}{\gamma}.
\end{align*}
Together, we find that for any $\tau'$ such that \cref{eq:mu large} holds, we must have
\begin{align*}
  \tau' < \tau + \frac{1}{4 \gamma \norm{\tilde\mu(S(\tau))}}.
\end{align*}
Accordingly, there must exist some minimal
\begin{align}\label{eq:tau bound}
  \tau' \leq \tau + \frac{1}{4 \gamma \norm{\tilde\mu(S(\tau))}}.
\end{align}
such that
\begin{align}\label{eq:gradsqr eq}
  \norm{\tilde\mu(S(\tau'))}^2 = \frac{\norm{\tilde\mu(S(\tau))}^2}2.
\end{align}
Moreover, for this $\tau'$ we have
\begin{align*}
  \norm{S(\tau') - S(\tau)}
&\leq \int_\tau^{\tau'} \norm{S'(t)} \, dt
= \int_\tau^{\tau'} \sqrt{ - \partial_t \norm{\tilde\mu\bigl(S(t)\bigr)}^2 } \, dt \\
&\leq \sqrt{ \int_\tau^{\tau'} - \partial_t \norm{\tilde\mu\bigl(S(t)\bigr)}^2 \, dt} \; \sqrt{\int_\tau^{\tau'} 1 \, dt} \\
&= \sqrt{ \norm{\tilde\mu\bigl(S(\tau)\bigr)}^2 - \norm{\tilde\mu\bigl(S(\tau')\bigr)}^2 } \, \sqrt{\tau'-\tau} \\
&\leq \frac{ \norm{\tilde\mu\bigl(S(\tau)\bigr)} }{\sqrt 2} \, \sqrt{\frac{1}{4 \gamma \norm{\tilde\mu(S(\tau))}}} \\
&= \frac{1}{2\sqrt{2}} \sqrt{ \frac {\norm{\tilde\mu\bigl(S(\tau)\bigr)}} {\gamma} },
\end{align*}
where we used the triangle inequality, then \cref{it:grad1} of \cref{lem:flow}, then the Cauchy-Schwarz inequality, and finally \cref{eq:tau bound,eq:gradsqr eq}.
\end{proof}

We now prove the desired relation between \cref{eq:eps error,eq:delta error}:

\begin{thm}\label{thm:eps to delta}
Let $T$ be a tensor with $\mcf T\neq0$ (for some and hence for any minimal canonical form) and let $S \in G \cdot T$.
Then there exists a minimal canonical form $\mcf T \in \overline{G \cdot T}$ such that
\begin{align*}
  \frac {\norm{S - \mcf T}} {\norm{S}} \leq \sqrt{\frac {2 \eps} {\gamma}}
\qquad\text{for}\qquad \eps := \frac{1}{\tr\sigma} \sqrt{ \sum_{k=1}^m \norm*{\sigma_{k,1} - \sigma_{k,2}^T}^2 },
\end{align*}
where $\gamma$ is the constant from \cref{def:gamma}.
\end{thm}
\begin{proof}
If $\mu(S) = 0$ then $\mcf T = S$ is a minimal canonical form of $T$ and there is nothing to prove.
Otherwise let us, for every $k\geq0$, denote by $\tau_k$ the first time when
\begin{align*}
  \norm{\tilde\mu(S(\tau_k))}^2 = \frac1{2^k} \norm{\tilde\mu(S)}^2,
\end{align*}
so $\tau_0 = 0$.
By \cref{key lemma},
\begin{align*}
  \tau_k
= \sum_{l=1}^k \left( \tau_l - \tau_{l-1} \right)
\leq \sum_{l=1}^k \frac{1}{4 \gamma \norm{\tilde\mu(S(\tau_{l-1}))}}
= \frac{1}{4 \gamma} \sum_{l=1}^k \frac1{\sqrt{\frac1{2^{l-1}}} \norm{\tilde\mu(S)}}
\leq \frac{2^{k/2}}{\gamma \norm{\tilde\mu(S)}}
\end{align*}
In particular, $\tilde\mu(S(t)) \to 0$ as $t\to\infty$, since we know from \cref{it:grad1} of \cref{lem:flow} that $\norm{\tilde\mu(S(t))}^2$ is monotonically decreasing.

Next, we prove that the subsequence $S(\tau_k)$ converges to a minimal canonical form of~$T$ with the desired properties.
We first show that the $S(\tau_k)$ form a Cauchy sequence.
Indeed, for any $k \leq l$, using \cref{key lemma},
\begin{align*}
  \norm{S(\tau_k) - S(\tau_l)}
&\leq \sum_{m=k+1}^l \norm{S(\tau_m) - S(\tau_{m-1})}
\leq \sum_{m=k+1}^l \frac1{2 \sqrt{2}} \sqrt{ \frac {\norm{\tilde\mu\bigl(S(\tau_{m-1})\bigr)}} { \gamma} } \\
&= \frac1{2 \sqrt{2}} \sqrt{ \frac {\norm{\tilde\mu\bigl(S\bigr)}} { \gamma} } \sum_{m=k+1}^l \sqrt{\frac1{2^{m-1}}}
\leq \sqrt{ \frac {2 \norm{\tilde\mu\bigl(S\bigr)}} {\gamma} } \sqrt{\frac1{2^k}},
\end{align*}
which shows that indeed $S(\tau_k)$ is a Cauchy sequence.
If we denote by $S'$ its limit, then $T' \in \overline{G \cdot S} = \overline{G \cdot T}$ (by \cref{it:grad3} of \cref{lem:flow}) and hence $T'\neq0$ (since $\mcf T \neq 0$ by assumption).
Moreover, $\tilde\mu(T')=0$ by the above, hence~$T'$ is a minimal canonical form of $T$.
Finally,
\begin{align*}
  \norm{S - T'}
= \lim_{l\to\infty} \norm{S(\tau_0) - S(\tau_l)}
\leq \sqrt{ \frac {2 \norm{\tilde\mu\bigl(S\bigr)}} {\gamma} }
= \norm S \sqrt{ \frac {2\eps} {\gamma} }
\end{align*}
using the preceding estimate and \cref{eq:mu tilde norm vs eps}
\end{proof}

By combining \cref{cor:peps scaling explicit guarantee,thm:eps to delta} it follows that using the first-order algorithm in \cref{algo:PEPS grad descent} with $\eps := \gamma \delta^2/2$ one can in time $\poly(\frac1\gamma, \frac1\delta, \text{input size})$ obtain a group element $\vec g \in G$ such that the tensor $S := \vec g \cdot T$ satisfies \cref{eq:delta error}, i.e.,
\begin{align*}
  \frac {\norm{S - \mcf T}} {\norm S} \leq \delta.
\end{align*}
In the next section we will see that the dependence on $\delta$ can be improved to $\log(1/\delta$), see \cref{cor:second order}.

\subsection{Second-order algorithm}
As promised earlier, there is also a second numerical method that one can use to approximate normal forms in our setting.
This is a more sophisticated \emph{second-order} method, which uses information about the Hessian of the Kempf--Ness function $F_v$ to determine the direction in which to move (as is done for instance in Newton's method), whereas the first-order method discussed in \cref{subsec:first order,algo:PEPS grad descent} before only use information about the gradient.

In \cite[Algo.~5.2]{burgisser2019theory}, a ``box-constrained Newton method'' is analyzed, which uses Newton steps constrained to a constant-sized box to make progress in the objective.
It naturally minimizes the norm of the resulting vector (as opposed to the size of the gradient).
Its guarantees applied to our setting are as follows:

\begin{thm}[{\cite[Thm.~8.12]{burgisser2019theory}}]\label{thm:second order}
  Let $T \in \Mat_{D_1 \dotsb D_m \times D_1 \dotsb D_m}^d$ be a tensor such that $\mcf T\neq0$ (for some and hence for any minimal canonical form).
  Assume that the entries of $T$ are in $\Q[i]$ and given by storing the numerators and denominators in binary.
  Then there exists an algorithm that, given $T$ and $0<\zeta<1$, returns a group element $\vec g \in \GL(D_1) \times \cdots \times \GL(D_m)$ such that the tensor $S := \vec g \cdot T$ satisfies $\norm S \leq \norm T$ and
  \begin{equation*}
    \log \frac{\norm{S}}{\norm{\mcf{T}}} \leq \zeta
  \qquad\text{and hence}\qquad
    \frac{\norm{\mcf{T}}}{\norm{S}} \geq 1 - \zeta
  \end{equation*}
  in time $\poly(\gamma^{-1}, D_1, \dotsc, D_m, \log(1/\zeta), \inpsize{T})$, where $\gamma$ is defined in \cref{def:gamma}, and $\inpsize{T}$ is the total number of bits used to represent~$T$.
\end{thm}

By combining \cref{thm:second order} with the results of \cref{subsec:relerr}, we arrive at the following result which was stated informally as \cref{prp:algorithm} in the introduction.

\begin{cor}\label{cor:second order}
  Let $T \in \Mat_{D_1 \dotsb D_m \times D_1 \dotsb D_m}^d$ be a tensor such that $\mcf T\neq0$ (for some and hence for any minimal canonical form).
  Assume that the entries of $T$ are in $\Q[i]$ and given by storing the numerators and denominators in binary.
  Then there exists an algorithm that, given $T$ and $0<\delta<1$, returns a group element $\vec g \in \GL(D_1) \times \cdots \times \GL(D_m)$ such that the tensor $S := \vec g \cdot T$ satisfies satisfies $\norm S \leq \norm T$ and
  \begin{align*}
    \frac{\norm{S - \mcf T}}{\norm S} \leq \delta,
  \end{align*}
  in time $\poly(\gamma^{-1}, D_1, \dotsc, D_m, \log(1/\delta), \inpsize{T})$, where $\gamma$ is defined in \cref{def:gamma}, and $\inpsize{T}$ is the total number of bits used to represent~$T$.
\end{cor}
\begin{proof}
Apply the algorithm of \cref{thm:second order} with
\begin{align}\label{eq:choice of zeta}
  \zeta := \frac{\gamma^2}{64 m} \delta^4
\end{align}
to obtain in the stated runtime a group element~$g\in G$ such that the tensor $S := g \cdot T$ satisfies $\norm S \leq \norm T$ and
\begin{align}\label{eq:from second order}
  \frac{\norm{\mcf{T}}}{\norm{S}} \geq 1 - \zeta
\qquad\text{and hence}\qquad
  \frac{\norm{\mcf{T}}^2}{\norm{S}^2} \geq 1 - 2\zeta.
\end{align}
We now check that $S$ satisfies the desired condition.
First, by \cref{thm:peps error conversion}, for $\sigma=\ket S\bra S$ we have that
\begin{align*}
  \frac{\norm{\mcf T}^2}{{\norm S}^2} \leq 1 - \frac{\eps^2}{8m}
\qquad\text{for}\qquad \eps := \frac{1}{\tr\sigma} \sqrt{ \sum_{k=1}^m \norm*{\sigma_{k,1} - \sigma_{k,2}^T}^2 },
\end{align*}
and hence, using \cref{eq:from second order},
\begin{align}\label{eq:eps upper bound in second order}
  \eps
\leq \sqrt{ 8m \left( 1 - \frac{\norm{\mcf T}^2}{{\norm S}^2} \right) }
\leq 4 \sqrt{ m \zeta }.
\end{align}
Finally, \cref{thm:eps to delta} implies that
\begin{align*}
  \frac {\norm{S - \mcf T}} {\norm{S}} \leq \sqrt{\frac {2 \eps} {\gamma}}
\qquad\text{and hence}\qquad
  \frac {\norm{S - \mcf T}} {\norm{S}}
\leq \sqrt{\frac {2 \eps} {\gamma}}
\leq \sqrt{\frac {8 \sqrt{ m \zeta }} {\gamma}}
\leq \delta,
\end{align*}
where used \cref{eq:eps upper bound in second order} and our choice of $\zeta$ in \cref{eq:choice of zeta}.
This concludes the proof.
\end{proof}

\begin{rem}\label{rem:abserr}
While \cref{cor:second order} uses relative $\ell^2$-error, which is most natural, we can also obtain a guarantee in absolute error, say
\begin{align*}
  \norm{S - \mcf T} \leq \delta',
\end{align*}
by applying \cref{cor:second order} with $\delta < \min(1, \delta' / \norm T)$.
As the second-order algorithm scales polynomially in~$\log(1/\delta)$, this still runs in time $\poly(\gamma^{-1}, D_1, \dotsc, D_m, \log(1/\delta'), \inpsize{T})$.
\end{rem}

\section{Conclusion and outlook}\label{sec:outlook}
The current work is a theoretical one, proposing a new canonical form and proving some of its key properties.
The fact that the minimal canonical form is rigorous in the sense that it can be proven to always exist as well as satisfy the basic properties discussed in \cref{sec:peps} sets it apart from other heuristic approaches \cite{phien2015fast,evenbly2018gauge}.
Besides this, we hope that the minimal canonical form will be of practical use in tensor network algorithms.
Below we outline four potential directions for application.
Detailed numerical study will be required to confirm the usefulness of these suggestions.

\begin{enumerate}
\item \textbf{Truncation of bond dimensions.}
In many tensor network algorithms \emph{truncation} of the bond dimension is a crucial step.
This is especially the case for ground state finding algorithms based on imaginary time evolution (Time Evolving Block Decimation, TEBD) in which each step consist of applying an operator to the tensor network which increases the ground state approximation accuracy but also the bond dimension, and then truncating the bond dimension.
One is given a tensor $T$ with a certain bond dimension $D$, and one would like to find a tensor $T'$ with a prescribed bond dimension $D' < D$ such that the tensor network state using $T$ is approximated as accurately as possible by the tensor network state using $T'$.
In one spatial dimension, for MPS, there is a natural way to do this using canonical forms.
For instance, one may use the left canonical form, in which case the reduced state $\rho_2$ on the right virtual dimension is maximally mixed.
Then one truncates to the subspace spanned by the eigenvectors of the $D'$ largest eigenvalues of the reduced state $\rho_1$ on the left virtual dimension.

The bond dimension truncation scheme for MPS is both computationally efficient and gives an optimal approximation given a prescribed bond dimension.
For two-dimensional PEPS there is no truncation scheme known which has both these desirable properties, which is closely related to the lacking of the equivalent of a left or right canonical form.
Various methods exist \cite{lubasch2014unifying, jiang2008accurate}, see for instance \cite{ran2020tensor} for an overview of different methods.
While these methods perform well in practice, in most cases good theoretical understanding is lacking.
Here, we propose the following natural truncation scheme: given a tensor $T$, compute its minimal canonical form $S$.
Then truncate to the subspace spanned by the eigenvectors corresponding to the $D'$ largest eigenvalues.

This proposal leads various questions which should be addressed in follow-up work.
First of all, it would be interesting to use such a truncation method in existing PEPS algorithms and study the performance of such schemes numerically.
Secondly, as our methods are designed for uniform (translation-invariant) systems one would hope that they are also of use to iPEPS methods, where precisely the absence of a canonical form has led to heuristic approaches to gauge-fixing \cite{phien2015infinite,phien2015fast} which work well in practice.
We would like to emphasize that especially the (non-rigorously defined) canonical form in \cite{phien2015fast} is fairly close in spirit to the minimal canonical form: it is defined by a condition similar (but different) to the characterization in \cref{thm:symmetric virtual rdm}.
This canonical form has been shown to indeed improve convergence of imaginary time evolution algorithms, which offers some hope for the prospect of using the minimal canonical form for truncation purposes.
Finally, a potential advantage of truncation schemes based on the minimal canonical form is that one could attempt to the framework of geometric invariant theory to prove that such a truncation scheme has good theoretical properties.

\item \textbf{Numerical stability.} Using minimal canonical forms in variational algorithms may be helpful, since appropriate gauge fixing is known to enhance the stability of variational algorithms \cite{lubasch2014algorithms, phien2015infinite}.

\item \textbf{Boundary-based approaches.}
PEPS have a very useful and explicit bulk-boundary correspondence \cite{cirac2011entanglement}, which allows one to map bulk properties in a region $R$ to properties of the associated boundary state $\rho_R$, defined essentially as the reduced density matrix in the virtual indices of the PEPS tensor $\ket{T_R}$ obtained after blocking the original PEPS tensor $T$ in the given region $R$. The key insight of \cite{cirac2011entanglement}, formalized later in \cite{kastoryano2019locality, perez2020locality}, is that if one interprets $\rho_R$ as a Gibbs state $\rho_R=e^{-H_E}$, the properties of $H_E$ (the so-called \emph{entanglement Hamiltonian}) encode the properties of the bulk of the system. This has led for instance to new numerical methods to detect topological phase transitions \cite{schuch2013topological}. Since $H_E$ and $\rho_R$ live in the virtual Hilbert spacec, it is crucial for this approach to be meaningful that the only gauge freedom one considers comes from unitaries, which do not change any of the relevant properties of $H_E$ or $\rho_R$, rather than arbitrary invertible matrices.
This is precisely what is ensured by working with the minimal canonical form.

\item  \textbf{Privacy in PEPS-based machine learning algorithms.}
Tensor networks, and PEPS and MPS in particular, have been used as variational Ans\"{a}tze in machine learning contexts \cite{stoudenmire2016supervised, cichocki2017tensor}.
This has the appeal that one can import known optimization techniques in condensed matter problems to machine learning.
Another potential advantage, compared to neural network-based approaches, lies in a higher interpretability; it is precisely the characterization of global properties in the local tensors of a tensor network which explains its success in quantum many-body problems.
In \cite{pozas2022physics} a new potential advantage of tensor networks in a machine learning context has been proposed, which we will now explain briefly.
There are two possible ways to look at a trained neural network or tensor network: as a black box in which one has only access to the input-output relation or as a \emph{white box} in which all internal parameters are provided. It is shown in \cite{pozas2022physics}, with machines trained in real data bases with medical records, that those internal parameters can reveal sensitive information from the training data set which however are not contained in the black-box picture. This white-box versus black-box scenario is the underlying problem behind obfuscation protocols\footnote{Though the inherent continuous nature of the variables makes the problem slightly different in this case.} and it is well known there that the perfect solution comes from the existence of a well-defined canonical form for the white-boxes that maps them one-to-one to the set of black boxes.
The basic idea in \cite{pozas2022physics} is that this can be done in MPS by defining a new canonical form which selects analytically and uniquely an element for each orbit of a normal MPS. However, as it is also discussed in \cite{pozas2022physics}, a way of sampling uniformly on all possible white-box representations of the same black-box function could equally do the job.

The minimal canonical form gives a way to extend this idea trivially to general PEPS. If the presentation (white-box) of the PEPS obtained in the training process is its minimal canonical form, sampling uniformly on all possible white-boxes amounts to sampling with the Haar measure on the unitary group, which can easily be done (as opposed to sampling on the whole general linear group). It is an interesting open question to see how this idea works in practice for PEPS. For MPS it is shown in \cite{pozas2022physics} that privacy improvements in practice are indeed dramatic.
\end{enumerate}

As alluded to in \cref{sec:tilings} another natural direction of inquiry is to find physically relevant models where there is topological order which is only revealed on manifolds other than a torus, and see how this relates to the minimal canonical form.
Finally, it would be interesting to connect to recent approaches that apply techniques from algebraic geometry and \emph{algebraic complexity theory} \cite{burgisser2013algebraic} to tensor network theory, for instance \cite{christandl2020tensor,christandl2021optimization}.
There are also various concrete follow-up questions concerning properties of the minimal canonical form and generalizations.

\begin{enumerate}
\item \textbf{Non-uniform PEPS} In this work we have mainly restricted to \emph{uniform} PEPS, where we consider contractions of copies of a single identical tensor. We also saw one example with non-uniform tensors, for MPS in \cref{sec:non ti mps}. In that case, we were able to recover the usual theory of canonical forms for MPS with open boundary conditions.
Clearly, an interesting direction for future research is to investigate generalizations of the minimal canonical form to non-uniform PEPS.
In this case we consider a fixed graph $\Gamma = (V,E)$, with a collection of tensors $(T_v)_{v \in V}$ at each vertex and where we contract along the edges $E$.
We now have a group $\GL(D_e)$ acting on each edge $e$ in the graph, so the full gauge group $G$ is the product over all edges $e \in E$ of these groups.
This setup is very similar to the one described in \cref{sec:non ti mps}.
We would like to formulate an appropriate minimization problem over a group orbit.
There are two obvious ways to approach this.
The first option is to minimize
\begin{align*}
  \sum_{v \in V} \norm{\vec g \cdot T_v}^2
\end{align*}
and define a minimal canonical form $\mcf{((T_v)_{v \in V})}$ as satisfying
\begin{align*}
  \mcf{((T_v)_{v \in V})} = \argmin \left\{ \sum_{v \in V} \norm{S_v}^2 : (S_v)_{v \in V} \in \overline{G \cdot (T_v)_{v \in V}} \right\}
\end{align*}
In the case where all tensors are equal, this should reduce to the minimal canonical form for uniform PEPS.
A second option (which is similar to the MPS construction in \cref{sec:non ti mps}) would be to consider for each edge $e$ the tensor network state $\ket{T_e}$ where we have contracted all edges except $e$.
We have a group action of $\GL(D_e)$ on this state, and we may minimize over its orbit.
We will report on these generalizations in future work.
\item \textbf{Algorithms for deciding gauge equivalence} While we have addressed the issue of computing a minimal canonical form for a given tensor, we have not extensively discussed algorithms for deciding whether two tensors $S$ and $T$ are gauge equivalent.
One approach is given by \cref{thm:fundamental-theorem}: one may simply check that $\ket{S_{\vec \pi}} = \ket{T_{\vec \pi}}$ for all $\vec \pi \in S_n^{m}$ with $n \leq n_{\max} = \exp(\mathcal O(mD^2 \log(mD)))$ (or in the case of MPS, for $n \leq D^2$).
However, an alternative strategy is as follows.
By \cref{thm:oci peps}, it suffices to first compute minimal canonical forms $\mcf{S}$ and $\mcf{T}$ (for which we have already provided algorithms) and then determine whether these are related by \emph{unitary} gauge transformations (which is rather nontrivial).
For $m = 1$, this strategy has been implemented in \cite{allenzhu2018operator}, while for $m\geq2$ we defer to future work.
\item \textbf{Computational complexity.}
It would be interesting to relate the computation of minimal canonical forms and of checking gauge equivalence to other orbit problems that have recently been studied intensely in the theoretical computer science literature, in order to get a better understanding of the computational complexity of the problem (see \cite{burgisser2019theory} and references therein).
\end{enumerate}

\section*{Acknowledgements}
D.~Perez-Garcia acknowledges support by the Spanish Ministry of Science and Innovation (``Severo Ochoa Programme for Centres of Excellence in R\&D'' CEX2019-000904-S and grant  PID2020-113523GB-I00), the Spanish Ministry of Economic Affairs and Digital Transformation (project QUANTUM ENIA, as part of the Recovery, Transformation and Resilience Plan, funded by EU program NextGenerationEU), Comunidad de Madrid (QUITEMAD-CM P2018/TCS-4342), and the CSIC Quantum Technologies Platform PTI-001.
H.~Nieuwboer and M.~Walter acknowledge NWO grant OCENW.KLEIN.267.
M.~Walter also acknowledges support by the Deutsche Forschungsgemeinschaft (DFG, German Research Foundation) under Germany's Excellence Strategy - EXC\ 2092\ CASA - 390781972, by the BMBF through project QuBRA, and by the European Research Council~(ERC) through ERC Starting Grant 101040907-SYMOPTIC.

\bibliographystyle{alpha}
\bibliography{peps-scaling}

\end{document}